\documentclass[10pt, aps, pra, amsmath, amssymb, amsfonts, floatfix, twocolumn, superscriptaddress]{revtex4-2}

\usepackage[T1]{fontenc}
\usepackage[utf8]{inputenc}
\usepackage{placeins}
\usepackage{graphicx}
\usepackage{bm}
\usepackage[breaklinks=true]{hyperref}
\usepackage{textgreek}
\usepackage{comment}
\hypersetup{
colorlinks=true,
linkcolor=blue,
citecolor=blue,
urlcolor=blue
}
\usepackage[dvipsnames]{xcolor}

\newcommand{\sectionmajor}[1]{
\vspace{1em}
\noindent
{\large\textbf{#1}}
}
\newcommand{\sectionminor}[1]{
\vspace{1em}
\noindent
{\textbf{#1}}
}

\begin{document}

\title{Pair beams unlock beyond‑terawatt attosecond free‑electron laser pulses}

\author{\c{C}a\u{g}r{\i} Erciyes}
\email{cagri.erciyes@mpi-hd.mpg.de}
\affiliation{Max-Planck-Institut f\"{u}r Kernphysik, Saupfercheckweg 1, 69117 Heidelberg, Germany}

\author{Christoph~H.~Keitel}
\affiliation{Max-Planck-Institut f\"{u}r Kernphysik, Saupfercheckweg 1, 69117 Heidelberg, Germany}

\author{Matteo~Tamburini}
\email{matteo.tamburini@mpi-hd.mpg.de}
\affiliation{Max-Planck-Institut f\"{u}r Kernphysik, Saupfercheckweg 1, 69117 Heidelberg, Germany}

\date{\today}

\begin{abstract}
\noindent
Free-electron lasers (FELs) generate the brightest coherent X-ray pulses available, enabling atomic-resolution and femtosecond-timescale studies across physics, chemistry, and biology. Realising their full potential at extreme peak powers and attosecond pulse durations critically depends on sustaining coherent gain across the full bunch length. Yet, the quasi-static longitudinal space-charge field in the ultrahigh-current regime imprints a slice-dependent energy detuning that quenches gain growth, so that current schemes typically sustain efficient lasing only across a limited fraction of the bunch. Here we demonstrate that a quasi-neutral electron-positron pair beam cancels this self-field and enables full-bunch high-gain lasing in ultracompressed beams without external compensation. Three-dimensional particle-in-cell simulations in a single-pass, untapered undulator confirm the mechanism across operating regimes: in the soft X-ray regime, the pair beam reaches $1.85\,\mathrm{TW}$ at $345\,\mathrm{as}$ with enhanced odd-harmonic emission and improved spatial coherence, while the electron-only beam fails to saturate; and a high-harmonic pair-cascade configuration yields ${\sim}10\,\mathrm{TW}$ in isolated ${\sim}3.5\,\mathrm{as}$ spikes with coherent amplification extending to photon energies of ${\sim}177\,\mathrm{keV}$. These results establish a new operating regime for ultrahigh-power attosecond light sources and open a direct route to coherent gamma-ray emission ($\geq\!100\,\mathrm{keV}$) currently inaccessible to magnetic-undulator FELs, with broad implications for ultrafast structural, electronic, and nuclear sciences.
\end{abstract}

\maketitle


\begin{figure*}[!t]
\centering
\includegraphics[width=0.99\linewidth]{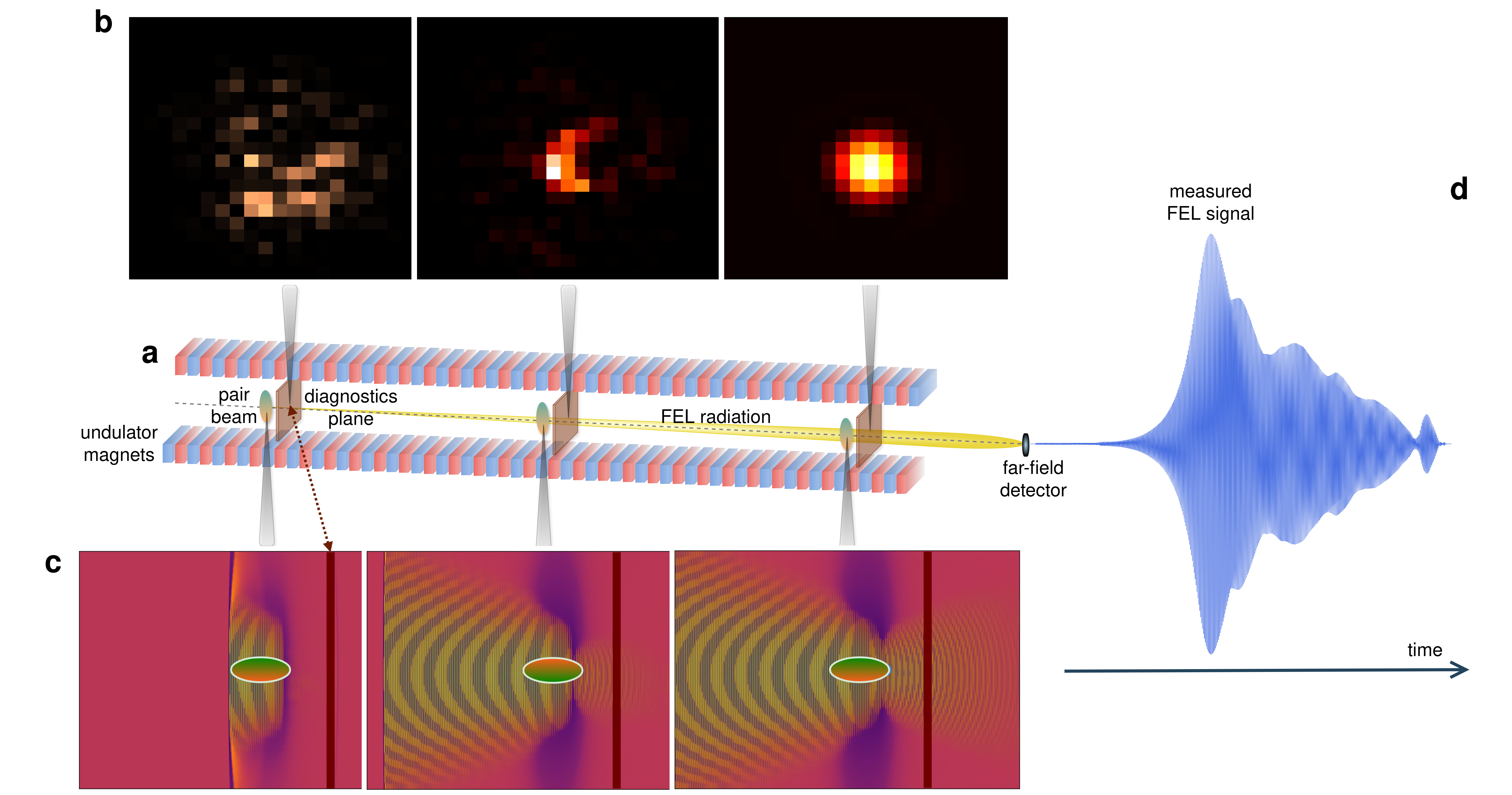}
\caption{\textbf{SASE FEL simulation scheme and diagnostic layout for the $\mathrm{e^-/e^+}$ pair beam.}
\textbf{a}~Schematic of the single-pass undulator stage.
\textbf{b}~Transverse profile of the Poynting flux amplitude, sampled in the laboratory frame at three stations along the undulator axis corresponding to initialisation, linear gain, and saturation.
\textbf{c}~Longitudinal profile of the electric field amplitude recorded in the bunch rest frame at the same three stations; the one-to-one correspondence with panel~\textbf{b} follows directly from the inverse Lorentz transformation.
\textbf{d}~Far-field radiation signal amplitude at the downstream detector.
Together, panels \textbf{b}--\textbf{d} form a complementary diagnostic set (near-field intensity, rest-frame field structure, and far-field emission) used consistently across all FEL configurations in this work.}
\label{fig:FEL_scheme}
\end{figure*}

\sectionmajor{Introduction}
\smallskip

\noindent
Free-electron lasers (FELs) generate bright, coherent, and ultrashort light from the infrared to hard X-rays. In both self-amplified spontaneous emission (SASE) and seeded operation, performance hinges on maintaining the resonance condition across each accelerator section, from injector through undulator. Recent advances target transform-limited bandwidths, attosecond timing, and terawatt (TW)-class peak power for time-resolved studies on electronic and nuclear timescales~\cite{yan2024terawatt, franz2024terawatt, Lentrodt2025Toward, Maroju2020AttosecondPulseShaping, Mayer2022FollowingExcitedStateChemicalShifts, Li2026NanoscaleUltrafastLatticeModulation}. One of the key obstacles at high peak current is the steady (DC) component of the longitudinal space-charge (LSC) field, which imprints a slice-dependent energy detuning acting as a linear chirp across the bunch (see Methods). This detuning remains benign only when lasing is confined to a narrow, low-charge window; beyond it, gain is suppressed and saturation power is capped. Modern attosecond TW schemes accordingly mitigate the resonance problem through either selective beam manipulation techniques such as fresh-slice and spoiler-based gating~\cite{duris2020tunable, lutman2018high}, or through active resonance control modes, such as chirp-taper compensation, cascaded superradiant amplification, and self-chirping~\cite{saldin2006chirptaper, franz2024terawatt, yan2024terawatt}. These approaches often restrict efficient lasing to only a limited fraction of the available bunch charge, which can constrain the attainable pulse power.

Eliminating the DC field at the source, rather than compensating it downstream, opens a direct path to ultrashort, high-contrast pulses at extreme peak power, including in single-spike operation~\cite{huang2017generating, duris2020tunable}. We pursue this route using a quasi-neutral electron-positron ($\mathrm{e^-/e^+}$) pair beam: charge neutralisation cancels the DC self-field, stabilises the resonance across the full bunch length, and preserves high peak current in every lasing slice without external gating. Building on our earlier laser-wiggler study~\cite{quin2025broadband}, practically limited to the XUV regime, the present work transplants the concept into the experimentally established magnetic-undulator, high-gain SASE regime, where DC space-charge chirp directly limits lasing under extreme compression. This makes quasi-neutrality a resonance-control mechanism independent of operating wavelength, extending pair-beam advantages across the full FEL spectrum while enhancing odd-harmonic output by orders of magnitude, beyond the terawatt threshold. At the theoretical level, we generalise high-gain FEL theory by retaining the finite-bunch longitudinal self-field as a coherent detuning term, an effect typically neglected in standard 1D treatments.

By cancelling the DC chirp, the pair beam removes the mechanism that truncates exponential gain at high current, enabling full-bunch lasing without charge loss through gating. Residual performance limits then arise from incoherent energy spread and transverse emittance, expanding the simultaneous peak-power and pulse-duration operating space in a controlled manner. Depending on the bunch length relative to the cooperation length, this supports either isolated single-spike or full-bunch spike-train emission. The same cancellation unlocks a beam geometry that has remained inaccessible for conventional FELs: the transversely extended, pancake-shaped bunch ($\sigma_\perp/\sigma_z \gg 1$, where $\sigma_\perp$ and $\sigma_z$ denote the transverse and longitudinal rms bunch sizes, respectively), which is well studied in high-brightness injector physics~\cite{serafini1997short, moody2009longitudinal}. In electron-only operation such bunches imprint a frozen, head-to-tail energy chirp that degrades brightness and quenches FEL gain before saturation~\cite{campbell2020analysis, hernandez2004longitudinal}; in a pair beam the DC field cancels by construction, enabling stable, gating-free operation at otherwise prohibitive peak currents. Optically, the pancake geometry corresponds to a larger Rayleigh range and beta function than a tightly focused beam, trading modest diffraction and coupling penalties for higher peak current density and, when emittance and mode overlap are preserved, an increased effective 3D Pierce parameter~\cite{huang2007review, Xie1995PAC3D}, thus yielding strongly bunched, compressed pulses with an enhanced odd-harmonic ladder within compact, few-metre undulator sections.

We test these advantages self-consistently by simulating the FEL interaction in the bunch rest frame with a Lorentz-boosted particle-in-cell (PIC) code. The diagnostic framework, illustrated in Fig.~\ref{fig:FEL_scheme}, comprises near-field radiation envelopes and power growth curves, far-field power spectra, bunching-factor evolution, and the temporal pulse structure at saturation. All configurations use a single-stage planar undulator without inter-module quadrupole lattices, reaching saturation within $\lesssim 15\,\mathrm{m}$. We focus on the short-bunch, ultrahigh-peak-current regime where the DC LSC chirp is comparable to the FEL gain bandwidth. Beyond the X-ray regime, the scheme opens a route to coherent $\gamma$-ray emission of direct interest for photonuclear reactions, isotope production, and high-contrast imaging~\cite{nedorezov2021nuclear}, with post-processed spectral narrowing techniques potentially enabling nuclear-resonance spectroscopy~\cite{heeg2017spectral}.

Experimental progress on several fronts supports the near-term feasibility of this concept. Pair beams are now routinely produced via the Bethe-Heitler mechanism with improving yield, brightness, and energy~\cite{chen2015scaling, arrowsmith2024laboratory, streeter2024narrow, noh2024charge}, while advances in beam transport and FEL-informed lattice design are pushing bunch compression, emittance, and energy spread toward FEL-grade targets~\cite{winkler2025active, wang2016high, habib2023attosecond, emma2025experimental}. The co-propagation precision required, femtosecond timing and micrometre-level transverse overlap, is closely analogous to that already demonstrated in PWFA driver-witness beamlines~\cite{Litos2014Nature, Yakimenko2019PRAB, Aschikhin2016NIMA} and cavity-based XFEL operation~\cite{Rauer2026cavity}, with positron transport established in dedicated PWFA lines~\cite{Corde2015Nature, Gessner2016NatCommun}. At FACET-II, successive compression stages have already produced ultrahigh-current femtosecond spikes~\cite{emma2025experimental}, directly demonstrating the compression performance required. Accordingly, all simulation parameters below lie within demonstrated or actively targeted performance envelopes of modern XFELs and advanced accelerators~\cite{habib2023attosecond, Yakimenko2019PRAB, emma2025experimental, prat2023x, prat2019generation, tomin2021accurate, rosenzweig2019next}.


\sectionmajor{Results}

\begin{figure*}[!t]
\centering
\includegraphics[width=0.99\linewidth]{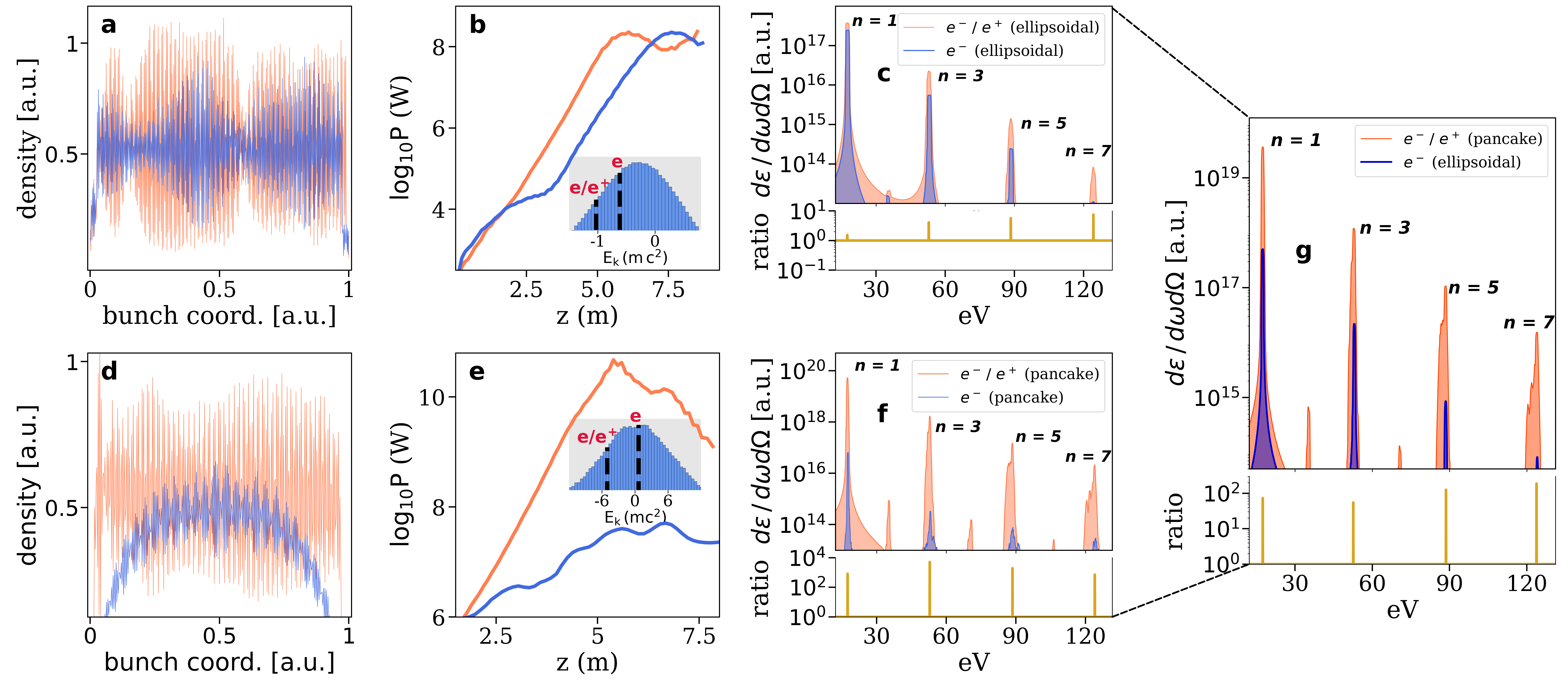}
\caption{\textbf{Ellipsoidal versus pancake bunch geometry in the UV FEL: PIC simulation results.}
All panels compare the quasineutral $\mathrm{e^-/e^+}$ pair beam (orange) with the $\mathrm{e^-}$-only beam (blue); the top row (\textbf{a}--\textbf{c}) shows the ellipsoidal bunch ($L_z = 100\,\mu\mathrm{m}$, $\sigma_\perp = 60\,\mu\mathrm{m}$) and the bottom row (\textbf{d}--\textbf{f}) the pancake bunch ($L_z = 10\,\mu\mathrm{m}$, $\sigma_\perp = 200\,\mu\mathrm{m}$).
\textbf{a,~d}~Longitudinal macro-particle density profiles at saturation, used as a proxy for microbunching extent.
\textbf{b,~e}~FEL power growth curves; insets show the kinetic energy change distribution at saturation, with dashed lines marking the mean energy shift per species, confirming synchronous energy transfer to the radiation field.
\textbf{c,~f}~Far-field power spectra up to the 7th harmonic with odd-harmonic intensity ratios below each spectrum. The legend in \textbf{c} and \textbf{f} applies to all panels in the respective row.
\textbf{g}~Cross-geometry comparison of the far-field spectrum between the $\mathrm{e^-}$-only ellipsoidal and $\mathrm{e^-/e^+}$ pancake beams, highlighting the harmonic enhancement from DC-field neutralisation.}
\label{fig:UV}
\end{figure*}

\sectionminor{Full bunch lasing with quasineutral pancake-shaped beams.}
\noindent
In all scenarios, the electron-only beam carries charge $Q_{\rm tot}$ while the pair beam carries $Q_{\rm tot}/2$ of each species, keeping the total number of radiating particles equal. Because electrons and positrons radiate constructively in a planar undulator (see Methods), the effective current entering the Pierce parameter is identical for both beams, $I_{\rm eff}^{\rm pair} = I_{\rm eff}^{e^-{\rm only}} = Q_{\rm tot}c/L_b$, so all observed gain differences arise solely from DC-field suppression.

We first establish the mechanism with a FLASH-like UV configuration: beam energy $350\,\mathrm{MeV}$, total charge $400\,\mathrm{pC}$, undulator period $\lambda_u \approx 31\,\mathrm{mm}$, $K = 1.5$, and fundamental wavelength $\lambda_r \approx 70\,\mathrm{nm}$. Conventional $\mathrm{e^-}$-only operation with an ellipsoidal bunch (encompassing round and prolate shapes with the criteria $\sigma_\perp/\sigma_z \lesssim \mathcal{O}(1)$) approaches GW-level peak powers and serves as our reference~\cite{schreiber2015flash, ackermann2007flash}. To isolate the role of LSC, the same total charge is then reshaped into a pancake bunch by compressing the bunch length by one order of magnitude while extending the transverse size proportionally.

Figure~\ref{fig:UV} compares the pair beam (orange) with the $\mathrm{e^-}$-only beam (blue) through longitudinal density profiles as a proxy for microbunching (Fig.~\ref{fig:UV}\,a,d), power growth curves with kinetic-energy distributions at saturation (Fig.~\ref{fig:UV}\,b,e), and far-field power spectra up to the 7th harmonic (Fig.~\ref{fig:UV}\,c,f). In the ellipsoidal case ($L_z = 100\,\mu\mathrm{m}$, $\sigma_\perp = 60\,\mu\mathrm{m}$), the DC-detuning metric (see Methods) remains well below unity: both beams lase identically, with full-window microbunching, overlapping power curves, and comparable harmonic content, confirming that this geometry is DC-free and providing a clean baseline. The picture changes sharply in the pancake case ($L_z = 10\,\mu\mathrm{m}$, $\sigma_\perp = 200\,\mu\mathrm{m}$): the $\mathrm{e^-}$-only beam becomes DC-limited, with bunching collapsing to a narrow central segment, power growth stalling before saturation, and the spectrum depleted at the fundamental and all odd harmonics, mirroring the behaviour that motivates selective slicing in current attosecond X-ray schemes. The pair beam, by contrast, sustains full-bunch lasing: microbunching spans the entire window, saturation power is markedly higher, and the odd-harmonic ladder extends cleanly to $n = 7$.

Crucially, this improvement originates from microbunching efficiency, not a power-duration trade-off. DC-field neutralisation preserves phase coherence across the full bunch (Fig.~\ref{fig:UV}\,d, orange), producing higher bunching factors $b_n = \langle e^{in\psi_j} \rangle$ (with $\psi_j$ the ponderomotive phase of the $j$-th particle) at $n = 1, 3, 5, 7$, which sustain synchronous energy transfer throughout the interaction (Fig.~\ref{fig:UV}\,e) and generate systematically stronger odd-harmonic emission (Fig.~\ref{fig:UV}\,c,f). Figure~\ref{fig:UV}\,g further shows that the $\mathrm{e^-/e^+}$ symmetry stabilises longitudinal phase-space oscillations through the post-saturation regime, with a precursor already visible in the ellipsoidal case (Fig.~\ref{fig:UV}\,c). With full-window lasing established in the UV, we turn to the more demanding question of whether a pair beam can reach the terawatt, attosecond class in a single, untapered undulator stage without self-seeding or any other auxiliary technique.

\begin{figure}[t!]
\centering
\includegraphics[width=0.99\linewidth]{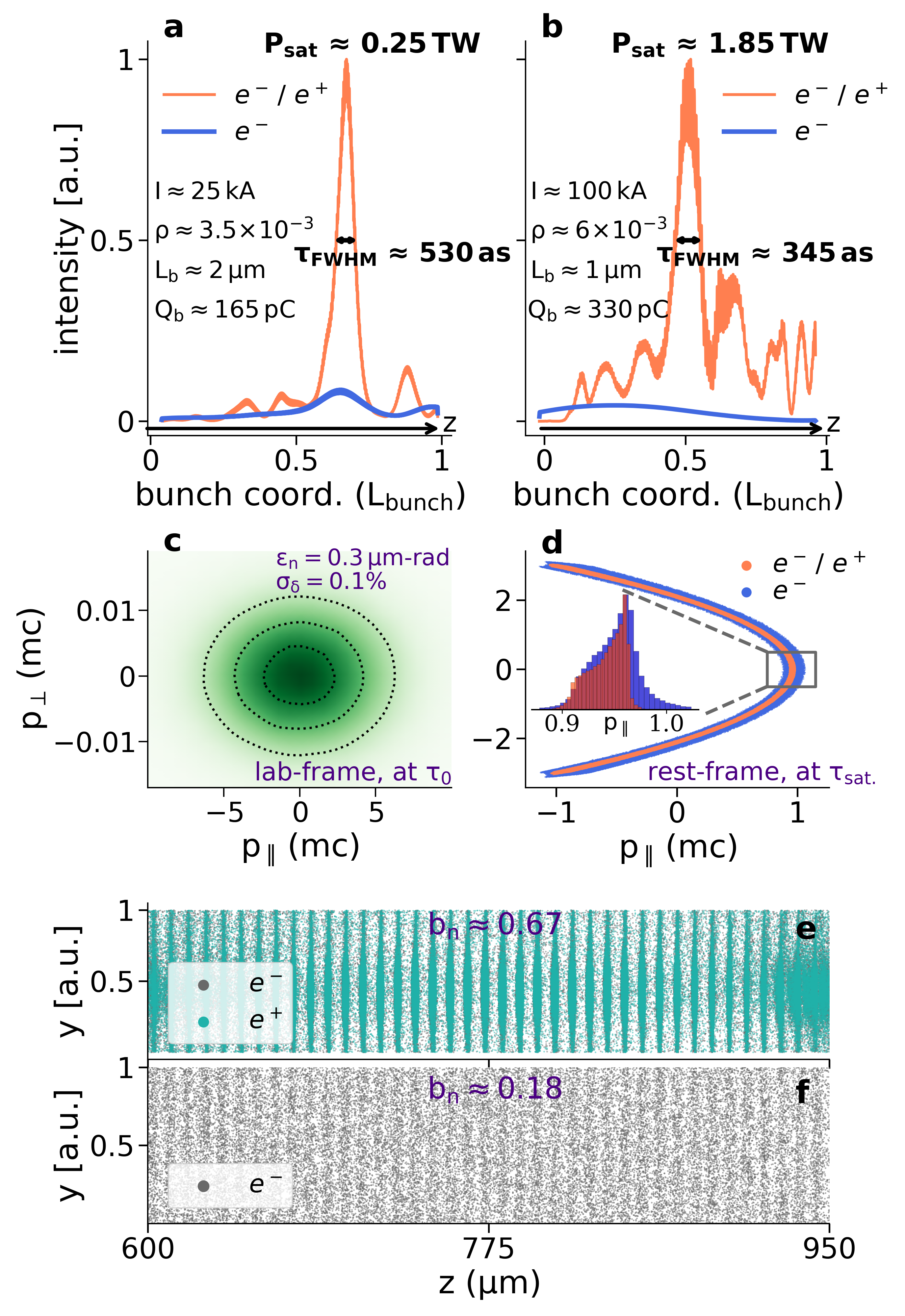}
\caption{\textbf{Pair-beam SASE in the soft X-ray regime.}
Fixed parameters: $E_b = 2\,\mathrm{GeV}$, $R_b \approx 90\,\mu\mathrm{m}$, $\varepsilon_n = 0.3\,\mu\mathrm{m\,rad}$, $\sigma_\delta = 0.1\%$, $\lambda_u = 3\,\mathrm{cm}$, $K = 3$ (single, untapered undulator; $L_{\mathrm{sat}} \lesssim 6\,\mathrm{m}$), $\lambda_{\mathrm{r}} \approx 5.4\,\mathrm{nm}$. Two charge--length settings are shown: $(Q_b,\,L_b) = (165\,\mathrm{pC},\,2\,\mu\mathrm{m})$ in panel~\textbf{a} and $(330\,\mathrm{pC},\,1\,\mu\mathrm{m})$ in panel~\textbf{b}, corresponding to peak currents of ${\sim}\,25\,\mathrm{kA}$ and ${\sim}\,100\,\mathrm{kA}$; all other parameters are held fixed.
\textbf{a,~b}~Temporal power envelope at the undulator exit for the $\mathrm{e^-/e^+}$ pair beam (orange) and the $\mathrm{e^-}$-only beam (blue).
\textbf{c}~Initial longitudinal momentum phase space in the laboratory frame, confirming beam quality within experimentally demonstrated ranges.
\textbf{d}~Longitudinal momentum phase space at saturation in the bunch rest frame; the pair beam retains a compact resonant core while the $\mathrm{e^-}$-only beam shows phase-space heating consistent with LSC-driven phase mixing.
\textbf{e,~f}~Transverse particle distribution in the central slice of the pair beam (\textbf{e}) and the $\mathrm{e^-}$-only beam (\textbf{f}) at saturation, with fundamental bunching factors $|b_1| \approx 0.67$ and $|b_1| \approx 0.18$, respectively; the pair beam yields terawatt-class attosecond SASE spikes.}
\label{fig:Xray}
\end{figure}

Figure~\ref{fig:Xray} compares a quasineutral pair beam with an otherwise identical $\mathrm{e^-}$-only beam at $2\,\mathrm{GeV}$, with parameters aligned to high-compression design targets at FACET-II~\cite{Yakimenko2019PRAB, emma2025experimental} and Athos~\cite{prat2023x}. At $Q_{\rm tot} = 165\,\mathrm{pC}$ and $L_b \simeq 2\,\mu\mathrm{m}$ (${\sim}\,25\,\mathrm{kA}$, Fig.~\ref{fig:Xray}\,a), multiple SASE spikes form on the cooperation-length scale~\cite{huang2007review}, with post-saturation mode competition selecting a dominant single spike. The pair beam reaches ${\sim}\,0.25\,\mathrm{TW}$ at ${\sim}\,530\,\mathrm{as}$ (FWHM); the $\mathrm{e^-}$-only beam is quenched before saturation. Halving the bunch length and doubling the charge raises the peak current to ${\sim}\,100\,\mathrm{kA}$ (Fig.~\ref{fig:Xray}\,b), extending recent tens-of-kA XFEL studies~\cite{yan2024terawatt, franz2024terawatt} into a regime where DC detuning is decisive. Here the pair beam sustains single-spike SASE to post-saturation, yielding ${\sim}\,1.85\,\mathrm{TW}$ at ${\sim}\,345\,\mathrm{as}$ (FWHM), with a time-bandwidth product approaching the Gaussian Fourier-transform limit, $\Delta t\,\Delta\nu / (\Delta t\,\Delta\nu)_{\rm FT} \approx 1.3$, while the $\mathrm{e^-}$-only beam fails to reach saturation entirely.

Both beams start from experimentally demonstrated quality: normalised emittance $\varepsilon_n = 0.3\,\mu\mathrm{m\,rad}$ and energy spread $\sigma_\delta = 0.1\%$~\cite{habib2023attosecond, prat2023x, franz2024terawatt, tomin2021accurate, prat2019generation} (Fig.~\ref{fig:Xray}\,c). By saturation (Fig.~\ref{fig:Xray}\,d), the $\mathrm{e^-}$-only beam shows marked phase-space heating consistent with LSC-driven phase mixing, while the pair beam retains a compact resonant core. The transverse slice distributions make the contrast most explicit (Fig.~\ref{fig:Xray}\,e,f): the pair-beam slice carries a well-developed microbunch train with $|b_1| \approx 0.67$ across the lasing window, compared with $|b_1| \approx 0.18$ and no discernible periodicity for the $\mathrm{e^-}$-only slice. Existing attosecond high-current schemes mitigate DC detuning by confining lasing to a pre-selected slice and applying chirp-taper or ESASE corrections~\cite{lutman2018high, ding2009generation, amann2012demonstration}; the pair-beam results probe the complementary regime of coherent full-window lasing, demonstrating that what DC detuning prohibits for an electron-only beam, quasi-neutrality restores.

\begin{figure*}[ht]
\centering
\includegraphics[width=0.99\linewidth]{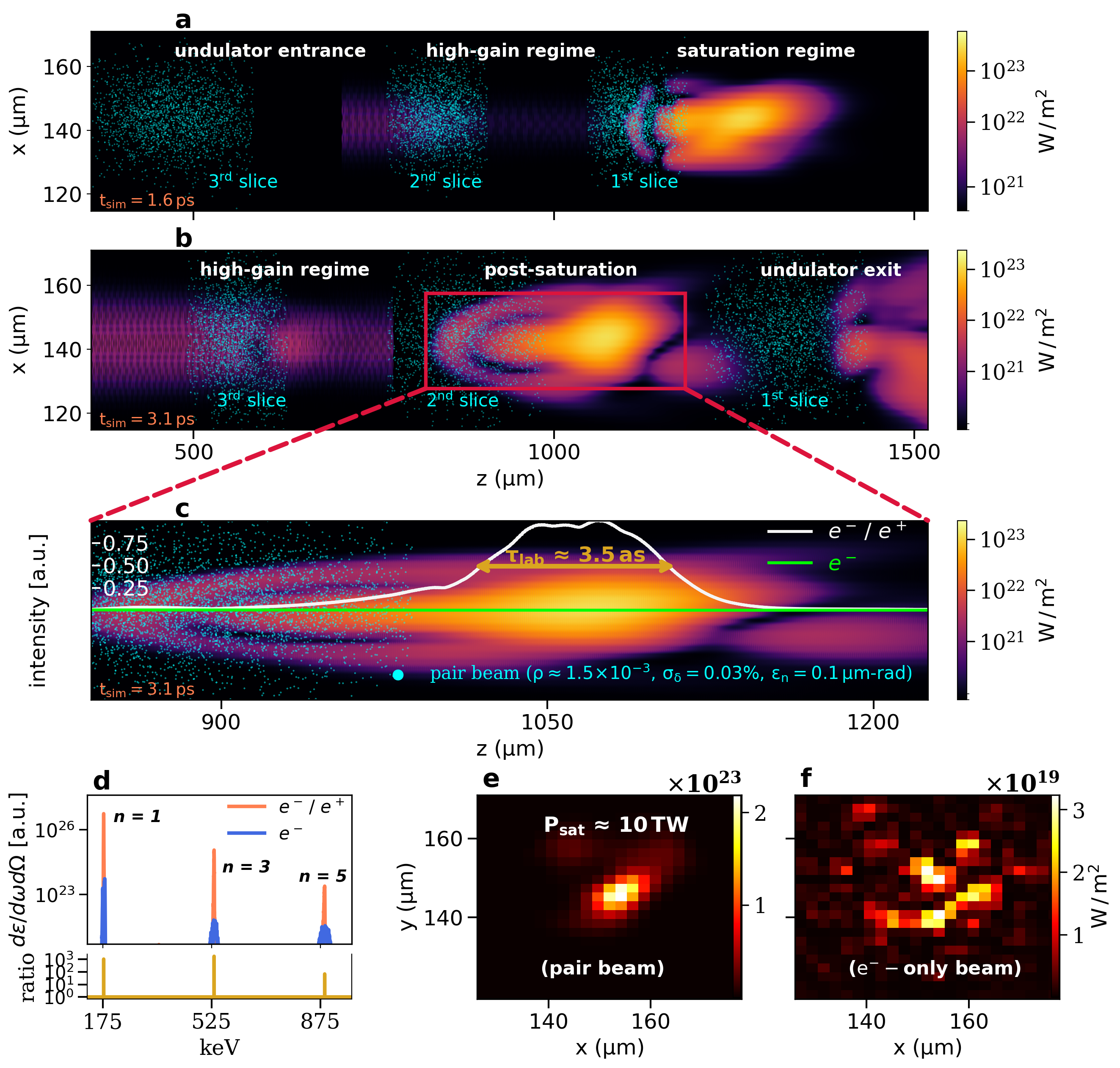}
\caption{\textbf{Coherent $\gamma$-ray lasing with the HHPC-FEL pair-beam concept.} Fixed parameters: $E_{\mathrm{slice}} = 48\,\mathrm{GeV}$, $Q_{\mathrm{slice}} \approx 5\,\mathrm{pC}$, $L_{\mathrm{slice}} \approx 4\,\mathrm{nm}$, $R_{\mathrm{slice}} \approx 15\,\mu\mathrm{m}$, $\varepsilon_n = 0.1\,\mu\mathrm{m\,rad}$, $\sigma_\delta = 0.03\%$, $\lambda_u = 3\,\mathrm{cm}$, $K = 2.5$ (single, untapered undulator; $L_{\mathrm{sat}} \approx 18\,\mathrm{m}$), $\lambda_{\mathrm{r}} \approx 7\,\mathrm{pm}$.
\textbf{a,~b}~Rest-frame radiation intensity snapshots at $t_{\mathrm{sim}} = 1.6\,\mathrm{ps}$ and $3.1\,\mathrm{ps}$, capturing sequential gain states of the $\mathrm{e^-/e^+}$ slice comb. At $t = 1.6\,\mathrm{ps}$ the leading slice has saturated while the trailing slice is still in exponential growth, confirming quasi-independent lasing; slice envelopes are traced by the cyan macro-particle sample. The apparent longitudinal compression of the slice train arises because the beam outside the undulator outruns the beam inside, an effect not observable in the laboratory frame.
\textbf{c}~Zoomed intensity snapshot of the mid-slice at $t_{\mathrm{sim}} = 3.1\,\mathrm{ps}$. The white envelope marks the slippage-broadened FEL pulse ($\mathrm{FWHM} \sim 3.5\,\mathrm{as}$ in the laboratory frame); the $\mathrm{e^-}$-only signal (green) fluctuates incoherently about zero, confirming the absence of gain without DC-field neutralisation.
\textbf{d}~Far-field power spectrum showing coherent amplification to the 5th harmonic, with the fundamental at ${\sim}\,177\,\mathrm{keV}$ and odd-harmonic intensity ratios.
\textbf{e,~f}~Transverse radiation intensity profiles downstream of the 2nd slice for the $\mathrm{e^-/e^+}$ pair beam (\textbf{e}) and the $\mathrm{e^-}$-only beam (\textbf{f}), illustrating the spatial-coherence contrast at saturation.}
\label{fig:Gamma}
\end{figure*}

\sectionminor{Towards the gamma-ray regime.}
\noindent
The leading current route to intense, narrow-band $\gamma$-ray emission is inverse Compton scattering, in which a laser pulse serves as the wiggler and whose coherence relies heavily on beam quality~\cite{quinPPCF25, filipescu2023spectral}; operating a conventional magnetic-undulator FEL in this regime remains largely conceptual~\cite{nedorezov2021nuclear}. The odd-harmonic boost of the SASE pair-beam results (Fig.~\ref{fig:UV}\,g) motivates a step in this direction. We propose the high-harmonic pair-cascade FEL (HHPC-FEL): DC-field neutralisation is used to structure the compressed bunch into a resonant lasing comb of nm-scale slices, each lasing quasi-independently in a single radiator undulator.

A modulator first imprints a dense nm-scale density modulation on the pair bunch, e.g. via echo-enabled harmonic generation~\cite{xiang2009echo} or diffraction-based nanopatterning~\cite{zhang2017experiments}, producing a comb of few-pC slices separated by field-free gaps. Each slice then undergoes high-gain SASE from shot noise in the downstream radiator, making the gain mechanism distinct from superradiant emission even though the resulting pulse train shares its short-emitter, single-wavepacket character~\cite{robles2024three, robles2025spectrotemporal}. Crucially, no charge is discarded, in contrast to prior cascade concepts relying on gating or selective slicing~\cite{franz2024terawatt, lutman2018high}: once DC detuning is cancelled, all particles in each slice contribute to lasing, provided the slice energy spread and scaled emittance remain below the Pierce parameter. Choosing slice lengths of order the cooperation length strongly favours single-spike-dominant operation without sacrificing peak power, while the intact odd-harmonic ladder can extend coherent emission toward the electron rest-mass energy (${\sim}\,0.511\,\mathrm{MeV}$) without further amplification.

As an illustrative case, we simulate nm-scale pair slices at $E_{\rm slice} = 48\,\mathrm{GeV}$ ($\lambda_u = 3\,\mathrm{cm}$, $K = 2.5$). This exceeds present FEL linac capabilities ($\lesssim 17.5\,\mathrm{GeV}$) and should be read as a next-generation projection, though it is modest relative to demonstrated high-energy accelerator technology~\cite{blumenfeld2007energy, emma2025experimental, habib2023attosecond}. The assumed slice quality ($Q_{\rm slice} \approx 5\,\mathrm{pC}$, $\varepsilon_n = 0.1\,\mu\mathrm{m\,rad}$, $\sigma_\delta = 0.03\%$) is consistent with emittance and energy spread demonstrated at modern FEL injectors and targeted for next-generation beams~\cite{tomin2021accurate, prat2019generation, rosenzweig2019next, emma2025experimental, Yakimenko2019PRAB}, extrapolating established XFEL beam quality to higher energy rather than invoking new physics. The initial bunch has transverse spot size $\lesssim 30\,\mu\mathrm{m}$ and total charge $Q_{\rm tot} \approx 2.5\,\mathrm{nC}$. Slice FWHM length $\approx 2\,\mathrm{nm}$ and inter-slice separation $\approx 6\,\mathrm{nm}$ are set so that the gap equals the saturation slippage length $L_{\rm slip} = N_{\rm sat}\,\lambda_r$, suppressing cross-talk and preserving quasi-independent lasing.

The results are summarised in Fig.~\ref{fig:Gamma}. Rest-frame intensity snapshots (Fig.~\ref{fig:Gamma}\,a,b) confirm quasi-independent lasing: at $t_{\rm sim} = 1.6\,\mathrm{ps}$ the leading slice has saturated while the trailing slice is still in exponential growth. At trailing-slice saturation (Fig.~\ref{fig:Gamma}\,c), an isolated spike of ${\rm FWHM} \sim 3.5\,\mathrm{as}$ is emitted; the matched $\mathrm{e^-}$-only control fluctuates incoherently about zero. The far-field spectrum (Fig.~\ref{fig:Gamma}\,d) shows coherent amplification to the 5th harmonic, with the fundamental at ${\sim}\,177\,\mathrm{keV}$. Transverse intensity maps (Fig.~\ref{fig:Gamma}\,e,f) underscore the spatial-coherence contrast: the pair-beam slice produces a bright, coherent mode; the $\mathrm{e^-}$-only slice remains DC-clipped and incoherent. The pair slice reaches ${\approx}\,10\,\mathrm{TW}$ single-spike power over an $18\,\mathrm{m}$ undulator at $\rho_{\rm eff} \approx 1.2\times10^{-3}$ (Eq.~\eqref{eq:rho}), which simultaneously relaxes the energy-spread tolerance and shortens $L_{\rm sat}$; quantum-diffusion-induced dephasing remains below $0.5\,\rho_{\rm eff}$ over the full interaction and is not expected to suppress $\gamma$-ray gain (see Methods)~\cite{huang2007review}.


\sectionmajor{Discussion}
\smallskip

\noindent
This proof-of-principle study aims to isolate and demonstrate the physical mechanism by which DC longitudinal space-charge cancellation unlocks full-bunch, high-current lasing in the pancake-beam geometry, together with the associated harmonic boost, not to constitute a facility design. The 3D PIC simulations therefore model a single-pass, untapered undulator, comparing electron-only and pair-beam configurations under identical conditions to expose the LSC effect in its purest form; multi-segment layouts with phase shifters, optical delay lines, or external focusing lattices are deliberately excluded to avoid conflating DC-field physics with auxiliary instrumentation effects.

The engineering challenges of a full realisation such as separate bunch compression, transport through dedicated magnetic chicanes, and high-precision merging at the undulator entrance, are real and non-trivial, and are left for specialised future studies. Beyond the core gain mechanism, however, the simulations establish a meaningful degree of practical robustness: DC-field suppression and full-bunch exponential gain are retained for charge imbalances up to ${\sim}\,30\%$ and for longitudinal and transverse offsets comparable to the beam envelope (see Methods), spanning error margins consistent with realistic experimental operation. Progress toward a full realisation is contingent on continued advances in pair-beam production (higher yield, reduced energy spread, and improved divergence) alongside beam quality control and FEL hardware development.

High-brightness relativistic pair beams can already be generated, compressed, and transported with ongoing quality improvement~\cite{sarri2015generation, arrowsmith2024laboratory, Yakimenko2019PRAB}, establishing individual building blocks of the proposed scheme. The outstanding missing element is a dedicated merger capable of forming a tightly overlapped, FEL-quality co-propagating beam, a precision requirement closely analogous to that demonstrated in PWFA driver-witness beamlines and cavity-based X-ray FEL operation, where two beams are synchronised with micrometre-level alignment using feedback-enabled diagnostics~\cite{Litos2014Nature, Yakimenko2019PRAB, Aschikhin2016NIMA, Corde2015Nature, Gessner2016NatCommun, Rauer2026cavity}. The present positron-beam quality remains below that routinely achieved for electron FEL operation, but current progress is encouraging.

A pragmatic integration path proceeds as follows. A target-based positron source with high-acceptance capture feeds a damping or conditioning stage to reach FEL-grade emittance and slice energy spread. Electrons and positrons are co-accelerated in a shared RF linac with the bunches interleaved at opposite RF phase ($\pi$-delay), analogous in spirit to fresh-bunch self-seeding architectures~\cite{emma2017compact}. A dipole-septum splitter downstream separates the two species into dedicated transport lines, each with chirp-chicane compression and transverse matching optics; a short, tunable path-length section in one arm removes the RF-induced half-period timing offset. The beams are then recombined at a feedback-stabilised Y-junction merger and directed into the undulator through charge-symmetric final focusing (e.g., solenoids or a plasma lens) to maintain transverse overlap at the entrance.

As a concrete near-term step, the pair-beam concept can be tested in the XUV regime using an LWFA-based pair beam in a compact undulator geometry. The larger Pierce parameter at XUV wavelengths shortens the gain length, relaxes beam-quality tolerances, and is compatible with pair-beam charge levels already achieved in LWFA experiments, offering a direct experimental path to the two key predictions: DC-field suppression and enhanced microbunching efficiency.


\sectionmajor{Methods}

\sectionminor{FEL simulation in Lorentz-boosted frame.}
\noindent
The FEL process spans an exceptionally wide range of length scales: nm-scale radiation wavelengths, $\mu$m-scale bunch dimensions, and m-scale undulator sections, three scales that cannot be resolved simultaneously in the laboratory frame without prohibitive computational cost. We therefore simulate the entire interaction in the bunch rest frame, which compresses the undulator length by a factor of $\gamma$ while stretching the radiation wavelength by the same factor, bringing all relevant scales into a tractable ratio. This Lorentz-boosted technique has been validated in laser-plasma accelerator studies, enabling speedups of several orders of magnitude~\cite{vay2007noninvariance, vay2021modeling}, and is adapted here to the FEL context following earlier implementations in \textsc{WARP}~\cite{fawley2009use} and \textsc{MITHRA}~\cite{fallahi2018mithra}.

In practice, both the undulator magnetostatic field and the initialised 6D beam phase space are Lorentz-transformed into the bunch rest frame, and the full $\mathrm{e^-/e^+}$ bunch-field system is then evolved self-consistently with the three-dimensional full-wave PIC code \textsc{Smilei}~\cite{derouillat2018smilei}. All diagnostics are subsequently inverse-transformed to the laboratory frame for analysis and comparison with analytical theory. The approach resolves the optical wave with sufficient spatial resolution while retaining the full undulator length needed to capture exponential gain and reach saturation within a single simulation run.

Figure~\ref{fig:FEL_scheme} illustrates the diagnostic layout used throughout this work. Synchronised stations are placed at three positions along the undulator axis, corresponding to the initialisation, linear-gain, and saturation points of the FEL process. At each station, upper and lower panels provide complementary views of the same physical state: the upper panel shows the radiation intensity distribution in the laboratory frame, while the lower panel shows the forward-radiation amplitude in the bunch rest frame. The one-to-one correspondence between these two representations follows directly from the inverse Lorentz transformation, since forward radiation amplified in the boosted frame maps unambiguously onto the FEL mode that a downstream detector would record. The figure thus serves a dual purpose: demonstrating the algorithmic pipeline for pair-beam FEL simulation and establishing the diagnostic framework applied consistently in all subsequent electron-only versus pair-beam comparisons. For each configuration, the key laboratory-frame PIC observables, principally the gain length and saturation power, are benchmarked against one-dimensional high-gain FEL theory corrected for three-dimensional effects via the Xie parametrisation~\cite{dohlus2008ultraviolet, saldin2013physics, Xie1995PAC3D}. As an independent cross-check, selected results are compared with \textsc{MITHRA}~\cite{fallahi2018mithra}, a boosted-frame self-consistent PIC code that has been validated against the widely used undulator-averaged code \textsc{GENESIS~1.3}~\cite{Reiche1999GENESIS13}; agreement is found to within shot-to-shot fluctuations in all tested configurations.

\sectionminor{Particle-in-cell simulation setup.}
\noindent
Full-wave finite-difference time-domain PIC codes have been validated against slowly-varying envelope approximation (SVEA) schemes in the FEL context~\cite{campbell2020analysis} and offer a more faithful representation of the field-particle interaction by avoiding the two principal limitations of SVEA codes: fast-timescale averaging and the inability to treat coherent synchrotron radiation (CSR) from first principles. Crucially, the PIC approach evolves the full Maxwell-Lorentz system for the entire bunch, retaining both the DC (bunch-scale) and AC (resonant) components of the space-charge field and their coupling to the radiation field without pre-selecting interaction modes. This complete electromagnetic treatment is essential here: envelope codes such as \textsc{GENESIS~1.3}~\cite{Reiche1999GENESIS13} include sophisticated time-dependent collective-effect modelling, among them long-range space charge over the full simulated bunch (version~4), but their wiggler-period-averaged formulation treats self-fields in a reduced description that does not capture the DC-field cancellation we seek to demonstrate.

All simulations are carried out in full 3D Cartesian geometry with \textsc{Smilei}~\cite{derouillat2018smilei}. Fields are initialised by a relativistic Poisson solver and advanced with the \textsc{M4} Maxwell solver~\cite{Lu2020}, which provides reduced, more isotropic numerical dispersion than the standard Yee scheme. Particle trajectories are integrated with the \textsc{HigueraCary} pusher~\cite{HigueraCary2017}, combining symplectic phase-space conservation with correct relativistic drift under full Lorentz-force dynamics; classical radiation reaction is included self-consistently via the Landau-Lifshitz model. Perfectly matched layers enforce absorbing boundary conditions for both fields and particles at all box boundaries.

The longitudinal cell size is set by iterative convergence tests, resolving the fundamental radiation wavelength with approximately 32 cells, sufficient to capture up to the 7th harmonic under the Nyquist criterion. The transverse resolution is chosen to suppress numerical heating, with the maximum permissible undulator parameter $K$ verified for each configuration. The longitudinal box length $L_z$ matches the rest-frame length of the stretched bunch; the transverse extents $L_x$ and $L_y$ are set large enough to accommodate the beam envelope and suppress artefacts from residual boundary reflections.

Each $\mathrm{e^-/e^+}$ bunch is represented by several million macro-particles, initialised with uniform weights and a Gaussian momentum distribution consistent with the specified energy spread and normalised emittance. Quasi-random loading of the longitudinal coordinates seeds the SASE shot-noise condition~\cite{reiche2000numerical}, ensuring a vanishingly small initial bunching factor. Far-field diagnostics are computed with the \textsc{RaDiO} algorithm~\cite{pardal2023radio}, integrated directly into \textsc{Smilei} to preserve parallel scalability; power spectra are obtained via Fourier transform using Parseval's theorem~\cite{pausch2018quantitatively}.

\sectionminor{Numerical robustness and tolerance analysis.}
\noindent
Because simulations are performed in the bunch rest frame, where no relativistic bulk drift is present, the dominant numerical requirement is an accurate electromagnetic dispersion relation rather than suppression of the numerical Cherenkov instability (NCI). This motivates the 4th-order \textsc{M4} Maxwell solver~\cite{Lu2020} as the production default, paired with the \textsc{HigueraCary} pusher~\cite{HigueraCary2017} for its symplectic phase-space conservation and correct relativistic drift.

To confirm independence from these specific choices, representative simulations were repeated with two alternative Maxwell solvers: the standard \textsc{Yee}~\cite{Yee1966} scheme, and the NCI-optimised \textsc{Lehe}~\cite{Lehe2013} scheme; and two alternative particle pushers: \textsc{Boris}~\cite{Boris1970} and \textsc{Vay}~\cite{Vay2008}. The key FEL observables, gain length, saturation power, spectral content, and microbunching statistics, remained consistent to within shot-to-shot fluctuations across all combinations. Resolution scans with refined grids and timesteps confirmed that 32 cells per fundamental wavelength is sufficient to capture up to the 7th harmonic under the Nyquist criterion, and that $\mathrm{CFL} = 0.70$ lies well within the convergent regime with no detectable effect on any diagnostic metric.

\begin{figure}[t]
\centering
\includegraphics[width=0.99\linewidth]{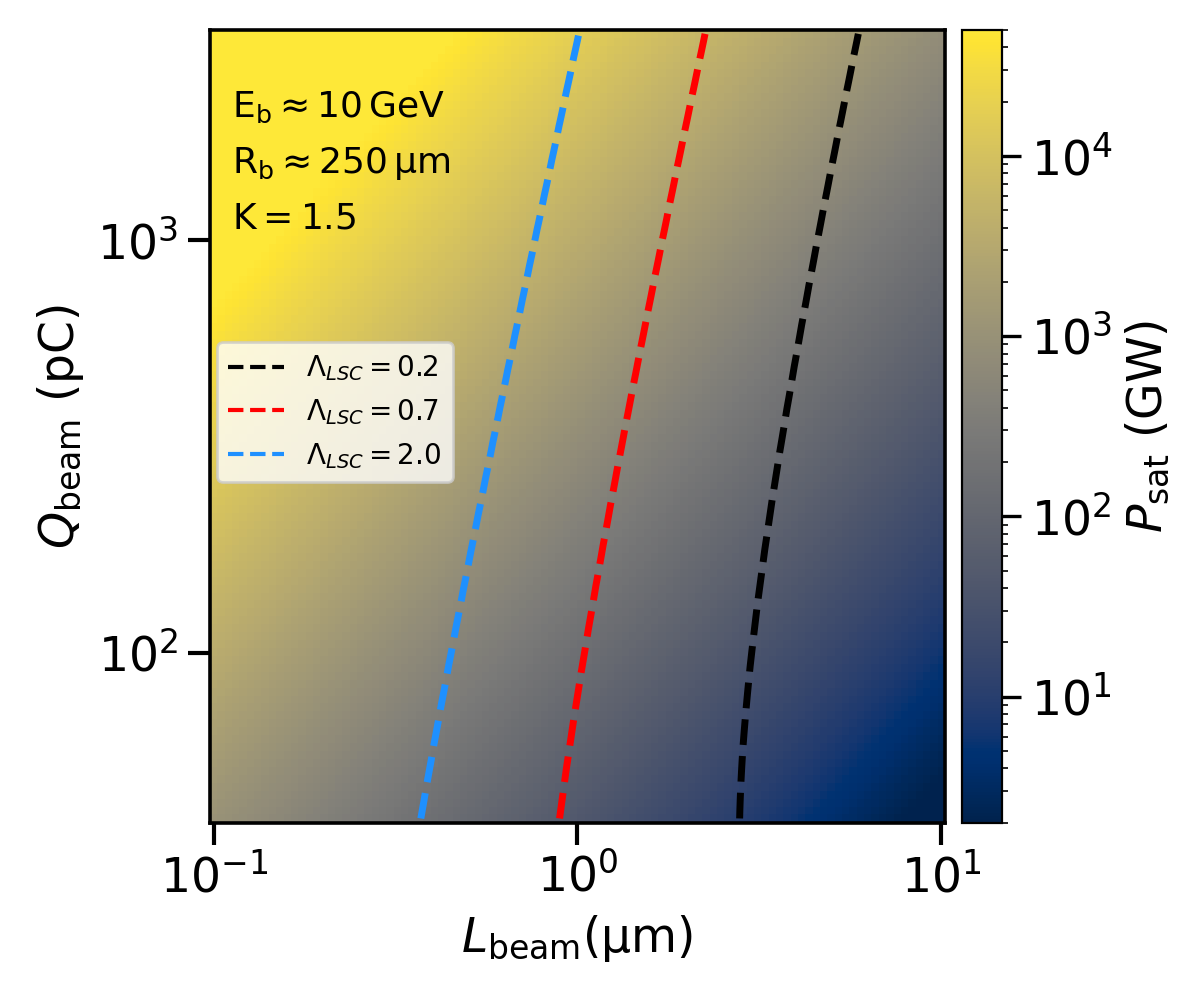}
\caption{\textbf{Saturation-power landscape as a function of total charge and bunch length.}
Colour map: estimated saturation power $P_{\mathrm{sat}}$ (Eq.~\eqref{eq:Psat}) across the $(Q_{\mathrm{beam}},\,L_{\mathrm{beam}})$ plane for fixed parameters $K = 1.5$, $R_b \simeq 250\,\mu\mathrm{m}$, $E_b \simeq 10\,\mathrm{GeV}$. Dashed contours show $\Lambda_{\mathrm{LSC}}$ at three thresholds: negligible suppression ($\Lambda_{\mathrm{LSC}} = 0.2$, black), onset of gain reduction ($\Lambda_{\mathrm{LSC}} = 0.7$, red), and breakdown of high-gain lasing ($\Lambda_{\mathrm{LSC}} = 2.0$, blue). Although $P_{\mathrm{sat}}$ increases with peak current toward the upper-left corner, the contours impose hard boundaries for an electron-only beam: beyond the blue contour, LSC detuning ultimately destroys coherent gain. A quasi-neutral pair beam cancels the DC self-field, driving $\Lambda_{\mathrm{LSC}} \to 0$ regardless of bunch charge or length, and unlocks the full high-power domain across the ultrashort, high-charge operating space.}
\label{fig:LSC_detuning}
\end{figure}

The robustness of pair-beam lasing against deviations from ideal quasi-neutrality and perfect spatial overlap was assessed using the soft X-ray configuration (Fig.~\ref{fig:Xray}). For charge balance, a controlled electron-positron abundance mismatch was introduced incrementally: full-bunch exponential gain and TW-scale saturation are essentially retained for imbalances up to ${\sim}\,30\%$, as the residual net charge keeps $\Lambda_{\rm LSC}$ well below the suppression threshold throughout the high-gain regime (Fig.~\ref{fig:LSC_detuning}). For spatial overlap, longitudinal and transverse offsets of approximately $10\%$ of the bunch length and beam radius were imposed to emulate realistic focusing and merging errors; these leave the exponential gain intact and produce only a modest reduction in saturation power, with harmonic content qualitatively unchanged. Together, these tests show that the key advantages of the pair-beam concept, DC-field suppression, full-bunch gain, and harmonic enhancement, are preserved over charge and overlap errors consistent with realistic experimental conditions, without requiring a fine-tuned, perfectly symmetric configuration.


\sectionminor{Modified FEL dispersion relation.}
\noindent
In standard FEL analysis, longitudinal space charge (LSC) enters the high-gain dispersion relation only as an oscillatory (AC) correction, frequently negligible at high $\gamma$~\cite{marcus2011gain}. Any steady (DC) background component is implicitly absorbed into the initial energy chirp and handled upstream through phase-space management in the bunch compressor and transport line. This separation is well justified for conventional operating conditions, though numerous studies have noted that the resulting frozen chirp, driven primarily by LSC in the compressor, can detune or severely limit FEL gain when not adequately linearised, particularly at lower beam energies or very high peak currents.

In the ultrahigh-current, pancake-bunch regime studied here, the DC contribution can no longer be absorbed implicitly: the quasi-static self-field of a short, transversely extended bunch imprints a slice-dependent energy chirp directly at the undulator entrance, driving the normalised DC detuning term toward order unity and suppressing saturation power across large fractions of the bunch. We therefore make this contribution explicit by incorporating it as a slice-dependent detuning offset into the linear FEL gain evolution, yielding a modified dispersion relation from which a compact diagnostic metric $\Lambda_{\rm LSC}$ quantifies the degree of DC-induced gain suppression for any beam and undulator configuration.

Following the standard 1D high-gain FEL formalism, we define the fractional energy detuning $\eta = (\gamma - \gamma_r)/\gamma_r$ and the ponderomotive phase advance $\mathrm{d}\psi/\mathrm{d}z = 2k_u\eta$, where $\gamma$ is the electron Lorentz factor, $\gamma_r$ its resonant value, $\psi$ the ponderomotive phase, and $k_u$ the undulator wavenumber. Introducing the Pierce parameter $\rho > 0$, gain wavenumber $\Gamma = 2k_u\rho$, plasma wavenumber $k_p$, normalised detuning $s = \eta/\rho$, and the dimensionless space-charge ratio $\mu = k_p/\Gamma$, the exponential growth ansatz $\tilde{E} \propto \exp(\lambda z)$ yields the linearised eigenvalue problem~\cite{dohlus2008ultraviolet}
\begin{equation}
  \lambda^3 + 2is\,\lambda^2 + \left(\mu^2 - s^2\right)\lambda - i = 0,
  \label{eq:cubic}
\end{equation}
which reduces to the familiar $\lambda^3 - i = 0$ in the ideal limit $\mu = 0$, $s = 0$. The 1D Pierce parameter for an elliptical Gaussian beam with rms transverse sizes $(\sigma_x, \sigma_y)$ is
\begin{equation}
  \rho \;=\;
  \left[
    \frac{I}{I_A}\;
    \frac{K^2[JJ]^2\lambda_u^2}{64\pi^2\gamma^3\sigma_x\sigma_y}
  \right]^{\!1/3},
  \label{eq:rho}
\end{equation}
where $I$ is the local slice current, $I_A \simeq 17.045\,\mathrm{kA}$ the Alfv\'{e}n current, $\lambda_u$ the undulator period, $K$ the deflection parameter, and $[JJ] \equiv J_0(\xi) - J_1(\xi)$ with $\xi = K^2/(4 + 2K^2)$ the Bessel coupling factor for a planar undulator.

In the coasting-beam idealisation, the equilibrium distribution $f_0$ is longitudinally uniform, so the homogeneous (DC/low-$k$) component of the bunch charge density produces a negligible on-axis longitudinal field: Lorentz contraction renders the bunch effectively infinite in the rest frame, leaving only the resonantly modulated density at the FEL scale to contribute, i.e. the $\mu$-term in Eq.~\eqref{eq:cubic}~\cite{saldin2013physics}.

For the ultrashort, high-current pancake bunches studied here this standard result is insufficient: the finite bunch length breaks the assumed translation invariance, and the resulting quasi-static low-$k$ self-field imprints an unavoidable, slice-dependent energy push on $f_0$. Retaining this contribution in the Vlasov-Maxwell linearisation shows that it enters the eigenvalue problem purely as a coherent detuning offset, with no resonant coupling to the FEL radiation field.

To formalise this, the longitudinal electric field is decomposed into a resonant AC component and a quasi-static DC self-field,
\begin{equation}
  E_z(z,\zeta) = E_z^{\mathrm{AC}}(z,\zeta) + E_z^{\mathrm{DC}}(\zeta),
\end{equation}
where $\zeta$ is the slice coordinate within the bunch. The DC field drives a slice-dependent detuning that is frozen over the gain length, motivating the split
\begin{equation}
  \frac{\mathrm{d}\eta}{\mathrm{d}z}\bigg|_{\mathrm{DC}}
  = -\frac{e\,E_z^{\mathrm{DC}}(\zeta)}{\gamma mc^2},
  \qquad
  \eta(z,\zeta)
  = \underbrace{\eta_{\mathrm{coh}}(\zeta)}_{\text{frozen}}
  + \underbrace{\eta_1(z,\zeta)}_{\text{slow}},
  \label{eq:eta-split}
\end{equation}
where $\eta_{\mathrm{coh}}(\zeta)$ is the stable, slice-dependent coherent detuning accumulated before and during early gain, and $\eta_1(z,\zeta)$ is the small FEL-driven modulation encompassing the initial energy spread. The ponderomotive phase equation becomes
\begin{equation}
  \frac{\mathrm{d}\psi}{\mathrm{d}z}
  = 2k_u\!\left[\eta_{\mathrm{coh}}(\zeta) + \eta_1(z,\zeta)\right].
  \label{eq:phase-shift}
\end{equation}
Since Eqs.~\eqref{eq:eta-split}--\eqref{eq:phase-shift} are structurally identical to the standard pendulum equations, the cubic eigenvalue equation~\eqref{eq:cubic} remains valid under the slowly-varying-amplitude approximation with the sole replacement
\begin{equation}
  s \;\equiv\; s_{\mathrm{tot}}(\zeta)
  \;=\; s_0 + s_{\mathrm{DC}}(\zeta),
  \qquad
  s_{\mathrm{DC}}(\zeta) = \frac{\eta_{\mathrm{coh}}(\zeta)}{\rho}.
  \label{eq:s-shift}
\end{equation}

The dispersion relation thus retains its standard cubic form; the DC self-field enters only as a frozen offset $s_{\mathrm{DC}}(\zeta)$ that shifts each slice's operating point independently, converting a single bunch-level solution into a slice-resolved family. Because the quasi-static self-field grows from zero at the bunch centre to its maximum at the head and tail, $s_{\mathrm{DC}}(\zeta)$ increases monotonically toward the bunch edges, progressively displacing those slices from resonance. Importantly, the DC field does not exchange energy resonantly with the radiation mode and does not grow along the undulator axis; instead, it degrades microbunching parametrically by imposing a fixed phase-advance rate on each slice through $\eta_{\mathrm{coh}}(\zeta)$.

This frozen-field model rests on three assumptions: $k_b \ll k_r$; negligible longitudinal reshaping of the bunch envelope inside the undulator; and negligible additional impedance sources such as CSR and chamber wakes. All three are well satisfied in the regimes studied here, where energy exchange is slow relative to the gain length and the beam envelope remains largely intact through saturation, making self-consistent evolution of the DC field unnecessary.

The standard LSC impedance $Z_\parallel(k)$ is derived as the linear response of a translation-invariant beam, in which the dominant bunch-scale Fourier component at $k_b \sim 2\pi/L_b$ produces a representative on-axis field
\begin{equation*}
  E_0^{\mathrm{AC}}(k_b)
  \;\simeq\;
  \frac{Z_0\,k_b\,I}{2\pi\gamma^2}\,
  F\!\left(\frac{k_b R}{\gamma}\right),
\end{equation*}
where $F$ encodes the transverse geometry factor~\cite{saldin2004longitudinal}. Because this impedance assumes an effectively infinite beam, it does not capture finite-length boundary effects: the $k \to 0$ (uniform, homogeneous) component of the charge density corresponds to a perfectly translation-invariant state and therefore generates no longitudinal self-field~\cite{venturini2008models}.

A finite bunch, however, is not translation invariant: its longitudinal boundaries generate a nonzero quasi-static field $E_z^{\mathrm{DC}}(\zeta)$ that imprints a coherent, head-to-tail energy chirp on the bunch before it enters the undulator. Once inside, this frozen chirp enters the FEL equations identically to a compressor-induced energy chirp as a slice-dependent detuning shift, formalised in Eq.~\eqref{eq:s-shift}. Because this low-$k$ contribution is absent from the standard impedance picture, $E_z^{\mathrm{DC}}(\zeta)$ is computed directly from the exact quasi-static on-axis electrostatic solution for a finite relativistic cylindrical bunch~\cite{Wiedemann2015PAP, fusco2004beam}, with the bunch geometry factor accounted for implicitly. Within this decomposition, the resonant (AC) microbunching response is represented by the FEL plasma wavenumber $k_p$, while the finite-bunch DC contribution enters as an additional slice-based electrostatic detuning.

Using the normalised slice coordinate $\zeta_n = |2\zeta/L_b|$ (with $\zeta_n = 0$ at the bunch centre), the coherent energy detuning accumulated over the interaction length $L_{\mathrm{inter}} \sim 18\,L_g$, where $L_g = \lambda_u/(4\pi\sqrt{3}\rho)$ is the power gain length, is
\begin{equation}
  \eta_{\mathrm{coh}}(\zeta_n)
  = \frac{e\,E^{\mathrm{DC}}_{\mathrm{eff}}(\zeta_n)\,L_{\mathrm{inter}}}
         {\gamma m c^2},
  \label{eq:eta-coh}
\end{equation}
which is normalised to the Pierce parameter as $s_{\mathrm{DC}}(\zeta_n) = \eta_{\mathrm{coh}}(\zeta_n)/\rho$.

To reduce DC-induced gain suppression to a single scalar diagnostic, we define the cumulative LSC detuning metric
\begin{equation}
  \Lambda_{\mathrm{LSC}}(\zeta_n)
  \;\equiv\;
  \sqrt{s_{\mathrm{DC}}(\zeta_n)^2 + \mu^2},
  \label{eq:Lambda}
\end{equation}
which combines the frozen DC detuning $s_{\mathrm{DC}}$ with the resonant plasma coupling $\mu$ into a compact, slice-resolved indicator. It is a heuristic intended for rapid regime identification; quantitative gain reduction for a given slice follows from the full eigenmode root analysis of Eq.~\eqref{eq:cubic}. Three regimes can be distinguished: $\Lambda_{\mathrm{LSC}} \ll 1$ indicates negligible suppression and near-ideal exponential growth; $\Lambda_{\mathrm{LSC}} \lesssim 1$ signals the onset of noticeable gain reduction; and $\Lambda_{\mathrm{LSC}} \gtrsim 2$ marks breakdown of the high-gain mode. Since both $s_{\mathrm{DC}}$ and $\mu$ scale with net charge and current, $\Lambda_{\mathrm{LSC}}$ grows rapidly with peak current for an electron-only beam.

For a quasi-neutral, co-located, density-matched pair beam, both contributions to $\Lambda_{\mathrm{LSC}}$ are simultaneously suppressed. The DC self-field is cancelled by the opposing net longitudinal currents of the two species ($E_z^{\mathrm{DC}} \approx 0$), driving $s_{\mathrm{DC}} \to 0$. The resonant AC coupling is suppressed by a related mechanism: FEL-driven microbunching is a symmetric, in-phase density modulation of both species ($\delta n_- \simeq \delta n_+$), so the net intra-slice longitudinal current vanishes and $\mu_{\mathrm{eff}} \ll 1$ (equivalently, $k_p \approx 0$). Pair plasma does support antisymmetric, charge-separated oscillations, but these require a relative density imbalance and are not resonantly driven by the FEL ponderomotive potential. Any residual mismatch or spatial offset reintroduces a small finite $\mu_{\mathrm{eff}}$, captured self-consistently in the PIC simulations. The combined result is $\Lambda_{\mathrm{LSC}} \approx 0$, while the radiative transverse current, and therefore the Pierce parameter coupling, remains undiminished, decoupling gain performance from the LSC limitation and supporting robust exponential growth even for strongly pancake-shaped beams at extreme peak currents.

It remains to establish that electrons and positrons radiate constructively and lock to the same ponderomotive phase. In a planar undulator, reversing the charge sign $q \to -q$ reverses both the transverse Lorentz acceleration $a_\perp$ and the charge itself, but the far-field radiation scales as $q\,a_\perp$ via the Li\'{e}nard-Wiechert expression: the product is unchanged, so the radiated Fourier component carries the same phase for both species at any observation point. In the FEL process, the radiating source is the transverse current harmonic $j_\perp \propto q\,v_\perp$; with mirrored trajectories the electron and positron contributions add constructively, doubling the effective source strength for equal populations. This invariance is captured at the Hamiltonian level in the standard 1D formulation: the interaction potential reads $H_\psi \propto -(q\,E\cdot v)_{\rm res}\cos\psi$, so the coupling amplitude is proportional to $q\,v_\perp$, and the Bessel factor $[JJ]$, depending only on $K^2$, carries the same sign for both species. Any residual sign ambiguity is absorbed by the shift $\psi \to \psi + \pi$, which leaves the stable fixed point and trapping unchanged. Equivalently, in the pendulum-equation picture the transformation $q \to -q$ is absorbed by $x \to -x$ (i.e.\ $\psi \to \psi + \pi$), leaving the equations of motion invariant; both species therefore evolve under the same ponderomotive phase and microbunch in lockstep. Finally, the opposite transverse deflections of the two species inside the undulator do not disrupt quasi-neutrality: the oscillation amplitudes are far smaller than the beam width in all configurations studied, so the beams remain effectively overlapped throughout the interaction.

If a single scalar summary of the LSC state is needed, $\Lambda_{\mathrm{LSC}}(\zeta_n)$ can be reduced to an RMS value via a current-weighted average over $\zeta_n$, or simply evaluated at its maximum within the lasing core. The local exponential growth rate at each slice is obtained by solving Eq.~\eqref{eq:cubic} numerically for the growing eigenmode root $\lambda(s_{\mathrm{tot}}(\zeta), \mu)$.

The standard saturation power estimate $P_{\mathrm{sat}} = \rho\,P_{\mathrm{beam}}$, where $P_{\mathrm{beam}} = I\gamma_0 m_e c^2$ and $I = Qc/L_{\mathrm{lab}}$, is generalised to include DC detuning by replacing $\rho$ with a slice-dependent effective Pierce parameter:
\begin{equation}
  P_{\mathrm{sat}}^{\mathrm{DC}}(\zeta) = \rho_{\mathrm{eff}}(\zeta)\,P_{\mathrm{beam}},
  \label{eq:Psat}
\end{equation}
where $\rho_{\mathrm{eff}}$ is defined by matching the DC-detuned linear gain rate to its unperturbed value,
\begin{equation}
  \rho_{\mathrm{eff}}(\zeta)
  = \rho\,
  \frac{\Re\!\left[\lambda\!\left(s_{\mathrm{tot}}(\zeta),\mu\right)\right]}
       {\Re\!\left[\lambda(s_0,\mu)\right]},
  \label{eq:rhoeff}
\end{equation}
with $\lambda(s,\mu)$ the growing root of Eq.~\eqref{eq:cubic}. Equation~\eqref{eq:rhoeff} equivalently defines a slice-dependent effective gain length $L_g^{\mathrm{eff}}(\zeta) = L_g\,\Re[\lambda(s_0,\mu)] / \Re[\lambda(s_{\mathrm{tot}}(\zeta),\mu)]$, or in compact notation $\rho_{\mathrm{eff}} = \rho\,\lambda_{\mathrm{DC}}/\lambda_{\mathrm{AC}}$, where $\lambda_{\mathrm{DC}} = \langle\lambda(\zeta_n)\rangle$ is the longitudinally averaged growth rate including DC detuning and $\lambda_{\mathrm{AC}}$ its unperturbed counterpart.

Intra-slice energy spread is incorporated by averaging the local growth rate over the slice energy distribution. Treating each slice as having a Gaussian energy deviation $\hat{\eta} \sim \mathcal{N}(0, \sigma_{\hat{\eta}}^2)$ with $\sigma_{\hat{\eta}} = \sigma_E/\rho$, the averaged growth rate at slice $\zeta_i$ is
\begin{equation*}
  \bar{\lambda}(\zeta_i) = \sum_j f(\hat{\eta}_j)\,\lambda_{ij},
\end{equation*}
where $\{\hat{\eta}_j\}$ are samples drawn from the distribution, $\lambda_{ij}$ is the growing root of Eq.~\eqref{eq:cubic} evaluated at $(\hat{\eta}_j, |\zeta_i|)$, and $f(\hat{\eta}_j)$ the corresponding Gaussian weight. This averaged rate $\bar{\lambda}(\zeta_i)$ then replaces $\lambda(s_{\mathrm{tot}}(\zeta),\mu)$ in Eq.~\eqref{eq:rhoeff} to give the energy-spread-corrected effective Pierce parameter.

To assess how LSC detuning constrains the high-current operating space, Fig.~\ref{fig:LSC_detuning} maps the attainable saturation power $P_{\mathrm{sat}}$, computed via Eq.~\eqref{eq:Psat}, across the $(Q_{\mathrm{beam}},\, L_{\mathrm{beam}})$ plane for fixed beam radius $R_b$ and energy $E_b$, scanning toward the high-charge, short-bunch corner relevant to future hard X-ray beamlines such as FACET-II~\cite{Yakimenko2019PRAB}. Overlaid as dashed contours is the metric $\Lambda_{\mathrm{LSC}}$, which combines the accumulated low-$k$ coherent detuning $\eta_{\mathrm{coh}}$ and the resonant plasma term $k_p/\Gamma$ into a single regime indicator. The three labelled contours mark the operating regimes introduced earlier: negligible suppression ($\Lambda_{\mathrm{LSC}} = 0.2$), onset of gain reduction ($\Lambda_{\mathrm{LSC}} = 0.7$), and breakdown of high-gain lasing ($\Lambda_{\mathrm{LSC}} = 2.0$).

Although $P_{\mathrm{sat}}$ increases with peak current toward the upper-left corner, the $\Lambda_{\mathrm{LSC}}$ contours impose hard practical boundaries on this scaling for an electron-only beam: beyond each contour, growing LSC detuning progressively suppresses gain until coherent lasing is lost entirely. A quasi-neutral pair beam cancels the DC self-field at the source, reducing $\Lambda_{\mathrm{LSC}}$ to near zero regardless of bunch charge or length, and thereby removes these boundaries to preserve access to the high-power domain across the full ultrashort, high-charge operating space.

For the representative setups shown in the Results section, the corresponding $\Lambda_{\mathrm{LSC}}$ values are consistent with the observed transition from near-ideal gain to DC-limited lasing. In the UV benchmark of Fig.~\ref{fig:UV}, the ellipsoidal bunch remains in the weak-detuning regime, with $\Lambda_{\mathrm{LSC}} \approx 0.15$, consistent with the close saturation powers and microbunching efficiency between the $\mathrm{e^-}$ and $\mathrm{e^-/e^+}$ beams. Reshaping the same charge into the pancake geometry drives the metric upward to $\Lambda_{\mathrm{LSC}} \approx 4.5$, placing the electron-only beam in the gain-suppressed regime and matching the observed collapse of full-window bunching and significantly reduced saturation power. For the soft-X-ray cases of Fig.~\ref{fig:Xray}, the lower-current configuration (${\sim}\,25\,\mathrm{kA}$) yields $\Lambda_{\mathrm{LSC}} \approx 0.75$, indicating substantial but not yet catastrophic detuning, whereas the ${\sim}\,100\,\mathrm{kA}$ case reaches $\Lambda_{\mathrm{LSC}} \approx 2$, consistent with the complete failure of the electron-only beam to saturate. In the HHPC-FEL $\gamma$-ray configuration of Fig.~\ref{fig:Gamma}, the corresponding electron-only slice metric is likewise $\Lambda_{\mathrm{LSC}} \approx 3.5$, in line with the absence of coherent emission. Across all cases, the quasi-neutral pair beam eliminates the DC self-field and drives $\Lambda_{\mathrm{LSC}} \to 0$, consistent with the restored full-window lasing seen in the PIC simulations.

\sectionminor{Collective and quantum effects.}
\noindent
This section justifies the classical, collisionless treatment of the co-propagating $\mathrm{e^-/e^+}$ bunch, addressing in turn collective electromagnetic, quantum, and radiation-reaction effects.

\textit{Plasma screening and transverse instabilities.} In the bunch rest frame, the electromagnetic skin depth $\delta = c/\omega_p$, where $\omega_p = \sqrt{n'e^2/(\varepsilon_0 m_e)}$, amounts to hundreds of microns, which is comparable to or larger than the beam radius $R$ in all configurations studied ($R < \delta$). The pair beam therefore lies outside the bulk-plasma screening regime: radiation-field waves penetrate freely, ensuring good mode overlap and efficient beam--radiation coupling. Transverse space-charge forces are likewise negligible, as the undulator-driven oscillation amplitude is far smaller than the beam width at all energies considered. The analytical model consequently retains only the longitudinal space-charge waves entering the FEL dispersion via $k_p = \omega_p/c$, consistent with the cold, paraxial beam approximation.

\textit{Electron-positron annihilation and bound-state formation.} In all configurations, the mean interparticle distance in the bunch rest frame, $d' \sim n'^{-1/3}$, is orders of magnitude larger than the positronium Bohr radius $a_{\mathrm{ps}} = \hbar/(\mu c\alpha) = 2a_0 \approx 0.106\,\mathrm{nm}$. The highest formation rate occurs in the soft X-ray scenario, where $n' = n_{\mathrm{lab}}/\gamma \simeq 10^{19}\,\mathrm{m}^{-3}$ and $d'/a_{\mathrm{Ps}} \simeq 4.3\times10^3$. Using $\langle\sigma v\rangle_{\mathrm{ann}} \simeq \pi r_e^2 c \simeq 7.48\times10^{-21}\,\mathrm{m^3\,s^{-1}}$ (low-energy limit is considered to derive conservative upper bound~\cite{BerestetskiiLifshitzPitaevskiiQED1982}), the direct annihilation rate is $\Gamma_{\mathrm{ann}} = n'\langle\sigma v\rangle_{\mathrm{ann}} \simeq 7.7\times10^{-2}\,\mathrm{s^{-1}}$. Bounding the positronium formation rate conservatively at the same order as direct annihilation~\cite{Gould1989PsPlasma} gives $\Gamma_{\mathrm{tot}} = \Gamma_{\mathrm{ann}} + \Gamma_{\mathrm{Ps}} \lesssim 2\,\Gamma_{\mathrm{ann}} \simeq 1.55\times10^{-1}\,\mathrm{s^{-1}}$. For an undulator length $L_u \simeq 6\,\mathrm{m}$, the rest-frame interaction time is $\tau = L_u/(\gamma c) \simeq 5.1\times10^{-12}\,\mathrm{s}$, giving a total loss fraction $f_{\mathrm{tot}} \simeq \Gamma_{\mathrm{tot}}\,\tau \lesssim 10^{-10}\%$, entirely negligible over the full undulator interaction.

\textit{Quantum diffusion.} Quantum diffusion for the $\gamma$-ray parameters is estimated from the planar undulator result of \citet{agapov2014diffusion},
\begin{align*}
\frac{d\langle(\delta\gamma)^2\rangle}{d(ct)}
= & \frac{122\pi^3}{15}\lambda_c r_e \frac{\gamma^4}{\lambda_u^3} K^3 f(K),
\\
f(K) \simeq & 1.20 + \left(K + 1.50K^2 + 0.95K^3\right)^{-1}.
\end{align*}
For $E = 48\,\mathrm{GeV}$, $\lambda_u = 3\,\mathrm{cm}$, and $K = 2.5$, integrating to $L_{\mathrm{sat}} \sim 18\,\mathrm{m}$ yields $\sigma_{\delta,\mathrm{qd}} \simeq 4.4\times10^{-4}$. Combined in quadrature with the initial spread $\sigma_{\delta,0} = 3\times10^{-4}$,
\begin{equation*}
\sigma_{\delta,\mathrm{tot}} = \left(\sigma_{\delta,0}^2 + \sigma_{\delta,\mathrm{qd}}^2\right)^{1/2} \simeq 5.35\times10^{-4} \simeq 0.45\,\rho_{\mathrm{eff}},
\end{equation*}
for $\rho_{\mathrm{eff}} \simeq 1.2\times10^{-3}$. The total spread thus remains below the FEL bandwidth ($\sigma_{\delta,\mathrm{tot}} \lesssim 0.5\,\rho_{\mathrm{eff}}$) and is not expected to suppress $\gamma$-ray gain~\cite{agapov2014diffusion, huang2007review}.

\textit{Strong-field QED effects.} For peak undulator fields $B_0 \sim 1\,\mathrm{T}$, the strong-field parameter is $\chi = \gamma\,|E_\perp + v\times B|/E_{\mathrm{cr}} \approx \gamma c B_0/E_{\mathrm{cr}}$, where $E_{\mathrm{cr}} = m_e^2 c^3/(e\hbar) \approx 1.3\times10^{18}\,\mathrm{V/m}$ is the QED critical field~\cite{ritusJSLR85, Baier-book}. Across all scenarios $\chi \lesssim 10^{-5}$, confirming that the instantaneous rest-frame field is negligibly small compared to $E_{\mathrm{cr}}$; the classical description is therefore fully valid. Classical radiation reaction is included self-consistently in all PIC simulations via the Landau-Lifshitz model~\cite{landaulifshitz_vol2}.


\sectionmajor{Data availability}
\smallskip

\noindent
Simulation data and Python scripts required to reproduce the figures are available from \c{C}.E. upon reasonable request.

\sectionmajor{Code availability}
\smallskip

\noindent
The key results of this paper have been produced with the publicly available particle-in-cell code \textsc{Smilei}~\cite{derouillat2018smilei} (\textit{version 5.1}) extended with FEL-specific input decks and external algorithm \textsc{RaDiO}. In future, we aim to make this hybrid code available for public use; before then, it can be made available upon reasonable request to \c{C}.E.

\sectionmajor{References}
\bibliography{bibliography}

\begin{thebibliography}{78}%
\makeatletter
\providecommand \@ifxundefined [1]{%
 \@ifx{#1\undefined}
}%
\providecommand \@ifnum [1]{%
 \ifnum #1\expandafter \@firstoftwo
 \else \expandafter \@secondoftwo
 \fi
}%
\providecommand \@ifx [1]{%
 \ifx #1\expandafter \@firstoftwo
 \else \expandafter \@secondoftwo
 \fi
}%
\providecommand \natexlab [1]{#1}%
\providecommand \enquote  [1]{``#1''}%
\providecommand \bibnamefont  [1]{#1}%
\providecommand \bibfnamefont [1]{#1}%
\providecommand \citenamefont [1]{#1}%
\providecommand \href@noop [0]{\@secondoftwo}%
\providecommand \href [0]{\begingroup \@sanitize@url \@href}%
\providecommand \@href[1]{\@@startlink{#1}\@@href}%
\providecommand \@@href[1]{\endgroup#1\@@endlink}%
\providecommand \@sanitize@url [0]{\catcode `\\12\catcode `\$12\catcode
  `\&12\catcode `\#12\catcode `\^12\catcode `\_12\catcode `\%12\relax}%
\providecommand \@@startlink[1]{}%
\providecommand \@@endlink[0]{}%
\providecommand \url  [0]{\begingroup\@sanitize@url \@url }%
\providecommand \@url [1]{\endgroup\@href {#1}{\urlprefix }}%
\providecommand \urlprefix  [0]{URL }%
\providecommand \Eprint [0]{\href }%
\providecommand \doibase [0]{https://doi.org/}%
\providecommand \selectlanguage [0]{\@gobble}%
\providecommand \bibinfo  [0]{\@secondoftwo}%
\providecommand \bibfield  [0]{\@secondoftwo}%
\providecommand \translation [1]{[#1]}%
\providecommand \BibitemOpen [0]{}%
\providecommand \bibitemStop [0]{}%
\providecommand \bibitemNoStop [0]{.\EOS\space}%
\providecommand \EOS [0]{\spacefactor3000\relax}%
\providecommand \BibitemShut  [1]{\csname bibitem#1\endcsname}%
\let\auto@bib@innerbib\@empty
\bibitem [{\citenamefont {Yan}\ \emph {et~al.}(2024)\citenamefont {Yan},
  \citenamefont {Qin}, \citenamefont {Chen}, \citenamefont {Decking},
  \citenamefont {Dijkstal}, \citenamefont {Guetg}, \citenamefont {Inoue},
  \citenamefont {Kujala}, \citenamefont {Liu}, \citenamefont {Long} \emph
  {et~al.}}]{yan2024terawatt}%
  \BibitemOpen
  \bibfield  {author} {\bibinfo {author} {\bibfnamefont {J.}~\bibnamefont
  {Yan}}, \bibinfo {author} {\bibfnamefont {W.}~\bibnamefont {Qin}}, \bibinfo
  {author} {\bibfnamefont {Y.}~\bibnamefont {Chen}}, \bibinfo {author}
  {\bibfnamefont {W.}~\bibnamefont {Decking}}, \bibinfo {author} {\bibfnamefont
  {P.}~\bibnamefont {Dijkstal}}, \bibinfo {author} {\bibfnamefont
  {M.}~\bibnamefont {Guetg}}, \bibinfo {author} {\bibfnamefont
  {I.}~\bibnamefont {Inoue}}, \bibinfo {author} {\bibfnamefont
  {N.}~\bibnamefont {Kujala}}, \bibinfo {author} {\bibfnamefont
  {S.}~\bibnamefont {Liu}}, \bibinfo {author} {\bibfnamefont {T.}~\bibnamefont
  {Long}}, \emph {et~al.},\ }\bibfield  {title} {\bibinfo {title}
  {Terawatt-attosecond hard {X-ray} free-electron laser at high repetition
  rate},\ }\href {https://doi.org/10.1038/s41566-024-01566-0} {\bibfield
  {journal} {\bibinfo  {journal} {Nat. Photonics}\ }\textbf {\bibinfo {volume}
  {18}},\ \bibinfo {pages} {1293} (\bibinfo {year} {2024})}\BibitemShut
  {NoStop}%
\bibitem [{\citenamefont {Franz}\ \emph {et~al.}(2024)\citenamefont {Franz},
  \citenamefont {Li}, \citenamefont {Driver}, \citenamefont {Robles},
  \citenamefont {Cesar}, \citenamefont {Isele}, \citenamefont {Guo},
  \citenamefont {Wang}, \citenamefont {Duris}, \citenamefont {Larsen} \emph
  {et~al.}}]{franz2024terawatt}%
  \BibitemOpen
  \bibfield  {author} {\bibinfo {author} {\bibfnamefont {P.}~\bibnamefont
  {Franz}}, \bibinfo {author} {\bibfnamefont {S.}~\bibnamefont {Li}}, \bibinfo
  {author} {\bibfnamefont {T.}~\bibnamefont {Driver}}, \bibinfo {author}
  {\bibfnamefont {R.~R.}\ \bibnamefont {Robles}}, \bibinfo {author}
  {\bibfnamefont {D.}~\bibnamefont {Cesar}}, \bibinfo {author} {\bibfnamefont
  {E.}~\bibnamefont {Isele}}, \bibinfo {author} {\bibfnamefont
  {Z.}~\bibnamefont {Guo}}, \bibinfo {author} {\bibfnamefont {J.}~\bibnamefont
  {Wang}}, \bibinfo {author} {\bibfnamefont {J.~P.}\ \bibnamefont {Duris}},
  \bibinfo {author} {\bibfnamefont {K.}~\bibnamefont {Larsen}}, \emph
  {et~al.},\ }\bibfield  {title} {\bibinfo {title} {Terawatt-scale attosecond
  {X-ray} pulses from a cascaded superradiant free-electron laser},\ }\href
  {https://doi.org/10.1038/s41566-024-01427-w} {\bibfield  {journal} {\bibinfo
  {journal} {Nat. Photonics}\ }\textbf {\bibinfo {volume} {18}},\ \bibinfo
  {pages} {698} (\bibinfo {year} {2024})}\BibitemShut {NoStop}%
\bibitem [{\citenamefont {Lentrodt}\ \emph {et~al.}(2025)\citenamefont
  {Lentrodt}, \citenamefont {Keitel},\ and\ \citenamefont
  {Evers}}]{Lentrodt2025Toward}%
  \BibitemOpen
  \bibfield  {author} {\bibinfo {author} {\bibfnamefont {D.}~\bibnamefont
  {Lentrodt}}, \bibinfo {author} {\bibfnamefont {C.~H.}\ \bibnamefont
  {Keitel}},\ and\ \bibinfo {author} {\bibfnamefont {J.}~\bibnamefont
  {Evers}},\ }\bibfield  {title} {\bibinfo {title} {Toward nonlinear optics
  with {M}{\"o}ssbauer nuclei using {X-ray} cavities},\ }\href
  {https://doi.org/10.1103/PhysRevLett.135.033801} {\bibfield  {journal}
  {\bibinfo  {journal} {Phys. Rev. Lett.}\ }\textbf {\bibinfo {volume} {135}},\
  \bibinfo {pages} {033801} (\bibinfo {year} {2025})}\BibitemShut {NoStop}%
\bibitem [{\citenamefont {Maroju}\ \emph {et~al.}(2020)\citenamefont {Maroju},
  \citenamefont {Grazioli}, \citenamefont {Di~Fraia}, \citenamefont {Moioli},
  \citenamefont {Ertel}, \citenamefont {Ahmadi}, \citenamefont {Plekan},
  \citenamefont {Finetti}, \citenamefont {Allaria}, \citenamefont {Giannessi},
  \citenamefont {De~Ninno}, \citenamefont {Spezzani}, \citenamefont {Penco},
  \citenamefont {Spampinati}, \citenamefont {Demidovich}, \citenamefont
  {Danailov}, \citenamefont {Borghes}, \citenamefont {Kourousias},
  \citenamefont {Sanches Dos~Reis}, \citenamefont {Bill{\'e}}, \citenamefont
  {Lutman}, \citenamefont {Squibb}, \citenamefont {Feifel}, \citenamefont
  {Carpeggiani}, \citenamefont {Reduzzi}, \citenamefont {Mazza}, \citenamefont
  {Meyer}, \citenamefont {Bengtsson}, \citenamefont {Ibrakovic}, \citenamefont
  {Simpson}, \citenamefont {Mauritsson}, \citenamefont {Csizmadia},
  \citenamefont {Dumergue}, \citenamefont {K{\"u}hn}, \citenamefont
  {Harshitha}, \citenamefont {You}, \citenamefont {Ueda}, \citenamefont
  {Labeye}, \citenamefont {B{\ae}kh{\o}j}, \citenamefont {Schafer},
  \citenamefont {Gryzlova}, \citenamefont {Grum-Grzhimailo}, \citenamefont
  {Prince}, \citenamefont {Callegari},\ and\ \citenamefont
  {Sansone}}]{Maroju2020AttosecondPulseShaping}%
  \BibitemOpen
  \bibfield  {author} {\bibinfo {author} {\bibfnamefont {P.~K.}\ \bibnamefont
  {Maroju}}, \bibinfo {author} {\bibfnamefont {C.}~\bibnamefont {Grazioli}},
  \bibinfo {author} {\bibfnamefont {M.}~\bibnamefont {Di~Fraia}}, \bibinfo
  {author} {\bibfnamefont {M.}~\bibnamefont {Moioli}}, \bibinfo {author}
  {\bibfnamefont {D.}~\bibnamefont {Ertel}}, \bibinfo {author} {\bibfnamefont
  {H.}~\bibnamefont {Ahmadi}}, \bibinfo {author} {\bibfnamefont
  {O.}~\bibnamefont {Plekan}}, \bibinfo {author} {\bibfnamefont
  {P.}~\bibnamefont {Finetti}}, \bibinfo {author} {\bibfnamefont
  {E.}~\bibnamefont {Allaria}}, \bibinfo {author} {\bibfnamefont
  {L.}~\bibnamefont {Giannessi}}, \bibinfo {author} {\bibfnamefont
  {G.}~\bibnamefont {De~Ninno}}, \bibinfo {author} {\bibfnamefont
  {C.}~\bibnamefont {Spezzani}}, \bibinfo {author} {\bibfnamefont
  {G.}~\bibnamefont {Penco}}, \bibinfo {author} {\bibfnamefont
  {S.}~\bibnamefont {Spampinati}}, \bibinfo {author} {\bibfnamefont
  {A.}~\bibnamefont {Demidovich}}, \bibinfo {author} {\bibfnamefont {M.~B.}\
  \bibnamefont {Danailov}}, \bibinfo {author} {\bibfnamefont {R.}~\bibnamefont
  {Borghes}}, \bibinfo {author} {\bibfnamefont {G.}~\bibnamefont {Kourousias}},
  \bibinfo {author} {\bibfnamefont {C.~E.}\ \bibnamefont {Sanches Dos~Reis}},
  \bibinfo {author} {\bibfnamefont {F.}~\bibnamefont {Bill{\'e}}}, \bibinfo
  {author} {\bibfnamefont {A.~A.}\ \bibnamefont {Lutman}}, \bibinfo {author}
  {\bibfnamefont {R.~J.}\ \bibnamefont {Squibb}}, \bibinfo {author}
  {\bibfnamefont {R.}~\bibnamefont {Feifel}}, \bibinfo {author} {\bibfnamefont
  {P.}~\bibnamefont {Carpeggiani}}, \bibinfo {author} {\bibfnamefont
  {M.}~\bibnamefont {Reduzzi}}, \bibinfo {author} {\bibfnamefont
  {T.}~\bibnamefont {Mazza}}, \bibinfo {author} {\bibfnamefont
  {M.}~\bibnamefont {Meyer}}, \bibinfo {author} {\bibfnamefont
  {S.}~\bibnamefont {Bengtsson}}, \bibinfo {author} {\bibfnamefont
  {N.}~\bibnamefont {Ibrakovic}}, \bibinfo {author} {\bibfnamefont {E.~R.}\
  \bibnamefont {Simpson}}, \bibinfo {author} {\bibfnamefont {J.}~\bibnamefont
  {Mauritsson}}, \bibinfo {author} {\bibfnamefont {T.}~\bibnamefont
  {Csizmadia}}, \bibinfo {author} {\bibfnamefont {M.}~\bibnamefont {Dumergue}},
  \bibinfo {author} {\bibfnamefont {S.}~\bibnamefont {K{\"u}hn}}, \bibinfo
  {author} {\bibfnamefont {N.~G.}\ \bibnamefont {Harshitha}}, \bibinfo {author}
  {\bibfnamefont {D.}~\bibnamefont {You}}, \bibinfo {author} {\bibfnamefont
  {K.}~\bibnamefont {Ueda}}, \bibinfo {author} {\bibfnamefont {M.}~\bibnamefont
  {Labeye}}, \bibinfo {author} {\bibfnamefont {J.~E.}\ \bibnamefont
  {B{\ae}kh{\o}j}}, \bibinfo {author} {\bibfnamefont {K.~J.}\ \bibnamefont
  {Schafer}}, \bibinfo {author} {\bibfnamefont {E.~V.}\ \bibnamefont
  {Gryzlova}}, \bibinfo {author} {\bibfnamefont {A.~N.}\ \bibnamefont
  {Grum-Grzhimailo}}, \bibinfo {author} {\bibfnamefont {K.~C.}\ \bibnamefont
  {Prince}}, \bibinfo {author} {\bibfnamefont {C.}~\bibnamefont {Callegari}},\
  and\ \bibinfo {author} {\bibfnamefont {G.}~\bibnamefont {Sansone}},\
  }\bibfield  {title} {\bibinfo {title} {Attosecond pulse shaping using a
  seeded free-electron laser},\ }\href
  {https://doi.org/10.1038/s41586-020-2005-6} {\bibfield  {journal} {\bibinfo
  {journal} {Nature}\ }\textbf {\bibinfo {volume} {578}},\ \bibinfo {pages}
  {386} (\bibinfo {year} {2020})}\BibitemShut {NoStop}%
\bibitem [{\citenamefont {Mayer}\ \emph {et~al.}(2022)\citenamefont {Mayer},
  \citenamefont {Lever}, \citenamefont {Picconi}, \citenamefont {Metje},
  \citenamefont {Alisauskas}, \citenamefont {Calegari}, \citenamefont
  {D{\"u}sterer}, \citenamefont {Ehlert}, \citenamefont {Feifel}, \citenamefont
  {Niebuhr}, \citenamefont {Manschwetus}, \citenamefont {Kuhlmann},
  \citenamefont {Mazza}, \citenamefont {Robinson}, \citenamefont {Squibb},
  \citenamefont {Trabattoni}, \citenamefont {Wallner}, \citenamefont
  {Saalfrank}, \citenamefont {Wolf},\ and\ \citenamefont
  {G{\"u}hr}}]{Mayer2022FollowingExcitedStateChemicalShifts}%
  \BibitemOpen
  \bibfield  {author} {\bibinfo {author} {\bibfnamefont {D.}~\bibnamefont
  {Mayer}}, \bibinfo {author} {\bibfnamefont {F.}~\bibnamefont {Lever}},
  \bibinfo {author} {\bibfnamefont {D.}~\bibnamefont {Picconi}}, \bibinfo
  {author} {\bibfnamefont {J.}~\bibnamefont {Metje}}, \bibinfo {author}
  {\bibfnamefont {S.}~\bibnamefont {Alisauskas}}, \bibinfo {author}
  {\bibfnamefont {F.}~\bibnamefont {Calegari}}, \bibinfo {author}
  {\bibfnamefont {S.}~\bibnamefont {D{\"u}sterer}}, \bibinfo {author}
  {\bibfnamefont {C.}~\bibnamefont {Ehlert}}, \bibinfo {author} {\bibfnamefont
  {R.}~\bibnamefont {Feifel}}, \bibinfo {author} {\bibfnamefont
  {M.}~\bibnamefont {Niebuhr}}, \bibinfo {author} {\bibfnamefont
  {B.}~\bibnamefont {Manschwetus}}, \bibinfo {author} {\bibfnamefont
  {M.}~\bibnamefont {Kuhlmann}}, \bibinfo {author} {\bibfnamefont
  {T.}~\bibnamefont {Mazza}}, \bibinfo {author} {\bibfnamefont {M.~S.}\
  \bibnamefont {Robinson}}, \bibinfo {author} {\bibfnamefont {R.~J.}\
  \bibnamefont {Squibb}}, \bibinfo {author} {\bibfnamefont {A.}~\bibnamefont
  {Trabattoni}}, \bibinfo {author} {\bibfnamefont {M.}~\bibnamefont {Wallner}},
  \bibinfo {author} {\bibfnamefont {P.}~\bibnamefont {Saalfrank}}, \bibinfo
  {author} {\bibfnamefont {T.~J.~A.}\ \bibnamefont {Wolf}},\ and\ \bibinfo
  {author} {\bibfnamefont {M.}~\bibnamefont {G{\"u}hr}},\ }\bibfield  {title}
  {\bibinfo {title} {Following excited-state chemical shifts in molecular
  ultrafast x-ray photoelectron spectroscopy},\ }\href
  {https://doi.org/10.1038/s41467-021-27908-y} {\bibfield  {journal} {\bibinfo
  {journal} {Nat. Commun.}\ }\textbf {\bibinfo {volume} {13}},\ \bibinfo
  {pages} {198} (\bibinfo {year} {2022})}\BibitemShut {NoStop}%
\bibitem [{\citenamefont {Li}\ \emph {et~al.}(2026)\citenamefont {Li},
  \citenamefont {Wang}, \citenamefont {Zhang}, \citenamefont {Song},
  \citenamefont {Sun}, \citenamefont {Ng}, \citenamefont {Sato}, \citenamefont
  {Hanlon}, \citenamefont {Dahal}, \citenamefont {Balcazar}, \citenamefont
  {Esposito}, \citenamefont {She}, \citenamefont {Ornelas-Skarin},
  \citenamefont {Vila-Comamala}, \citenamefont {David}, \citenamefont {Berndt},
  \citenamefont {Miedaner}, \citenamefont {Zhang}, \citenamefont {Ihme},
  \citenamefont {Trigo}, \citenamefont {Nelson}, \citenamefont {Hastings},
  \citenamefont {Maznev}, \citenamefont {Foglia}, \citenamefont {Teitelbaum},
  \citenamefont {Reis},\ and\ \citenamefont
  {Zhu}}]{Li2026NanoscaleUltrafastLatticeModulation}%
  \BibitemOpen
  \bibfield  {author} {\bibinfo {author} {\bibfnamefont {H.}~\bibnamefont
  {Li}}, \bibinfo {author} {\bibfnamefont {N.}~\bibnamefont {Wang}}, \bibinfo
  {author} {\bibfnamefont {L.}~\bibnamefont {Zhang}}, \bibinfo {author}
  {\bibfnamefont {S.}~\bibnamefont {Song}}, \bibinfo {author} {\bibfnamefont
  {Y.}~\bibnamefont {Sun}}, \bibinfo {author} {\bibfnamefont {M.-L.}\
  \bibnamefont {Ng}}, \bibinfo {author} {\bibfnamefont {T.}~\bibnamefont
  {Sato}}, \bibinfo {author} {\bibfnamefont {D.}~\bibnamefont {Hanlon}},
  \bibinfo {author} {\bibfnamefont {S.}~\bibnamefont {Dahal}}, \bibinfo
  {author} {\bibfnamefont {M.~D.}\ \bibnamefont {Balcazar}}, \bibinfo {author}
  {\bibfnamefont {V.}~\bibnamefont {Esposito}}, \bibinfo {author}
  {\bibfnamefont {S.}~\bibnamefont {She}}, \bibinfo {author} {\bibfnamefont
  {C.~C.}\ \bibnamefont {Ornelas-Skarin}}, \bibinfo {author} {\bibfnamefont
  {J.}~\bibnamefont {Vila-Comamala}}, \bibinfo {author} {\bibfnamefont
  {C.}~\bibnamefont {David}}, \bibinfo {author} {\bibfnamefont
  {N.}~\bibnamefont {Berndt}}, \bibinfo {author} {\bibfnamefont {P.~R.}\
  \bibnamefont {Miedaner}}, \bibinfo {author} {\bibfnamefont {Z.}~\bibnamefont
  {Zhang}}, \bibinfo {author} {\bibfnamefont {M.}~\bibnamefont {Ihme}},
  \bibinfo {author} {\bibfnamefont {M.}~\bibnamefont {Trigo}}, \bibinfo
  {author} {\bibfnamefont {K.~A.}\ \bibnamefont {Nelson}}, \bibinfo {author}
  {\bibfnamefont {J.~B.}\ \bibnamefont {Hastings}}, \bibinfo {author}
  {\bibfnamefont {A.~A.}\ \bibnamefont {Maznev}}, \bibinfo {author}
  {\bibfnamefont {L.}~\bibnamefont {Foglia}}, \bibinfo {author} {\bibfnamefont
  {S.}~\bibnamefont {Teitelbaum}}, \bibinfo {author} {\bibfnamefont {D.~A.}\
  \bibnamefont {Reis}},\ and\ \bibinfo {author} {\bibfnamefont
  {D.}~\bibnamefont {Zhu}},\ }\bibfield  {title} {\bibinfo {title} {Nanoscale
  ultrafast lattice modulation with a free-electron laser},\ }\bibfield
  {journal} {\bibinfo  {journal} {Nat. Phys.}\ }\href
  {https://doi.org/10.1038/s41567-025-03161-8} {10.1038/s41567-025-03161-8}
  (\bibinfo {year} {2026})\BibitemShut {NoStop}%
\bibitem [{\citenamefont {Duris}\ \emph {et~al.}(2020)\citenamefont {Duris},
  \citenamefont {Li}, \citenamefont {Driver}, \citenamefont {Champenois},
  \citenamefont {MacArthur}, \citenamefont {Lutman}, \citenamefont {Zhang},
  \citenamefont {Rosenberger}, \citenamefont {Aldrich}, \citenamefont {Coffee}
  \emph {et~al.}}]{duris2020tunable}%
  \BibitemOpen
  \bibfield  {author} {\bibinfo {author} {\bibfnamefont {J.}~\bibnamefont
  {Duris}}, \bibinfo {author} {\bibfnamefont {S.}~\bibnamefont {Li}}, \bibinfo
  {author} {\bibfnamefont {T.}~\bibnamefont {Driver}}, \bibinfo {author}
  {\bibfnamefont {E.~G.}\ \bibnamefont {Champenois}}, \bibinfo {author}
  {\bibfnamefont {J.~P.}\ \bibnamefont {MacArthur}}, \bibinfo {author}
  {\bibfnamefont {A.~A.}\ \bibnamefont {Lutman}}, \bibinfo {author}
  {\bibfnamefont {Z.}~\bibnamefont {Zhang}}, \bibinfo {author} {\bibfnamefont
  {P.}~\bibnamefont {Rosenberger}}, \bibinfo {author} {\bibfnamefont {J.~W.}\
  \bibnamefont {Aldrich}}, \bibinfo {author} {\bibfnamefont {R.}~\bibnamefont
  {Coffee}}, \emph {et~al.},\ }\bibfield  {title} {\bibinfo {title} {Tunable
  isolated attosecond {X-ray} pulses with gigawatt peak power from a
  free-electron laser},\ }\href {https://doi.org/10.1038/s41566-019-0549-5}
  {\bibfield  {journal} {\bibinfo  {journal} {Nat. Photonics}\ }\textbf
  {\bibinfo {volume} {14}},\ \bibinfo {pages} {30} (\bibinfo {year}
  {2020})}\BibitemShut {NoStop}%
\bibitem [{\citenamefont {Lutman}\ \emph {et~al.}(2018)\citenamefont {Lutman},
  \citenamefont {Guetg}, \citenamefont {Maxwell}, \citenamefont {MacArthur},
  \citenamefont {Ding}, \citenamefont {Emma}, \citenamefont {Krzywinski},
  \citenamefont {Marinelli},\ and\ \citenamefont {Huang}}]{lutman2018high}%
  \BibitemOpen
  \bibfield  {author} {\bibinfo {author} {\bibfnamefont {A.~A.}\ \bibnamefont
  {Lutman}}, \bibinfo {author} {\bibfnamefont {M.~W.}\ \bibnamefont {Guetg}},
  \bibinfo {author} {\bibfnamefont {T.~J.}\ \bibnamefont {Maxwell}}, \bibinfo
  {author} {\bibfnamefont {J.~P.}\ \bibnamefont {MacArthur}}, \bibinfo {author}
  {\bibfnamefont {Y.}~\bibnamefont {Ding}}, \bibinfo {author} {\bibfnamefont
  {C.}~\bibnamefont {Emma}}, \bibinfo {author} {\bibfnamefont {J.}~\bibnamefont
  {Krzywinski}}, \bibinfo {author} {\bibfnamefont {A.}~\bibnamefont
  {Marinelli}},\ and\ \bibinfo {author} {\bibfnamefont {Z.}~\bibnamefont
  {Huang}},\ }\bibfield  {title} {\bibinfo {title} {High-power femtosecond soft
  x rays from fresh-slice multistage free-electron lasers},\ }\href
  {https://doi.org/10.1103/PhysRevLett.120.264801} {\bibfield  {journal}
  {\bibinfo  {journal} {Phys. Rev. Lett.}\ }\textbf {\bibinfo {volume} {120}},\
  \bibinfo {pages} {264801} (\bibinfo {year} {2018})}\BibitemShut {NoStop}%
\bibitem [{\citenamefont {Saldin}\ \emph {et~al.}(2006)\citenamefont {Saldin},
  \citenamefont {Schneidmiller},\ and\ \citenamefont
  {Yurkov}}]{saldin2006chirptaper}%
  \BibitemOpen
  \bibfield  {author} {\bibinfo {author} {\bibfnamefont {E.~L.}\ \bibnamefont
  {Saldin}}, \bibinfo {author} {\bibfnamefont {E.~A.}\ \bibnamefont
  {Schneidmiller}},\ and\ \bibinfo {author} {\bibfnamefont {M.~V.}\
  \bibnamefont {Yurkov}},\ }\bibfield  {title} {\bibinfo {title}
  {Self-amplified spontaneous emission fel with energy-chirped electron beam
  and its application for generation of attosecond {X-ray} pulses},\ }\href
  {https://doi.org/10.1103/PhysRevSTAB.9.050702} {\bibfield  {journal}
  {\bibinfo  {journal} {Phys. Rev. ST Accel. Beams}\ }\textbf {\bibinfo
  {volume} {9}},\ \bibinfo {pages} {050702} (\bibinfo {year}
  {2006})}\BibitemShut {NoStop}%
\bibitem [{\citenamefont {Huang}\ \emph {et~al.}(2017)\citenamefont {Huang},
  \citenamefont {Ding}, \citenamefont {Feng}, \citenamefont {Hemsing},
  \citenamefont {Huang}, \citenamefont {Krzywinski}, \citenamefont {Lutman},
  \citenamefont {Marinelli}, \citenamefont {Maxwell},\ and\ \citenamefont
  {Zhu}}]{huang2017generating}%
  \BibitemOpen
  \bibfield  {author} {\bibinfo {author} {\bibfnamefont {S.}~\bibnamefont
  {Huang}}, \bibinfo {author} {\bibfnamefont {Y.}~\bibnamefont {Ding}},
  \bibinfo {author} {\bibfnamefont {Y.}~\bibnamefont {Feng}}, \bibinfo {author}
  {\bibfnamefont {E.}~\bibnamefont {Hemsing}}, \bibinfo {author} {\bibfnamefont
  {Z.}~\bibnamefont {Huang}}, \bibinfo {author} {\bibfnamefont
  {J.}~\bibnamefont {Krzywinski}}, \bibinfo {author} {\bibfnamefont {A.~A.}\
  \bibnamefont {Lutman}}, \bibinfo {author} {\bibfnamefont {A.}~\bibnamefont
  {Marinelli}}, \bibinfo {author} {\bibfnamefont {T.~J.}\ \bibnamefont
  {Maxwell}},\ and\ \bibinfo {author} {\bibfnamefont {D.}~\bibnamefont {Zhu}},\
  }\bibfield  {title} {\bibinfo {title} {Generating single-spike hard {X-ray}
  pulses with nonlinear bunch compression in free-electron lasers},\ }\href
  {https://doi.org/10.1103/PhysRevLett.119.154801} {\bibfield  {journal}
  {\bibinfo  {journal} {Phys. Rev. Lett.}\ }\textbf {\bibinfo {volume} {119}},\
  \bibinfo {pages} {154801} (\bibinfo {year} {2017})}\BibitemShut {NoStop}%
\bibitem [{\citenamefont {Quin}\ \emph
  {et~al.}(2025{\natexlab{a}})\citenamefont {Quin}, \citenamefont {Di~Piazza},
  \citenamefont {Erciyes}, \citenamefont {Keitel},\ and\ \citenamefont
  {Tamburini}}]{quin2025broadband}%
  \BibitemOpen
  \bibfield  {author} {\bibinfo {author} {\bibfnamefont {M.~J.}\ \bibnamefont
  {Quin}}, \bibinfo {author} {\bibfnamefont {A.}~\bibnamefont {Di~Piazza}},
  \bibinfo {author} {\bibfnamefont {{\c{C}}.}~\bibnamefont {Erciyes}}, \bibinfo
  {author} {\bibfnamefont {C.~H.}\ \bibnamefont {Keitel}},\ and\ \bibinfo
  {author} {\bibfnamefont {M.}~\bibnamefont {Tamburini}},\ }\bibfield  {title}
  {\bibinfo {title} {Broadband coherent {XUV} light from e$^{-}$/e$^{+}$
  microbunching in an intense laser pulse},\ }\href
  {https://doi.org/10.1038/s42005-025-02209-8} {\bibfield  {journal} {\bibinfo
  {journal} {Commun. Phys.}\ }\textbf {\bibinfo {volume} {8}},\ \bibinfo
  {pages} {282} (\bibinfo {year} {2025}{\natexlab{a}})}\BibitemShut {NoStop}%
\bibitem [{\citenamefont {Serafini}(1997)}]{serafini1997short}%
  \BibitemOpen
  \bibfield  {author} {\bibinfo {author} {\bibfnamefont {L.}~\bibnamefont
  {Serafini}},\ }\bibfield  {title} {\bibinfo {title} {The short bunch blow-out
  regime in {RF} photoinjectors},\ }in\ \href {https://doi.org/10.1063/1.54425}
  {\emph {\bibinfo {booktitle} {AIP Conf. Proc.}}},\ Vol.\ \bibinfo {volume}
  {413}\ (\bibinfo {year} {1997})\ pp.\ \bibinfo {pages} {321--334}\BibitemShut
  {NoStop}%
\bibitem [{\citenamefont {Moody}\ \emph {et~al.}(2009)\citenamefont {Moody},
  \citenamefont {Musumeci}, \citenamefont {Gutierrez}, \citenamefont
  {Rosenzweig},\ and\ \citenamefont {Scoby}}]{moody2009longitudinal}%
  \BibitemOpen
  \bibfield  {author} {\bibinfo {author} {\bibfnamefont {J.~T.}\ \bibnamefont
  {Moody}}, \bibinfo {author} {\bibfnamefont {P.}~\bibnamefont {Musumeci}},
  \bibinfo {author} {\bibfnamefont {M.~S.}\ \bibnamefont {Gutierrez}}, \bibinfo
  {author} {\bibfnamefont {J.~B.}\ \bibnamefont {Rosenzweig}},\ and\ \bibinfo
  {author} {\bibfnamefont {C.~M.}\ \bibnamefont {Scoby}},\ }\bibfield  {title}
  {\bibinfo {title} {Longitudinal phase space characterization of the blow-out
  regime of rf photoinjector operation},\ }\href
  {https://doi.org/10.1103/PhysRevSTAB.12.070704} {\bibfield  {journal}
  {\bibinfo  {journal} {Phys. Rev. ST Accel. Beams}\ }\textbf {\bibinfo
  {volume} {12}},\ \bibinfo {pages} {070704} (\bibinfo {year}
  {2009})}\BibitemShut {NoStop}%
\bibitem [{\citenamefont {Campbell}\ \emph {et~al.}(2020)\citenamefont
  {Campbell}, \citenamefont {Freund}, \citenamefont {Henderson}, \citenamefont
  {McNeil}, \citenamefont {Traczykowski},\ and\ \citenamefont {van~der
  Slot}}]{campbell2020analysis}%
  \BibitemOpen
  \bibfield  {author} {\bibinfo {author} {\bibfnamefont {L.~T.}\ \bibnamefont
  {Campbell}}, \bibinfo {author} {\bibfnamefont {H.~P.}\ \bibnamefont
  {Freund}}, \bibinfo {author} {\bibfnamefont {J.~R.}\ \bibnamefont
  {Henderson}}, \bibinfo {author} {\bibfnamefont {B.~W.~J.}\ \bibnamefont
  {McNeil}}, \bibinfo {author} {\bibfnamefont {P.}~\bibnamefont
  {Traczykowski}},\ and\ \bibinfo {author} {\bibfnamefont {P.~J.~M.}\
  \bibnamefont {van~der Slot}},\ }\bibfield  {title} {\bibinfo {title}
  {Analysis of ultra-short bunches in free-electron lasers},\ }\href
  {https://doi.org/10.1088/1367-2630/ab9850} {\bibfield  {journal} {\bibinfo
  {journal} {New J. Phys.}\ }\textbf {\bibinfo {volume} {22}},\ \bibinfo
  {pages} {073031} (\bibinfo {year} {2020})}\BibitemShut {NoStop}%
\bibitem [{\citenamefont {Hernandez-Garcia}\ \emph {et~al.}(2004)\citenamefont
  {Hernandez-Garcia}, \citenamefont {Beard}, \citenamefont {Behre},
  \citenamefont {Benson}, \citenamefont {Biallas}, \citenamefont {Boyce},
  \citenamefont {Douglas}, \citenamefont {Dylla}, \citenamefont {Evans},
  \citenamefont {Grippo} \emph {et~al.}}]{hernandez2004longitudinal}%
  \BibitemOpen
  \bibfield  {author} {\bibinfo {author} {\bibfnamefont {C.}~\bibnamefont
  {Hernandez-Garcia}}, \bibinfo {author} {\bibfnamefont {K.}~\bibnamefont
  {Beard}}, \bibinfo {author} {\bibfnamefont {C.}~\bibnamefont {Behre}},
  \bibinfo {author} {\bibfnamefont {S.}~\bibnamefont {Benson}}, \bibinfo
  {author} {\bibfnamefont {G.}~\bibnamefont {Biallas}}, \bibinfo {author}
  {\bibfnamefont {J.}~\bibnamefont {Boyce}}, \bibinfo {author} {\bibfnamefont
  {D.}~\bibnamefont {Douglas}}, \bibinfo {author} {\bibfnamefont {H.~F.}\
  \bibnamefont {Dylla}}, \bibinfo {author} {\bibfnamefont {R.}~\bibnamefont
  {Evans}}, \bibinfo {author} {\bibfnamefont {A.}~\bibnamefont {Grippo}}, \emph
  {et~al.},\ }\bibfield  {title} {\bibinfo {title} {Longitudinal space charge
  effects in the {JLAB IR FEL SRF} linac},\ }in\ \href
  {https://accelconf.web.cern.ch/f04/papers/TUBOS02/TUBOS02.PDF} {\emph
  {\bibinfo {booktitle} {Proc. FEL 2004 (26th Int. Free-Electron Laser
  Conf.)}}}\ (\bibinfo {year} {2004})\ \bibinfo {note} {paper
  TUBOS02}\BibitemShut {NoStop}%
\bibitem [{\citenamefont {Huang}\ and\ \citenamefont
  {Kim}(2007)}]{huang2007review}%
  \BibitemOpen
  \bibfield  {author} {\bibinfo {author} {\bibfnamefont {Z.}~\bibnamefont
  {Huang}}\ and\ \bibinfo {author} {\bibfnamefont {K.-J.}\ \bibnamefont
  {Kim}},\ }\bibfield  {title} {\bibinfo {title} {Review of {X-ray}
  free-electron laser theory},\ }\href
  {https://doi.org/10.1103/PhysRevSTAB.10.034801} {\bibfield  {journal}
  {\bibinfo  {journal} {Phys. Rev. ST Accel. Beams}\ }\textbf {\bibinfo
  {volume} {10}},\ \bibinfo {pages} {034801} (\bibinfo {year}
  {2007})}\BibitemShut {NoStop}%
\bibitem [{\citenamefont {Xie}(1995)}]{Xie1995PAC3D}%
  \BibitemOpen
  \bibfield  {author} {\bibinfo {author} {\bibfnamefont {M.}~\bibnamefont
  {Xie}},\ }\bibfield  {title} {\bibinfo {title} {Design optimization for an
  {X-ray} free electron laser driven by {SLAC} linac},\ }in\ \href
  {https://doi.org/10.1109/PAC.1995.504603} {\emph {\bibinfo {booktitle} {Proc.
  1995 Particle Accelerator Conf. (PAC'95)}}}\ (\bibinfo {address} {Dallas,
  Texas, USA},\ \bibinfo {year} {1995})\ pp.\ \bibinfo {pages}
  {183--185}\BibitemShut {NoStop}%
\bibitem [{\citenamefont {Nedorezov}\ \emph {et~al.}(2021)\citenamefont
  {Nedorezov}, \citenamefont {Rykovanov},\ and\ \citenamefont
  {Savel'ev}}]{nedorezov2021nuclear}%
  \BibitemOpen
  \bibfield  {author} {\bibinfo {author} {\bibfnamefont {V.~G.}\ \bibnamefont
  {Nedorezov}}, \bibinfo {author} {\bibfnamefont {S.~G.}\ \bibnamefont
  {Rykovanov}},\ and\ \bibinfo {author} {\bibfnamefont {A.~B.}\ \bibnamefont
  {Savel'ev}},\ }\bibfield  {title} {\bibinfo {title} {Nuclear photonics:
  results and prospects},\ }\href {https://doi.org/10.3367/UFNe.2021.03.038960}
  {\bibfield  {journal} {\bibinfo  {journal} {Phys.-Usp.}\ }\textbf {\bibinfo
  {volume} {64}},\ \bibinfo {pages} {1214} (\bibinfo {year}
  {2021})}\BibitemShut {NoStop}%
\bibitem [{\citenamefont {Heeg}\ \emph {et~al.}(2017)\citenamefont {Heeg},
  \citenamefont {Kaldun}, \citenamefont {Strohm}, \citenamefont {Reiser},
  \citenamefont {Ott}, \citenamefont {Subramanian}, \citenamefont {Lentrodt},
  \citenamefont {Haber}, \citenamefont {Wille}, \citenamefont {Goerttler} \emph
  {et~al.}}]{heeg2017spectral}%
  \BibitemOpen
  \bibfield  {author} {\bibinfo {author} {\bibfnamefont {K.~P.}\ \bibnamefont
  {Heeg}}, \bibinfo {author} {\bibfnamefont {A.}~\bibnamefont {Kaldun}},
  \bibinfo {author} {\bibfnamefont {C.}~\bibnamefont {Strohm}}, \bibinfo
  {author} {\bibfnamefont {P.}~\bibnamefont {Reiser}}, \bibinfo {author}
  {\bibfnamefont {C.}~\bibnamefont {Ott}}, \bibinfo {author} {\bibfnamefont
  {R.}~\bibnamefont {Subramanian}}, \bibinfo {author} {\bibfnamefont
  {D.}~\bibnamefont {Lentrodt}}, \bibinfo {author} {\bibfnamefont
  {J.}~\bibnamefont {Haber}}, \bibinfo {author} {\bibfnamefont {H.-C.}\
  \bibnamefont {Wille}}, \bibinfo {author} {\bibfnamefont {S.}~\bibnamefont
  {Goerttler}}, \emph {et~al.},\ }\bibfield  {title} {\bibinfo {title}
  {Spectral narrowing of {X-ray} pulses for precision spectroscopy with nuclear
  resonances},\ }\href {https://doi.org/10.1126/science.aan3512} {\bibfield
  {journal} {\bibinfo  {journal} {Science}\ }\textbf {\bibinfo {volume}
  {357}},\ \bibinfo {pages} {375} (\bibinfo {year} {2017})}\BibitemShut
  {NoStop}%
\bibitem [{\citenamefont {Chen}\ \emph {et~al.}(2015)\citenamefont {Chen},
  \citenamefont {Fiuza}, \citenamefont {Link}, \citenamefont {Hazi},
  \citenamefont {Hill}, \citenamefont {Hoarty}, \citenamefont {James},
  \citenamefont {Kerr}, \citenamefont {Meyerhofer}, \citenamefont {Myatt} \emph
  {et~al.}}]{chen2015scaling}%
  \BibitemOpen
  \bibfield  {author} {\bibinfo {author} {\bibfnamefont {H.}~\bibnamefont
  {Chen}}, \bibinfo {author} {\bibfnamefont {F.}~\bibnamefont {Fiuza}},
  \bibinfo {author} {\bibfnamefont {A.}~\bibnamefont {Link}}, \bibinfo {author}
  {\bibfnamefont {A.}~\bibnamefont {Hazi}}, \bibinfo {author} {\bibfnamefont
  {M.}~\bibnamefont {Hill}}, \bibinfo {author} {\bibfnamefont {D.}~\bibnamefont
  {Hoarty}}, \bibinfo {author} {\bibfnamefont {S.}~\bibnamefont {James}},
  \bibinfo {author} {\bibfnamefont {S.}~\bibnamefont {Kerr}}, \bibinfo {author}
  {\bibfnamefont {D.~D.}\ \bibnamefont {Meyerhofer}}, \bibinfo {author}
  {\bibfnamefont {J.}~\bibnamefont {Myatt}}, \emph {et~al.},\ }\bibfield
  {title} {\bibinfo {title} {Scaling the yield of laser-driven
  electron-positron jets to laboratory astrophysical applications},\ }\href
  {https://doi.org/10.1103/PhysRevLett.114.215001} {\bibfield  {journal}
  {\bibinfo  {journal} {Phys. Rev. Lett.}\ }\textbf {\bibinfo {volume} {114}},\
  \bibinfo {pages} {215001} (\bibinfo {year} {2015})}\BibitemShut {NoStop}%
\bibitem [{\citenamefont {Arrowsmith}\ \emph {et~al.}(2024)\citenamefont
  {Arrowsmith}, \citenamefont {Simon}, \citenamefont {Bilbao}, \citenamefont
  {Bott}, \citenamefont {Burger}, \citenamefont {Chen}, \citenamefont {Cruz},
  \citenamefont {Davenne}, \citenamefont {Efthymiopoulos}, \citenamefont
  {Froula} \emph {et~al.}}]{arrowsmith2024laboratory}%
  \BibitemOpen
  \bibfield  {author} {\bibinfo {author} {\bibfnamefont {C.~D.}\ \bibnamefont
  {Arrowsmith}}, \bibinfo {author} {\bibfnamefont {P.}~\bibnamefont {Simon}},
  \bibinfo {author} {\bibfnamefont {P.~J.}\ \bibnamefont {Bilbao}}, \bibinfo
  {author} {\bibfnamefont {A.~F.~A.}\ \bibnamefont {Bott}}, \bibinfo {author}
  {\bibfnamefont {S.}~\bibnamefont {Burger}}, \bibinfo {author} {\bibfnamefont
  {H.}~\bibnamefont {Chen}}, \bibinfo {author} {\bibfnamefont {F.~D.}\
  \bibnamefont {Cruz}}, \bibinfo {author} {\bibfnamefont {T.}~\bibnamefont
  {Davenne}}, \bibinfo {author} {\bibfnamefont {I.}~\bibnamefont
  {Efthymiopoulos}}, \bibinfo {author} {\bibfnamefont {D.~H.}\ \bibnamefont
  {Froula}}, \emph {et~al.},\ }\bibfield  {title} {\bibinfo {title} {Laboratory
  realization of relativistic pair-plasma beams},\ }\href
  {https://doi.org/10.1038/s41467-024-49346-2} {\bibfield  {journal} {\bibinfo
  {journal} {Nat. Commun.}\ }\textbf {\bibinfo {volume} {15}},\ \bibinfo
  {pages} {5029} (\bibinfo {year} {2024})}\BibitemShut {NoStop}%
\bibitem [{\citenamefont {Streeter}\ \emph {et~al.}(2024)\citenamefont
  {Streeter}, \citenamefont {Colgan}, \citenamefont {Carderelli}, \citenamefont
  {Ma}, \citenamefont {Cavanagh}, \citenamefont {Los}, \citenamefont {Ahmed},
  \citenamefont {Antoine}, \citenamefont {Audet}, \citenamefont {Balcazar}
  \emph {et~al.}}]{streeter2024narrow}%
  \BibitemOpen
  \bibfield  {author} {\bibinfo {author} {\bibfnamefont {M.~J.~V.}\
  \bibnamefont {Streeter}}, \bibinfo {author} {\bibfnamefont {C.}~\bibnamefont
  {Colgan}}, \bibinfo {author} {\bibfnamefont {J.}~\bibnamefont {Carderelli}},
  \bibinfo {author} {\bibfnamefont {Y.}~\bibnamefont {Ma}}, \bibinfo {author}
  {\bibfnamefont {N.}~\bibnamefont {Cavanagh}}, \bibinfo {author}
  {\bibfnamefont {E.~E.}\ \bibnamefont {Los}}, \bibinfo {author} {\bibfnamefont
  {H.}~\bibnamefont {Ahmed}}, \bibinfo {author} {\bibfnamefont {A.~F.}\
  \bibnamefont {Antoine}}, \bibinfo {author} {\bibfnamefont {T.}~\bibnamefont
  {Audet}}, \bibinfo {author} {\bibfnamefont {M.~D.}\ \bibnamefont {Balcazar}},
  \emph {et~al.},\ }\bibfield  {title} {\bibinfo {title} {Narrow bandwidth,
  low-emittance positron beams from a laser-wakefield accelerator},\ }\href
  {https://doi.org/10.1038/s41598-024-56281-1} {\bibfield  {journal} {\bibinfo
  {journal} {Sci. Rep.}\ }\textbf {\bibinfo {volume} {14}},\ \bibinfo {pages}
  {6001} (\bibinfo {year} {2024})}\BibitemShut {NoStop}%
\bibitem [{\citenamefont {Noh}\ \emph {et~al.}(2024)\citenamefont {Noh},
  \citenamefont {Song}, \citenamefont {Mirzaie}, \citenamefont {Hojbota},
  \citenamefont {Kim}, \citenamefont {Lee}, \citenamefont {Won}, \citenamefont
  {Song}, \citenamefont {Song}, \citenamefont {Ryu} \emph
  {et~al.}}]{noh2024charge}%
  \BibitemOpen
  \bibfield  {author} {\bibinfo {author} {\bibfnamefont {Y.}~\bibnamefont
  {Noh}}, \bibinfo {author} {\bibfnamefont {J.}~\bibnamefont {Song}}, \bibinfo
  {author} {\bibfnamefont {M.}~\bibnamefont {Mirzaie}}, \bibinfo {author}
  {\bibfnamefont {C.~I.}\ \bibnamefont {Hojbota}}, \bibinfo {author}
  {\bibfnamefont {H.-i.}\ \bibnamefont {Kim}}, \bibinfo {author} {\bibfnamefont
  {S.}~\bibnamefont {Lee}}, \bibinfo {author} {\bibfnamefont {J.}~\bibnamefont
  {Won}}, \bibinfo {author} {\bibfnamefont {H.}~\bibnamefont {Song}}, \bibinfo
  {author} {\bibfnamefont {C.}~\bibnamefont {Song}}, \bibinfo {author}
  {\bibfnamefont {C.-M.}\ \bibnamefont {Ryu}}, \emph {et~al.},\ }\bibfield
  {title} {\bibinfo {title} {Charge-neutral, {GeV}-scale electron-positron pair
  beams produced using bremsstrahlung gamma rays},\ }\href
  {https://doi.org/10.1038/s42005-024-01527-7} {\bibfield  {journal} {\bibinfo
  {journal} {Commun. Phys.}\ }\textbf {\bibinfo {volume} {7}},\ \bibinfo
  {pages} {44} (\bibinfo {year} {2024})}\BibitemShut {NoStop}%
\bibitem [{\citenamefont {Winkler}\ \emph {et~al.}(2025)\citenamefont
  {Winkler}, \citenamefont {Trunk}, \citenamefont {H{\"u}bner}, \citenamefont
  {Martinez de~la Ossa}, \citenamefont {Jalas}, \citenamefont {Kirchen},
  \citenamefont {Agapov}, \citenamefont {Antipov}, \citenamefont {Brinkmann},
  \citenamefont {Eichner} \emph {et~al.}}]{winkler2025active}%
  \BibitemOpen
  \bibfield  {author} {\bibinfo {author} {\bibfnamefont {P.}~\bibnamefont
  {Winkler}}, \bibinfo {author} {\bibfnamefont {M.}~\bibnamefont {Trunk}},
  \bibinfo {author} {\bibfnamefont {L.}~\bibnamefont {H{\"u}bner}}, \bibinfo
  {author} {\bibfnamefont {A.}~\bibnamefont {Martinez de~la Ossa}}, \bibinfo
  {author} {\bibfnamefont {S.}~\bibnamefont {Jalas}}, \bibinfo {author}
  {\bibfnamefont {M.}~\bibnamefont {Kirchen}}, \bibinfo {author} {\bibfnamefont
  {I.}~\bibnamefont {Agapov}}, \bibinfo {author} {\bibfnamefont {S.~A.}\
  \bibnamefont {Antipov}}, \bibinfo {author} {\bibfnamefont {R.}~\bibnamefont
  {Brinkmann}}, \bibinfo {author} {\bibfnamefont {T.}~\bibnamefont {Eichner}},
  \emph {et~al.},\ }\bibfield  {title} {\bibinfo {title} {Active energy
  compression of a laser-plasma electron beam},\ }\href
  {https://doi.org/10.1038/s41586-025-08772-y} {\bibfield  {journal} {\bibinfo
  {journal} {Nature}\ ,\ \bibinfo {pages} {1}} (\bibinfo {year}
  {2025})}\BibitemShut {NoStop}%
\bibitem [{\citenamefont {Wang}\ \emph {et~al.}(2016)\citenamefont {Wang},
  \citenamefont {Li}, \citenamefont {Liu}, \citenamefont {Zhang}, \citenamefont
  {Qi}, \citenamefont {Yu}, \citenamefont {Liu}, \citenamefont {Fang},
  \citenamefont {Qin}, \citenamefont {Wang} \emph {et~al.}}]{wang2016high}%
  \BibitemOpen
  \bibfield  {author} {\bibinfo {author} {\bibfnamefont {W.~T.}\ \bibnamefont
  {Wang}}, \bibinfo {author} {\bibfnamefont {W.~T.}\ \bibnamefont {Li}},
  \bibinfo {author} {\bibfnamefont {J.~S.}\ \bibnamefont {Liu}}, \bibinfo
  {author} {\bibfnamefont {Z.~J.}\ \bibnamefont {Zhang}}, \bibinfo {author}
  {\bibfnamefont {R.}~\bibnamefont {Qi}}, \bibinfo {author} {\bibfnamefont
  {C.~H.}\ \bibnamefont {Yu}}, \bibinfo {author} {\bibfnamefont {J.~Q.}\
  \bibnamefont {Liu}}, \bibinfo {author} {\bibfnamefont {M.}~\bibnamefont
  {Fang}}, \bibinfo {author} {\bibfnamefont {Z.~Y.}\ \bibnamefont {Qin}},
  \bibinfo {author} {\bibfnamefont {C.}~\bibnamefont {Wang}}, \emph {et~al.},\
  }\bibfield  {title} {\bibinfo {title} {High-brightness high-energy electron
  beams from a laser wakefield accelerator via energy chirp control},\ }\href
  {https://doi.org/10.1103/PhysRevLett.117.124801} {\bibfield  {journal}
  {\bibinfo  {journal} {Phys. Rev. Lett.}\ }\textbf {\bibinfo {volume} {117}},\
  \bibinfo {pages} {124801} (\bibinfo {year} {2016})}\BibitemShut {NoStop}%
\bibitem [{\citenamefont {Habib}\ \emph {et~al.}(2023)\citenamefont {Habib},
  \citenamefont {Manahan}, \citenamefont {Scherkl}, \citenamefont {Heinemann},
  \citenamefont {Sutherland}, \citenamefont {Altuiri}, \citenamefont
  {Alotaibi}, \citenamefont {Litos}, \citenamefont {Cary}, \citenamefont
  {Raubenheimer} \emph {et~al.}}]{habib2023attosecond}%
  \BibitemOpen
  \bibfield  {author} {\bibinfo {author} {\bibfnamefont {A.~F.}\ \bibnamefont
  {Habib}}, \bibinfo {author} {\bibfnamefont {G.~G.}\ \bibnamefont {Manahan}},
  \bibinfo {author} {\bibfnamefont {P.}~\bibnamefont {Scherkl}}, \bibinfo
  {author} {\bibfnamefont {T.}~\bibnamefont {Heinemann}}, \bibinfo {author}
  {\bibfnamefont {A.}~\bibnamefont {Sutherland}}, \bibinfo {author}
  {\bibfnamefont {R.}~\bibnamefont {Altuiri}}, \bibinfo {author} {\bibfnamefont
  {B.~M.}\ \bibnamefont {Alotaibi}}, \bibinfo {author} {\bibfnamefont
  {M.}~\bibnamefont {Litos}}, \bibinfo {author} {\bibfnamefont
  {J.}~\bibnamefont {Cary}}, \bibinfo {author} {\bibfnamefont {T.}~\bibnamefont
  {Raubenheimer}}, \emph {et~al.},\ }\bibfield  {title} {\bibinfo {title}
  {Attosecond-\aa{}ngstrom free-electron-laser towards the cold beam limit},\
  }\href {https://doi.org/10.1038/s41467-023-36592-z} {\bibfield  {journal}
  {\bibinfo  {journal} {Nat. Commun.}\ }\textbf {\bibinfo {volume} {14}},\
  \bibinfo {pages} {1054} (\bibinfo {year} {2023})}\BibitemShut {NoStop}%
\bibitem [{\citenamefont {Emma}\ \emph {et~al.}(2025)\citenamefont {Emma},
  \citenamefont {Majernik}, \citenamefont {Swanson}, \citenamefont {Ariniello},
  \citenamefont {Gessner}, \citenamefont {Hessami}, \citenamefont {Hogan},
  \citenamefont {Knetsch}, \citenamefont {Larsen}, \citenamefont {Marinelli}
  \emph {et~al.}}]{emma2025experimental}%
  \BibitemOpen
  \bibfield  {author} {\bibinfo {author} {\bibfnamefont {C.}~\bibnamefont
  {Emma}}, \bibinfo {author} {\bibfnamefont {N.}~\bibnamefont {Majernik}},
  \bibinfo {author} {\bibfnamefont {K.~K.}\ \bibnamefont {Swanson}}, \bibinfo
  {author} {\bibfnamefont {R.}~\bibnamefont {Ariniello}}, \bibinfo {author}
  {\bibfnamefont {S.}~\bibnamefont {Gessner}}, \bibinfo {author} {\bibfnamefont
  {R.}~\bibnamefont {Hessami}}, \bibinfo {author} {\bibfnamefont {M.~J.}\
  \bibnamefont {Hogan}}, \bibinfo {author} {\bibfnamefont {A.}~\bibnamefont
  {Knetsch}}, \bibinfo {author} {\bibfnamefont {K.~A.}\ \bibnamefont {Larsen}},
  \bibinfo {author} {\bibfnamefont {A.}~\bibnamefont {Marinelli}}, \emph
  {et~al.},\ }\bibfield  {title} {\bibinfo {title} {Experimental generation of
  extreme electron beams for advanced accelerator applications},\ }\href
  {https://doi.org/10.1103/PhysRevLett.134.085001} {\bibfield  {journal}
  {\bibinfo  {journal} {Phys. Rev. Lett.}\ }\textbf {\bibinfo {volume} {134}},\
  \bibinfo {pages} {085001} (\bibinfo {year} {2025})}\BibitemShut {NoStop}%
\bibitem [{\citenamefont {Litos}\ \emph {et~al.}(2014)\citenamefont {Litos},
  \citenamefont {Adli}, \citenamefont {An}, \citenamefont {Clarke},
  \citenamefont {Clayton} \emph {et~al.}}]{Litos2014Nature}%
  \BibitemOpen
  \bibfield  {author} {\bibinfo {author} {\bibfnamefont {M.}~\bibnamefont
  {Litos}}, \bibinfo {author} {\bibfnamefont {E.}~\bibnamefont {Adli}},
  \bibinfo {author} {\bibfnamefont {W.}~\bibnamefont {An}}, \bibinfo {author}
  {\bibfnamefont {C.~I.}\ \bibnamefont {Clarke}}, \bibinfo {author}
  {\bibfnamefont {C.~E.}\ \bibnamefont {Clayton}}, \emph {et~al.},\ }\bibfield
  {title} {\bibinfo {title} {High-efficiency acceleration of an electron beam
  in a plasma wakefield accelerator},\ }\href
  {https://doi.org/10.1038/nature13882} {\bibfield  {journal} {\bibinfo
  {journal} {Nature}\ }\textbf {\bibinfo {volume} {515}},\ \bibinfo {pages}
  {92} (\bibinfo {year} {2014})}\BibitemShut {NoStop}%
\bibitem [{\citenamefont {Yakimenko}\ \emph {et~al.}(2019)\citenamefont
  {Yakimenko} \emph {et~al.}}]{Yakimenko2019PRAB}%
  \BibitemOpen
  \bibfield  {author} {\bibinfo {author} {\bibfnamefont {V.}~\bibnamefont
  {Yakimenko}} \emph {et~al.},\ }\bibfield  {title} {\bibinfo {title}
  {{FACET-II} facility for advanced accelerator experimental tests},\ }\href
  {https://doi.org/10.1103/PhysRevAccelBeams.22.101301} {\bibfield  {journal}
  {\bibinfo  {journal} {Phys. Rev. Accel. Beams}\ }\textbf {\bibinfo {volume}
  {22}},\ \bibinfo {pages} {101301} (\bibinfo {year} {2019})}\BibitemShut
  {NoStop}%
\bibitem [{\citenamefont {Aschikhin}\ \emph {et~al.}(2016)\citenamefont
  {Aschikhin}, \citenamefont {Behrens}, \citenamefont {Foster} \emph
  {et~al.}}]{Aschikhin2016NIMA}%
  \BibitemOpen
  \bibfield  {author} {\bibinfo {author} {\bibfnamefont {A.}~\bibnamefont
  {Aschikhin}}, \bibinfo {author} {\bibfnamefont {C.}~\bibnamefont {Behrens}},
  \bibinfo {author} {\bibfnamefont {B.}~\bibnamefont {Foster}}, \emph
  {et~al.},\ }\bibfield  {title} {\bibinfo {title} {The {FLASHForward} facility
  at {DESY}},\ }\href {https://doi.org/10.1016/j.nima.2015.10.005} {\bibfield
  {journal} {\bibinfo  {journal} {Nucl. Instrum. Methods Phys. Res. A}\
  }\textbf {\bibinfo {volume} {806}},\ \bibinfo {pages} {175} (\bibinfo {year}
  {2016})}\BibitemShut {NoStop}%
\bibitem [{\citenamefont {Rauer}\ \emph {et~al.}(2026)\citenamefont {Rauer},
  \citenamefont {Bahns}, \citenamefont {Friedrich}, \citenamefont {Casalbuoni},
  \citenamefont {Di~Felice}, \citenamefont {Dommach}, \citenamefont
  {Freijo~Martin}, \citenamefont {Freund}, \citenamefont {Gr{\"u}nert},
  \citenamefont {Guetg}, \citenamefont {Karpics}, \citenamefont {Karabekyan},
  \citenamefont {Koch}, \citenamefont {Kujala}, \citenamefont {La~Civita},
  \citenamefont {Liu}, \citenamefont {Maltezopoulos}, \citenamefont {Makita},
  \citenamefont {Mayet}, \citenamefont {M{\"u}ller}, \citenamefont {Rio},
  \citenamefont {Samoylova}, \citenamefont {Schmidtchen}, \citenamefont
  {Scholz}, \citenamefont {Silenzi}, \citenamefont {Strauch}, \citenamefont
  {Thoden}, \citenamefont {Wohlenberg}, \citenamefont {Vannoni}, \citenamefont
  {Yang}, \citenamefont {Decking}, \citenamefont {Rossbach},\ and\
  \citenamefont {Sinn}}]{Rauer2026cavity}%
  \BibitemOpen
  \bibfield  {author} {\bibinfo {author} {\bibfnamefont {P.}~\bibnamefont
  {Rauer}}, \bibinfo {author} {\bibfnamefont {I.}~\bibnamefont {Bahns}},
  \bibinfo {author} {\bibfnamefont {B.}~\bibnamefont {Friedrich}}, \bibinfo
  {author} {\bibfnamefont {S.}~\bibnamefont {Casalbuoni}}, \bibinfo {author}
  {\bibfnamefont {M.}~\bibnamefont {Di~Felice}}, \bibinfo {author}
  {\bibfnamefont {M.}~\bibnamefont {Dommach}}, \bibinfo {author} {\bibfnamefont
  {I.}~\bibnamefont {Freijo~Martin}}, \bibinfo {author} {\bibfnamefont
  {W.}~\bibnamefont {Freund}}, \bibinfo {author} {\bibfnamefont
  {J.}~\bibnamefont {Gr{\"u}nert}}, \bibinfo {author} {\bibfnamefont
  {M.}~\bibnamefont {Guetg}}, \bibinfo {author} {\bibfnamefont
  {I.}~\bibnamefont {Karpics}}, \bibinfo {author} {\bibfnamefont
  {S.}~\bibnamefont {Karabekyan}}, \bibinfo {author} {\bibfnamefont
  {A.}~\bibnamefont {Koch}}, \bibinfo {author} {\bibfnamefont {N.}~\bibnamefont
  {Kujala}}, \bibinfo {author} {\bibfnamefont {D.}~\bibnamefont {La~Civita}},
  \bibinfo {author} {\bibfnamefont {J.}~\bibnamefont {Liu}}, \bibinfo {author}
  {\bibfnamefont {T.}~\bibnamefont {Maltezopoulos}}, \bibinfo {author}
  {\bibfnamefont {M.}~\bibnamefont {Makita}}, \bibinfo {author} {\bibfnamefont
  {F.}~\bibnamefont {Mayet}}, \bibinfo {author} {\bibfnamefont
  {L.}~\bibnamefont {M{\"u}ller}}, \bibinfo {author} {\bibfnamefont
  {B.}~\bibnamefont {Rio}}, \bibinfo {author} {\bibfnamefont {L.}~\bibnamefont
  {Samoylova}}, \bibinfo {author} {\bibfnamefont {S.}~\bibnamefont
  {Schmidtchen}}, \bibinfo {author} {\bibfnamefont {M.}~\bibnamefont {Scholz}},
  \bibinfo {author} {\bibfnamefont {A.}~\bibnamefont {Silenzi}}, \bibinfo
  {author} {\bibfnamefont {V.}~\bibnamefont {Strauch}}, \bibinfo {author}
  {\bibfnamefont {D.}~\bibnamefont {Thoden}}, \bibinfo {author} {\bibfnamefont
  {T.}~\bibnamefont {Wohlenberg}}, \bibinfo {author} {\bibfnamefont
  {M.}~\bibnamefont {Vannoni}}, \bibinfo {author} {\bibfnamefont
  {F.}~\bibnamefont {Yang}}, \bibinfo {author} {\bibfnamefont {W.}~\bibnamefont
  {Decking}}, \bibinfo {author} {\bibfnamefont {J.}~\bibnamefont {Rossbach}},\
  and\ \bibinfo {author} {\bibfnamefont {H.}~\bibnamefont {Sinn}},\ }\bibfield
  {title} {\bibinfo {title} {Lasing of a cavity-based {X-ray} source},\ }\href
  {https://doi.org/10.1038/s41586-025-10025-x} {\bibfield  {journal} {\bibinfo
  {journal} {Nature}\ }\textbf {\bibinfo {volume} {650}},\ \bibinfo {pages}
  {93} (\bibinfo {year} {2026})}\BibitemShut {NoStop}%
\bibitem [{\citenamefont {Corde}\ \emph {et~al.}(2015)\citenamefont {Corde},
  \citenamefont {Adli}, \citenamefont {Allen} \emph
  {et~al.}}]{Corde2015Nature}%
  \BibitemOpen
  \bibfield  {author} {\bibinfo {author} {\bibfnamefont {S.}~\bibnamefont
  {Corde}}, \bibinfo {author} {\bibfnamefont {E.}~\bibnamefont {Adli}},
  \bibinfo {author} {\bibfnamefont {J.~M.}\ \bibnamefont {Allen}}, \emph
  {et~al.},\ }\bibfield  {title} {\bibinfo {title} {Multi-gigaelectronvolt
  acceleration of positrons in a self-loaded plasma wakefield},\ }\href
  {https://doi.org/10.1038/nature14890} {\bibfield  {journal} {\bibinfo
  {journal} {Nature}\ }\textbf {\bibinfo {volume} {524}},\ \bibinfo {pages}
  {442} (\bibinfo {year} {2015})}\BibitemShut {NoStop}%
\bibitem [{\citenamefont {Gessner}\ \emph {et~al.}(2016)\citenamefont
  {Gessner}, \citenamefont {Adli}, \citenamefont {Allen}, \citenamefont {An},
  \citenamefont {Clarke}, \citenamefont {Clayton}, \citenamefont {Corde} \emph
  {et~al.}}]{Gessner2016NatCommun}%
  \BibitemOpen
  \bibfield  {author} {\bibinfo {author} {\bibfnamefont {S.}~\bibnamefont
  {Gessner}}, \bibinfo {author} {\bibfnamefont {E.}~\bibnamefont {Adli}},
  \bibinfo {author} {\bibfnamefont {J.~M.}\ \bibnamefont {Allen}}, \bibinfo
  {author} {\bibfnamefont {W.}~\bibnamefont {An}}, \bibinfo {author}
  {\bibfnamefont {C.~I.}\ \bibnamefont {Clarke}}, \bibinfo {author}
  {\bibfnamefont {C.~E.}\ \bibnamefont {Clayton}}, \bibinfo {author}
  {\bibfnamefont {S.}~\bibnamefont {Corde}}, \emph {et~al.},\ }\bibfield
  {title} {\bibinfo {title} {Demonstration of a positron beam-driven hollow
  channel plasma wakefield accelerator},\ }\href
  {https://doi.org/10.1038/ncomms11785} {\bibfield  {journal} {\bibinfo
  {journal} {Nat. Commun.}\ }\textbf {\bibinfo {volume} {7}},\ \bibinfo {pages}
  {11785} (\bibinfo {year} {2016})}\BibitemShut {NoStop}%
\bibitem [{\citenamefont {Prat}\ \emph {et~al.}(2023)\citenamefont {Prat},
  \citenamefont {Al~Haddad}, \citenamefont {Arrell}, \citenamefont {Augustin},
  \citenamefont {Boll}, \citenamefont {Bostedt}, \citenamefont {Calvi},
  \citenamefont {Cavalieri}, \citenamefont {Craievich}, \citenamefont {Dax}
  \emph {et~al.}}]{prat2023x}%
  \BibitemOpen
  \bibfield  {author} {\bibinfo {author} {\bibfnamefont {E.}~\bibnamefont
  {Prat}}, \bibinfo {author} {\bibfnamefont {A.}~\bibnamefont {Al~Haddad}},
  \bibinfo {author} {\bibfnamefont {C.}~\bibnamefont {Arrell}}, \bibinfo
  {author} {\bibfnamefont {S.}~\bibnamefont {Augustin}}, \bibinfo {author}
  {\bibfnamefont {M.}~\bibnamefont {Boll}}, \bibinfo {author} {\bibfnamefont
  {C.}~\bibnamefont {Bostedt}}, \bibinfo {author} {\bibfnamefont
  {M.}~\bibnamefont {Calvi}}, \bibinfo {author} {\bibfnamefont {A.~L.}\
  \bibnamefont {Cavalieri}}, \bibinfo {author} {\bibfnamefont {P.}~\bibnamefont
  {Craievich}}, \bibinfo {author} {\bibfnamefont {A.}~\bibnamefont {Dax}},
  \emph {et~al.},\ }\bibfield  {title} {\bibinfo {title} {An {X-ray}
  free-electron laser with a highly configurable undulator and integrated
  chicanes for tailored pulse properties},\ }\href
  {https://doi.org/10.1038/s41467-023-40759-z} {\bibfield  {journal} {\bibinfo
  {journal} {Nat. Commun.}\ }\textbf {\bibinfo {volume} {14}},\ \bibinfo
  {pages} {5069} (\bibinfo {year} {2023})}\BibitemShut {NoStop}%
\bibitem [{\citenamefont {Prat}\ \emph {et~al.}(2019)\citenamefont {Prat},
  \citenamefont {Dijkstal}, \citenamefont {Aiba}, \citenamefont {Bettoni},
  \citenamefont {Craievich}, \citenamefont {Ferrari}, \citenamefont
  {Ischebeck}, \citenamefont {L{\"o}hl}, \citenamefont {Malyzhenkov},
  \citenamefont {Orlandi} \emph {et~al.}}]{prat2019generation}%
  \BibitemOpen
  \bibfield  {author} {\bibinfo {author} {\bibfnamefont {E.}~\bibnamefont
  {Prat}}, \bibinfo {author} {\bibfnamefont {P.}~\bibnamefont {Dijkstal}},
  \bibinfo {author} {\bibfnamefont {M.}~\bibnamefont {Aiba}}, \bibinfo {author}
  {\bibfnamefont {S.}~\bibnamefont {Bettoni}}, \bibinfo {author} {\bibfnamefont
  {P.}~\bibnamefont {Craievich}}, \bibinfo {author} {\bibfnamefont
  {E.}~\bibnamefont {Ferrari}}, \bibinfo {author} {\bibfnamefont
  {R.}~\bibnamefont {Ischebeck}}, \bibinfo {author} {\bibfnamefont
  {F.}~\bibnamefont {L{\"o}hl}}, \bibinfo {author} {\bibfnamefont
  {A.}~\bibnamefont {Malyzhenkov}}, \bibinfo {author} {\bibfnamefont {G.~L.}\
  \bibnamefont {Orlandi}}, \emph {et~al.},\ }\bibfield  {title} {\bibinfo
  {title} {Generation and characterization of intense ultralow-emittance
  electron beams for compact {X-ray} free-electron lasers},\ }\href
  {https://doi.org/10.1103/PhysRevLett.123.234801} {\bibfield  {journal}
  {\bibinfo  {journal} {Phys. Rev. Lett.}\ }\textbf {\bibinfo {volume} {123}},\
  \bibinfo {pages} {234801} (\bibinfo {year} {2019})}\BibitemShut {NoStop}%
\bibitem [{\citenamefont {Tomin}\ \emph {et~al.}(2021)\citenamefont {Tomin},
  \citenamefont {Zagorodnov}, \citenamefont {Decking}, \citenamefont
  {Golubeva},\ and\ \citenamefont {Scholz}}]{tomin2021accurate}%
  \BibitemOpen
  \bibfield  {author} {\bibinfo {author} {\bibfnamefont {S.}~\bibnamefont
  {Tomin}}, \bibinfo {author} {\bibfnamefont {I.}~\bibnamefont {Zagorodnov}},
  \bibinfo {author} {\bibfnamefont {W.}~\bibnamefont {Decking}}, \bibinfo
  {author} {\bibfnamefont {N.}~\bibnamefont {Golubeva}},\ and\ \bibinfo
  {author} {\bibfnamefont {M.}~\bibnamefont {Scholz}},\ }\bibfield  {title}
  {\bibinfo {title} {Accurate measurement of uncorrelated energy spread in
  electron beam},\ }\href {https://doi.org/10.1103/PhysRevAccelBeams.24.064201}
  {\bibfield  {journal} {\bibinfo  {journal} {Phys. Rev. Accel. Beams}\
  }\textbf {\bibinfo {volume} {24}},\ \bibinfo {pages} {064201} (\bibinfo
  {year} {2021})}\BibitemShut {NoStop}%
\bibitem [{\citenamefont {Rosenzweig}\ \emph {et~al.}(2019)\citenamefont
  {Rosenzweig}, \citenamefont {Cahill}, \citenamefont {Dolgashev},
  \citenamefont {Emma}, \citenamefont {Fukasawa}, \citenamefont {Li},
  \citenamefont {Limborg}, \citenamefont {Maxson}, \citenamefont {Musumeci},
  \citenamefont {Nause} \emph {et~al.}}]{rosenzweig2019next}%
  \BibitemOpen
  \bibfield  {author} {\bibinfo {author} {\bibfnamefont {J.~B.}\ \bibnamefont
  {Rosenzweig}}, \bibinfo {author} {\bibfnamefont {A.}~\bibnamefont {Cahill}},
  \bibinfo {author} {\bibfnamefont {V.}~\bibnamefont {Dolgashev}}, \bibinfo
  {author} {\bibfnamefont {C.}~\bibnamefont {Emma}}, \bibinfo {author}
  {\bibfnamefont {A.}~\bibnamefont {Fukasawa}}, \bibinfo {author}
  {\bibfnamefont {R.}~\bibnamefont {Li}}, \bibinfo {author} {\bibfnamefont
  {C.}~\bibnamefont {Limborg}}, \bibinfo {author} {\bibfnamefont
  {J.}~\bibnamefont {Maxson}}, \bibinfo {author} {\bibfnamefont
  {P.}~\bibnamefont {Musumeci}}, \bibinfo {author} {\bibfnamefont
  {A.}~\bibnamefont {Nause}}, \emph {et~al.},\ }\bibfield  {title} {\bibinfo
  {title} {Next generation high brightness electron beams from ultrahigh field
  cryogenic {RF} photocathode sources},\ }\href
  {https://doi.org/10.1103/PhysRevAccelBeams.22.023403} {\bibfield  {journal}
  {\bibinfo  {journal} {Phys. Rev. Accel. Beams}\ }\textbf {\bibinfo {volume}
  {22}},\ \bibinfo {pages} {023403} (\bibinfo {year} {2019})}\BibitemShut
  {NoStop}%
\bibitem [{\citenamefont {Schreiber}\ and\ \citenamefont
  {Faatz}(2015)}]{schreiber2015flash}%
  \BibitemOpen
  \bibfield  {author} {\bibinfo {author} {\bibfnamefont {S.}~\bibnamefont
  {Schreiber}}\ and\ \bibinfo {author} {\bibfnamefont {B.}~\bibnamefont
  {Faatz}},\ }\bibfield  {title} {\bibinfo {title} {The free-electron laser
  {FLASH}},\ }\href {https://doi.org/10.1017/hpl.2015.16} {\bibfield  {journal}
  {\bibinfo  {journal} {High Power Laser Sci. Eng.}\ }\textbf {\bibinfo
  {volume} {3}},\ \bibinfo {pages} {e20} (\bibinfo {year} {2015})}\BibitemShut
  {NoStop}%
\bibitem [{\citenamefont {Ackermann}\ \emph {et~al.}(2007)\citenamefont
  {Ackermann}, \citenamefont {Asova}, \citenamefont {Ayvazyan}, \citenamefont
  {Azima}, \citenamefont {Baboi}, \citenamefont {Baehr}, \citenamefont
  {Behrens}, \citenamefont {Bohnet}, \citenamefont {Bonfigt}, \citenamefont
  {Brambrink} \emph {et~al.}}]{ackermann2007flash}%
  \BibitemOpen
  \bibfield  {author} {\bibinfo {author} {\bibfnamefont {W.}~\bibnamefont
  {Ackermann}}, \bibinfo {author} {\bibfnamefont {G.}~\bibnamefont {Asova}},
  \bibinfo {author} {\bibfnamefont {V.}~\bibnamefont {Ayvazyan}}, \bibinfo
  {author} {\bibfnamefont {A.}~\bibnamefont {Azima}}, \bibinfo {author}
  {\bibfnamefont {N.}~\bibnamefont {Baboi}}, \bibinfo {author} {\bibfnamefont
  {J.}~\bibnamefont {Baehr}}, \bibinfo {author} {\bibfnamefont
  {C.}~\bibnamefont {Behrens}}, \bibinfo {author} {\bibfnamefont
  {I.}~\bibnamefont {Bohnet}}, \bibinfo {author} {\bibfnamefont
  {S.}~\bibnamefont {Bonfigt}}, \bibinfo {author} {\bibfnamefont
  {E.}~\bibnamefont {Brambrink}}, \emph {et~al.},\ }\bibfield  {title}
  {\bibinfo {title} {Operation of a free-electron laser from the extreme
  ultraviolet to the water window},\ }\href
  {https://doi.org/10.1038/nphoton.2007.76} {\bibfield  {journal} {\bibinfo
  {journal} {Nat. Photonics}\ }\textbf {\bibinfo {volume} {1}},\ \bibinfo
  {pages} {336} (\bibinfo {year} {2007})}\BibitemShut {NoStop}%
\bibitem [{\citenamefont {Ding}\ \emph {et~al.}(2009)\citenamefont {Ding},
  \citenamefont {Huang}, \citenamefont {Ratner}, \citenamefont {Bucksbaum},\
  and\ \citenamefont {Merdji}}]{ding2009generation}%
  \BibitemOpen
  \bibfield  {author} {\bibinfo {author} {\bibfnamefont {Y.}~\bibnamefont
  {Ding}}, \bibinfo {author} {\bibfnamefont {Z.}~\bibnamefont {Huang}},
  \bibinfo {author} {\bibfnamefont {D.}~\bibnamefont {Ratner}}, \bibinfo
  {author} {\bibfnamefont {P.}~\bibnamefont {Bucksbaum}},\ and\ \bibinfo
  {author} {\bibfnamefont {H.}~\bibnamefont {Merdji}},\ }\bibfield  {title}
  {\bibinfo {title} {Generation of attosecond {X-ray} pulses with a multicycle
  two-color enhanced self-amplified spontaneous emission scheme},\ }\href
  {https://doi.org/10.1103/PhysRevSTAB.12.060703} {\bibfield  {journal}
  {\bibinfo  {journal} {Phys. Rev. ST Accel. Beams}\ }\textbf {\bibinfo
  {volume} {12}},\ \bibinfo {pages} {060703} (\bibinfo {year}
  {2009})}\BibitemShut {NoStop}%
\bibitem [{\citenamefont {Amann}\ \emph {et~al.}(2012)\citenamefont {Amann},
  \citenamefont {Berg}, \citenamefont {Blank}, \citenamefont {Decker},
  \citenamefont {Ding}, \citenamefont {Emma}, \citenamefont {Feng},
  \citenamefont {Frisch}, \citenamefont {Fritz}, \citenamefont {Hastings} \emph
  {et~al.}}]{amann2012demonstration}%
  \BibitemOpen
  \bibfield  {author} {\bibinfo {author} {\bibfnamefont {J.}~\bibnamefont
  {Amann}}, \bibinfo {author} {\bibfnamefont {W.}~\bibnamefont {Berg}},
  \bibinfo {author} {\bibfnamefont {V.}~\bibnamefont {Blank}}, \bibinfo
  {author} {\bibfnamefont {F.-J.}\ \bibnamefont {Decker}}, \bibinfo {author}
  {\bibfnamefont {Y.}~\bibnamefont {Ding}}, \bibinfo {author} {\bibfnamefont
  {P.}~\bibnamefont {Emma}}, \bibinfo {author} {\bibfnamefont {Y.}~\bibnamefont
  {Feng}}, \bibinfo {author} {\bibfnamefont {J.}~\bibnamefont {Frisch}},
  \bibinfo {author} {\bibfnamefont {D.}~\bibnamefont {Fritz}}, \bibinfo
  {author} {\bibfnamefont {J.}~\bibnamefont {Hastings}}, \emph {et~al.},\
  }\bibfield  {title} {\bibinfo {title} {Demonstration of self-seeding in a
  hard-{X-ray} free-electron laser},\ }\href
  {https://doi.org/10.1038/nphoton.2012.180} {\bibfield  {journal} {\bibinfo
  {journal} {Nat. Photonics}\ }\textbf {\bibinfo {volume} {6}},\ \bibinfo
  {pages} {693} (\bibinfo {year} {2012})}\BibitemShut {NoStop}%
\bibitem [{\citenamefont {Quin}\ \emph
  {et~al.}(2025{\natexlab{b}})\citenamefont {Quin}, \citenamefont {Di~Piazza},\
  and\ \citenamefont {Tamburini}}]{quinPPCF25}%
  \BibitemOpen
  \bibfield  {author} {\bibinfo {author} {\bibfnamefont {M.~J.}\ \bibnamefont
  {Quin}}, \bibinfo {author} {\bibfnamefont {A.}~\bibnamefont {Di~Piazza}},\
  and\ \bibinfo {author} {\bibfnamefont {M.}~\bibnamefont {Tamburini}},\
  }\bibfield  {title} {\bibinfo {title} {Coherent frequency combs from
  electrons colliding with a laser pulse},\ }\href
  {https://doi.org/10.1088/1361-6587/adc59c} {\bibfield  {journal} {\bibinfo
  {journal} {Plasma Phys. Control. Fusion}\ }\textbf {\bibinfo {volume} {67}},\
  \bibinfo {pages} {055008} (\bibinfo {year} {2025}{\natexlab{b}})}\BibitemShut
  {NoStop}%
\bibitem [{\citenamefont {Filipescu}\ \emph {et~al.}(2023)\citenamefont
  {Filipescu}, \citenamefont {Gheorghe}, \citenamefont {Stopani}, \citenamefont
  {Belyshev}, \citenamefont {Hashimoto}, \citenamefont {Miyamoto},\ and\
  \citenamefont {Utsunomiya}}]{filipescu2023spectral}%
  \BibitemOpen
  \bibfield  {author} {\bibinfo {author} {\bibfnamefont {D.}~\bibnamefont
  {Filipescu}}, \bibinfo {author} {\bibfnamefont {I.}~\bibnamefont {Gheorghe}},
  \bibinfo {author} {\bibfnamefont {K.}~\bibnamefont {Stopani}}, \bibinfo
  {author} {\bibfnamefont {S.}~\bibnamefont {Belyshev}}, \bibinfo {author}
  {\bibfnamefont {S.}~\bibnamefont {Hashimoto}}, \bibinfo {author}
  {\bibfnamefont {S.}~\bibnamefont {Miyamoto}},\ and\ \bibinfo {author}
  {\bibfnamefont {H.}~\bibnamefont {Utsunomiya}},\ }\bibfield  {title}
  {\bibinfo {title} {Spectral distribution and flux of $\gamma$-ray beams
  produced through {Compton} scattering of unsynchronized laser and electron
  beams},\ }\href {https://doi.org/10.1016/j.nima.2022.167885} {\bibfield
  {journal} {\bibinfo  {journal} {Nucl. Instrum. Methods Phys. Res. A}\
  }\textbf {\bibinfo {volume} {1047}},\ \bibinfo {pages} {167885} (\bibinfo
  {year} {2023})}\BibitemShut {NoStop}%
\bibitem [{\citenamefont {Xiang}\ and\ \citenamefont
  {Stupakov}(2009)}]{xiang2009echo}%
  \BibitemOpen
  \bibfield  {author} {\bibinfo {author} {\bibfnamefont {D.}~\bibnamefont
  {Xiang}}\ and\ \bibinfo {author} {\bibfnamefont {G.}~\bibnamefont
  {Stupakov}},\ }\bibfield  {title} {\bibinfo {title} {Echo-enabled harmonic
  generation free electron laser},\ }\href
  {https://doi.org/10.1103/PhysRevSTAB.12.030702} {\bibfield  {journal}
  {\bibinfo  {journal} {Phys. Rev. ST Accel. Beams}\ }\textbf {\bibinfo
  {volume} {12}},\ \bibinfo {pages} {030702} (\bibinfo {year}
  {2009})}\BibitemShut {NoStop}%
\bibitem [{\citenamefont {Zhang}\ \emph {et~al.}(2017)\citenamefont {Zhang},
  \citenamefont {Graves}, \citenamefont {Malin}, \citenamefont {Spence},
  \citenamefont {Li}, \citenamefont {Nanni}, \citenamefont {Shen},
  \citenamefont {Weathersby}, \citenamefont {Yang}, \citenamefont {Cesar} \emph
  {et~al.}}]{zhang2017experiments}%
  \BibitemOpen
  \bibfield  {author} {\bibinfo {author} {\bibfnamefont {C.}~\bibnamefont
  {Zhang}}, \bibinfo {author} {\bibfnamefont {W.~S.}\ \bibnamefont {Graves}},
  \bibinfo {author} {\bibfnamefont {L.~E.}\ \bibnamefont {Malin}}, \bibinfo
  {author} {\bibfnamefont {J.~C.~H.}\ \bibnamefont {Spence}}, \bibinfo {author}
  {\bibfnamefont {R.~K.}\ \bibnamefont {Li}}, \bibinfo {author} {\bibfnamefont
  {E.~A.}\ \bibnamefont {Nanni}}, \bibinfo {author} {\bibfnamefont
  {X.}~\bibnamefont {Shen}}, \bibinfo {author} {\bibfnamefont {S.~P.}\
  \bibnamefont {Weathersby}}, \bibinfo {author} {\bibfnamefont
  {J.}~\bibnamefont {Yang}}, \bibinfo {author} {\bibfnamefont {D.}~\bibnamefont
  {Cesar}}, \emph {et~al.},\ }\bibfield  {title} {\bibinfo {title} {Experiments
  in electron beam nanopatterning},\ }in\ \href
  {https://doi.org/10.18429/JACoW-FEL2017-TUP038} {\emph {\bibinfo {booktitle}
  {Proc. 38th Int. Free-Electron Laser Conf. (FEL'17)}}}\ (\bibinfo {year}
  {2017})\BibitemShut {NoStop}%
\bibitem [{\citenamefont {Robles}\ \emph {et~al.}(2024)\citenamefont {Robles},
  \citenamefont {Giannessi},\ and\ \citenamefont
  {Marinelli}}]{robles2024three}%
  \BibitemOpen
  \bibfield  {author} {\bibinfo {author} {\bibfnamefont {R.~R.}\ \bibnamefont
  {Robles}}, \bibinfo {author} {\bibfnamefont {L.}~\bibnamefont {Giannessi}},\
  and\ \bibinfo {author} {\bibfnamefont {A.}~\bibnamefont {Marinelli}},\
  }\bibfield  {title} {\bibinfo {title} {Three-dimensional theory of
  superradiant free-electron lasers},\ }\href
  {https://doi.org/10.1103/PhysRevResearch.6.033158} {\bibfield  {journal}
  {\bibinfo  {journal} {Phys. Rev. Res.}\ }\textbf {\bibinfo {volume} {6}},\
  \bibinfo {pages} {033158} (\bibinfo {year} {2024})}\BibitemShut {NoStop}%
\bibitem [{\citenamefont {Robles}\ \emph {et~al.}(2025)\citenamefont {Robles},
  \citenamefont {Larsen}, \citenamefont {Cesar}, \citenamefont {Driver},
  \citenamefont {Duris}, \citenamefont {Franz}, \citenamefont {Garratt},
  \citenamefont {Guo}, \citenamefont {Just}, \citenamefont {Lemons} \emph
  {et~al.}}]{robles2025spectrotemporal}%
  \BibitemOpen
  \bibfield  {author} {\bibinfo {author} {\bibfnamefont {R.~R.}\ \bibnamefont
  {Robles}}, \bibinfo {author} {\bibfnamefont {K.~A.}\ \bibnamefont {Larsen}},
  \bibinfo {author} {\bibfnamefont {D.}~\bibnamefont {Cesar}}, \bibinfo
  {author} {\bibfnamefont {T.}~\bibnamefont {Driver}}, \bibinfo {author}
  {\bibfnamefont {J.}~\bibnamefont {Duris}}, \bibinfo {author} {\bibfnamefont
  {P.}~\bibnamefont {Franz}}, \bibinfo {author} {\bibfnamefont
  {D.}~\bibnamefont {Garratt}}, \bibinfo {author} {\bibfnamefont
  {V.}~\bibnamefont {Guo}}, \bibinfo {author} {\bibfnamefont {G.}~\bibnamefont
  {Just}}, \bibinfo {author} {\bibfnamefont {R.}~\bibnamefont {Lemons}}, \emph
  {et~al.},\ }\bibfield  {title} {\bibinfo {title} {Spectrotemporal shaping of
  attosecond {X-ray} pulses with a fresh-slice free-electron laser},\ }\href
  {https://doi.org/10.1103/PhysRevLett.134.115001} {\bibfield  {journal}
  {\bibinfo  {journal} {Phys. Rev. Lett.}\ }\textbf {\bibinfo {volume} {134}},\
  \bibinfo {pages} {115001} (\bibinfo {year} {2025})}\BibitemShut {NoStop}%
\bibitem [{\citenamefont {Blumenfeld}\ \emph {et~al.}(2007)\citenamefont
  {Blumenfeld}, \citenamefont {Clayton}, \citenamefont {Decker}, \citenamefont
  {Hogan}, \citenamefont {Huang}, \citenamefont {Ischebeck}, \citenamefont
  {Iverson}, \citenamefont {Joshi}, \citenamefont {Katsouleas}, \citenamefont
  {Kirby} \emph {et~al.}}]{blumenfeld2007energy}%
  \BibitemOpen
  \bibfield  {author} {\bibinfo {author} {\bibfnamefont {I.}~\bibnamefont
  {Blumenfeld}}, \bibinfo {author} {\bibfnamefont {C.~E.}\ \bibnamefont
  {Clayton}}, \bibinfo {author} {\bibfnamefont {F.-J.}\ \bibnamefont {Decker}},
  \bibinfo {author} {\bibfnamefont {M.~J.}\ \bibnamefont {Hogan}}, \bibinfo
  {author} {\bibfnamefont {C.}~\bibnamefont {Huang}}, \bibinfo {author}
  {\bibfnamefont {R.}~\bibnamefont {Ischebeck}}, \bibinfo {author}
  {\bibfnamefont {R.}~\bibnamefont {Iverson}}, \bibinfo {author} {\bibfnamefont
  {C.}~\bibnamefont {Joshi}}, \bibinfo {author} {\bibfnamefont
  {T.}~\bibnamefont {Katsouleas}}, \bibinfo {author} {\bibfnamefont
  {N.}~\bibnamefont {Kirby}}, \emph {et~al.},\ }\bibfield  {title} {\bibinfo
  {title} {Energy doubling of 42 gev electrons in a metre-scale plasma
  wakefield accelerator},\ }\href {https://doi.org/10.1038/nature05538}
  {\bibfield  {journal} {\bibinfo  {journal} {Nature}\ }\textbf {\bibinfo
  {volume} {445}},\ \bibinfo {pages} {741} (\bibinfo {year}
  {2007})}\BibitemShut {NoStop}%
\bibitem [{\citenamefont {Sarri}\ \emph {et~al.}(2015)\citenamefont {Sarri},
  \citenamefont {Poder}, \citenamefont {Cole}, \citenamefont {Schumaker},
  \citenamefont {Di~Piazza}, \citenamefont {Reville}, \citenamefont
  {Dzelzainis}, \citenamefont {Doria}, \citenamefont {Gizzi}, \citenamefont
  {Grittani} \emph {et~al.}}]{sarri2015generation}%
  \BibitemOpen
  \bibfield  {author} {\bibinfo {author} {\bibfnamefont {G.}~\bibnamefont
  {Sarri}}, \bibinfo {author} {\bibfnamefont {K.}~\bibnamefont {Poder}},
  \bibinfo {author} {\bibfnamefont {J.~M.}\ \bibnamefont {Cole}}, \bibinfo
  {author} {\bibfnamefont {W.}~\bibnamefont {Schumaker}}, \bibinfo {author}
  {\bibfnamefont {A.}~\bibnamefont {Di~Piazza}}, \bibinfo {author}
  {\bibfnamefont {B.}~\bibnamefont {Reville}}, \bibinfo {author} {\bibfnamefont
  {T.}~\bibnamefont {Dzelzainis}}, \bibinfo {author} {\bibfnamefont
  {D.}~\bibnamefont {Doria}}, \bibinfo {author} {\bibfnamefont {L.~A.}\
  \bibnamefont {Gizzi}}, \bibinfo {author} {\bibfnamefont {G.}~\bibnamefont
  {Grittani}}, \emph {et~al.},\ }\bibfield  {title} {\bibinfo {title}
  {Generation of neutral and high-density electron--positron pair plasmas in
  the laboratory},\ }\href {https://doi.org/10.1038/ncomms7747} {\bibfield
  {journal} {\bibinfo  {journal} {Nat. Commun.}\ }\textbf {\bibinfo {volume}
  {6}},\ \bibinfo {pages} {6747} (\bibinfo {year} {2015})}\BibitemShut
  {NoStop}%
\bibitem [{\citenamefont {Emma}\ \emph {et~al.}(2017)\citenamefont {Emma},
  \citenamefont {Feng}, \citenamefont {Nguyen}, \citenamefont {Ratti},\ and\
  \citenamefont {Pellegrini}}]{emma2017compact}%
  \BibitemOpen
  \bibfield  {author} {\bibinfo {author} {\bibfnamefont {C.}~\bibnamefont
  {Emma}}, \bibinfo {author} {\bibfnamefont {Y.}~\bibnamefont {Feng}}, \bibinfo
  {author} {\bibfnamefont {D.~C.}\ \bibnamefont {Nguyen}}, \bibinfo {author}
  {\bibfnamefont {A.}~\bibnamefont {Ratti}},\ and\ \bibinfo {author}
  {\bibfnamefont {C.}~\bibnamefont {Pellegrini}},\ }\bibfield  {title}
  {\bibinfo {title} {Compact double-bunch {X-ray} free electron lasers for
  fresh bunch self-seeding and harmonic lasing},\ }\href
  {https://doi.org/10.1103/PhysRevAccelBeams.20.030701} {\bibfield  {journal}
  {\bibinfo  {journal} {Phys. Rev. Accel. Beams}\ }\textbf {\bibinfo {volume}
  {20}},\ \bibinfo {pages} {030701} (\bibinfo {year} {2017})}\BibitemShut
  {NoStop}%
\bibitem [{\citenamefont {Vay}(2007)}]{vay2007noninvariance}%
  \BibitemOpen
  \bibfield  {author} {\bibinfo {author} {\bibfnamefont {J.-L.}\ \bibnamefont
  {Vay}},\ }\bibfield  {title} {\bibinfo {title} {Noninvariance of space- and
  time-scale ranges under a {Lorentz} transformation and the implications for
  the study of relativistic interactions},\ }\href
  {https://doi.org/10.1103/PhysRevLett.98.130405} {\bibfield  {journal}
  {\bibinfo  {journal} {Phys. Rev. Lett.}\ }\textbf {\bibinfo {volume} {98}},\
  \bibinfo {pages} {130405} (\bibinfo {year} {2007})}\BibitemShut {NoStop}%
\bibitem [{\citenamefont {Vay}\ \emph {et~al.}(2021)\citenamefont {Vay},
  \citenamefont {Huebl}, \citenamefont {Almgren}, \citenamefont {Amorim},
  \citenamefont {Bell}, \citenamefont {Fedeli}, \citenamefont {Ge},
  \citenamefont {Gott}, \citenamefont {Grote}, \citenamefont {Hogan} \emph
  {et~al.}}]{vay2021modeling}%
  \BibitemOpen
  \bibfield  {author} {\bibinfo {author} {\bibfnamefont {J.-L.}\ \bibnamefont
  {Vay}}, \bibinfo {author} {\bibfnamefont {A.}~\bibnamefont {Huebl}}, \bibinfo
  {author} {\bibfnamefont {A.}~\bibnamefont {Almgren}}, \bibinfo {author}
  {\bibfnamefont {L.~D.}\ \bibnamefont {Amorim}}, \bibinfo {author}
  {\bibfnamefont {J.}~\bibnamefont {Bell}}, \bibinfo {author} {\bibfnamefont
  {L.}~\bibnamefont {Fedeli}}, \bibinfo {author} {\bibfnamefont
  {L.}~\bibnamefont {Ge}}, \bibinfo {author} {\bibfnamefont {K.}~\bibnamefont
  {Gott}}, \bibinfo {author} {\bibfnamefont {D.~P.}\ \bibnamefont {Grote}},
  \bibinfo {author} {\bibfnamefont {M.}~\bibnamefont {Hogan}}, \emph {et~al.},\
  }\bibfield  {title} {\bibinfo {title} {Modeling of a chain of three plasma
  accelerator stages with the {WarpX} electromagnetic {PIC} code on {GPUs}},\
  }\bibfield  {journal} {\bibinfo  {journal} {Phys. Plasmas}\ }\textbf
  {\bibinfo {volume} {28}},\ \href {https://doi.org/10.1063/5.0028512}
  {10.1063/5.0028512} (\bibinfo {year} {2021})\BibitemShut {NoStop}%
\bibitem [{\citenamefont {Fawley}\ and\ \citenamefont
  {Vay}(2009)}]{fawley2009use}%
  \BibitemOpen
  \bibfield  {author} {\bibinfo {author} {\bibfnamefont {W.~M.}\ \bibnamefont
  {Fawley}}\ and\ \bibinfo {author} {\bibfnamefont {J.-L.}\ \bibnamefont
  {Vay}},\ }\bibfield  {title} {\bibinfo {title} {Use of the lorentz-boosted
  frame transformation to simulate free-electron laser amplifier physics},\
  }in\ \href {https://doi.org/10.1063/1.3089537} {\emph {\bibinfo {booktitle}
  {AIP Conf. Proc.}}},\ Vol.\ \bibinfo {volume} {1086}\ (\bibinfo {year}
  {2009})\ pp.\ \bibinfo {pages} {346--350}\BibitemShut {NoStop}%
\bibitem [{\citenamefont {Fallahi}\ \emph {et~al.}(2018)\citenamefont
  {Fallahi}, \citenamefont {Yahaghi},\ and\ \citenamefont
  {K{\"a}rtner}}]{fallahi2018mithra}%
  \BibitemOpen
  \bibfield  {author} {\bibinfo {author} {\bibfnamefont {A.}~\bibnamefont
  {Fallahi}}, \bibinfo {author} {\bibfnamefont {A.}~\bibnamefont {Yahaghi}},\
  and\ \bibinfo {author} {\bibfnamefont {F.~X.}\ \bibnamefont {K{\"a}rtner}},\
  }\bibfield  {title} {\bibinfo {title} {{MITHRA} 1.0: a full-wave simulation
  tool for free electron lasers},\ }\href
  {https://doi.org/10.1016/j.cpc.2018.03.011} {\bibfield  {journal} {\bibinfo
  {journal} {Comput. Phys. Commun.}\ }\textbf {\bibinfo {volume} {228}},\
  \bibinfo {pages} {192} (\bibinfo {year} {2018})},\ \bibinfo {note} {code
  repository: \url{https://github.com/aryafallahi/mithra}}\BibitemShut
  {NoStop}%
\bibitem [{\citenamefont {Derouillat}\ \emph {et~al.}(2018)\citenamefont
  {Derouillat}, \citenamefont {Beck}, \citenamefont {P{\'e}rez}, \citenamefont
  {Vinci}, \citenamefont {Chiaramello}, \citenamefont {Grassi}, \citenamefont
  {Fl{\'e}}, \citenamefont {Bouchard}, \citenamefont {Plotnikov}, \citenamefont
  {Aunai} \emph {et~al.}}]{derouillat2018smilei}%
  \BibitemOpen
  \bibfield  {author} {\bibinfo {author} {\bibfnamefont {J.}~\bibnamefont
  {Derouillat}}, \bibinfo {author} {\bibfnamefont {A.}~\bibnamefont {Beck}},
  \bibinfo {author} {\bibfnamefont {F.}~\bibnamefont {P{\'e}rez}}, \bibinfo
  {author} {\bibfnamefont {T.}~\bibnamefont {Vinci}}, \bibinfo {author}
  {\bibfnamefont {M.}~\bibnamefont {Chiaramello}}, \bibinfo {author}
  {\bibfnamefont {A.}~\bibnamefont {Grassi}}, \bibinfo {author} {\bibfnamefont
  {M.}~\bibnamefont {Fl{\'e}}}, \bibinfo {author} {\bibfnamefont
  {G.}~\bibnamefont {Bouchard}}, \bibinfo {author} {\bibfnamefont
  {I.}~\bibnamefont {Plotnikov}}, \bibinfo {author} {\bibfnamefont
  {N.}~\bibnamefont {Aunai}}, \emph {et~al.},\ }\bibfield  {title} {\bibinfo
  {title} {Smilei: a collaborative, open-source, multi-purpose particle-in-cell
  code for plasma simulation},\ }\href
  {https://doi.org/10.1016/j.cpc.2017.09.024} {\bibfield  {journal} {\bibinfo
  {journal} {Comput. Phys. Commun.}\ }\textbf {\bibinfo {volume} {222}},\
  \bibinfo {pages} {351} (\bibinfo {year} {2018})},\ \bibinfo {note} {code
  repository: \url{https://github.com/SmileiPIC/Smilei}}\BibitemShut {NoStop}%
\bibitem [{\citenamefont {Dohlus}\ \emph {et~al.}(2008)\citenamefont {Dohlus},
  \citenamefont {Rossbach},\ and\ \citenamefont
  {Schm{\"u}ser}}]{dohlus2008ultraviolet}%
  \BibitemOpen
  \bibfield  {author} {\bibinfo {author} {\bibfnamefont {M.}~\bibnamefont
  {Dohlus}}, \bibinfo {author} {\bibfnamefont {J.}~\bibnamefont {Rossbach}},\
  and\ \bibinfo {author} {\bibfnamefont {P.}~\bibnamefont {Schm{\"u}ser}},\
  }\href@noop {} {\emph {\bibinfo {title} {Ultraviolet and Soft {X-ray}
  Free-Electron Lasers}}}\ (\bibinfo  {publisher} {Springer},\ \bibinfo
  {address} {New York},\ \bibinfo {year} {2008})\BibitemShut {NoStop}%
\bibitem [{\citenamefont {Saldin}\ \emph {et~al.}(2013)\citenamefont {Saldin},
  \citenamefont {Schneidmiller},\ and\ \citenamefont
  {Yurkov}}]{saldin2013physics}%
  \BibitemOpen
  \bibfield  {author} {\bibinfo {author} {\bibfnamefont {E.~L.}\ \bibnamefont
  {Saldin}}, \bibinfo {author} {\bibfnamefont {E.~V.}\ \bibnamefont
  {Schneidmiller}},\ and\ \bibinfo {author} {\bibfnamefont {M.~V.}\
  \bibnamefont {Yurkov}},\ }\href@noop {} {\emph {\bibinfo {title} {The Physics
  of Free Electron Lasers}}}\ (\bibinfo  {publisher} {Springer Science \&
  Business Media},\ \bibinfo {year} {2013})\BibitemShut {NoStop}%
\bibitem [{\citenamefont {Reiche}(1999)}]{Reiche1999GENESIS13}%
  \BibitemOpen
  \bibfield  {author} {\bibinfo {author} {\bibfnamefont {S.}~\bibnamefont
  {Reiche}},\ }\bibfield  {title} {\bibinfo {title} {{GENESIS} 1.3: a fully 3d
  time-dependent {FEL} simulation code},\ }\href
  {https://doi.org/10.1016/S0168-9002(99)00114-X} {\bibfield  {journal}
  {\bibinfo  {journal} {Nucl. Instrum. Methods Phys. Res. A}\ }\textbf
  {\bibinfo {volume} {429}},\ \bibinfo {pages} {243} (\bibinfo {year}
  {1999})},\ \bibinfo {note} {code repository:
  \url{https://github.com/svenreiche/Genesis-1.3-Version4}}\BibitemShut
  {NoStop}%
\bibitem [{\citenamefont {Lu}\ \emph {et~al.}(2020)\citenamefont {Lu},
  \citenamefont {Kilian}, \citenamefont {Guo}, \citenamefont {Li},\ and\
  \citenamefont {Liang}}]{Lu2020}%
  \BibitemOpen
  \bibfield  {author} {\bibinfo {author} {\bibfnamefont {Y.}~\bibnamefont
  {Lu}}, \bibinfo {author} {\bibfnamefont {P.}~\bibnamefont {Kilian}}, \bibinfo
  {author} {\bibfnamefont {F.}~\bibnamefont {Guo}}, \bibinfo {author}
  {\bibfnamefont {H.}~\bibnamefont {Li}},\ and\ \bibinfo {author}
  {\bibfnamefont {E.}~\bibnamefont {Liang}},\ }\bibfield  {title} {\bibinfo
  {title} {Time-step dependent force interpolation scheme for suppressing
  numerical cherenkov instability in relativistic particle-in-cell
  simulations},\ }\href {https://doi.org/10.1016/j.jcp.2020.109388} {\bibfield
  {journal} {\bibinfo  {journal} {J. Comput. Phys.}\ }\textbf {\bibinfo
  {volume} {413}},\ \bibinfo {pages} {109388} (\bibinfo {year}
  {2020})}\BibitemShut {NoStop}%
\bibitem [{\citenamefont {Higuera}\ and\ \citenamefont
  {Cary}(2017)}]{HigueraCary2017}%
  \BibitemOpen
  \bibfield  {author} {\bibinfo {author} {\bibfnamefont {A.~V.}\ \bibnamefont
  {Higuera}}\ and\ \bibinfo {author} {\bibfnamefont {J.~R.}\ \bibnamefont
  {Cary}},\ }\bibfield  {title} {\bibinfo {title} {Structure-preserving
  second-order integration of relativistic charged particle trajectories in
  electromagnetic fields},\ }\href {https://doi.org/10.1063/1.4979989}
  {\bibfield  {journal} {\bibinfo  {journal} {Phys. Plasmas}\ }\textbf
  {\bibinfo {volume} {24}},\ \bibinfo {pages} {052104} (\bibinfo {year}
  {2017})}\BibitemShut {NoStop}%
\bibitem [{\citenamefont {Reiche}(2000)}]{reiche2000numerical}%
  \BibitemOpen
  \bibfield  {author} {\bibinfo {author} {\bibfnamefont {S.}~\bibnamefont
  {Reiche}},\ }\emph {\bibinfo {title} {Numerical Studies for a Single Pass
  High Gain Free-Electron Laser}},\ \href@noop {} {\bibinfo {type} {Phd
  thesis}},\ \bibinfo  {school} {Universit{\"a}t Hamburg}, \bibinfo {address}
  {Hamburg, Germany} (\bibinfo {year} {2000}),\ \bibinfo {note} {dESY report
  no. DESY-THESIS-2000-012, 177 pages}\BibitemShut {NoStop}%
\bibitem [{\citenamefont {Pardal}\ \emph {et~al.}(2023)\citenamefont {Pardal},
  \citenamefont {Sainte-Marie}, \citenamefont {Reboul-Salze}, \citenamefont
  {Fonseca},\ and\ \citenamefont {Vieira}}]{pardal2023radio}%
  \BibitemOpen
  \bibfield  {author} {\bibinfo {author} {\bibfnamefont {M.}~\bibnamefont
  {Pardal}}, \bibinfo {author} {\bibfnamefont {A.}~\bibnamefont
  {Sainte-Marie}}, \bibinfo {author} {\bibfnamefont {A.}~\bibnamefont
  {Reboul-Salze}}, \bibinfo {author} {\bibfnamefont {R.~A.}\ \bibnamefont
  {Fonseca}},\ and\ \bibinfo {author} {\bibfnamefont {J.}~\bibnamefont
  {Vieira}},\ }\bibfield  {title} {\bibinfo {title} {Radio: an efficient
  spatiotemporal radiation diagnostic for particle-in-cell codes},\ }\href
  {https://doi.org/10.1016/j.cpc.2022.108634} {\bibfield  {journal} {\bibinfo
  {journal} {Comput. Phys. Commun.}\ }\textbf {\bibinfo {volume} {285}},\
  \bibinfo {pages} {108634} (\bibinfo {year} {2023})}\BibitemShut {NoStop}%
\bibitem [{\citenamefont {Pausch}\ \emph {et~al.}(2018)\citenamefont {Pausch},
  \citenamefont {Debus}, \citenamefont {Huebl}, \citenamefont {Schramm},
  \citenamefont {Steiniger}, \citenamefont {Widera},\ and\ \citenamefont
  {Bussmann}}]{pausch2018quantitatively}%
  \BibitemOpen
  \bibfield  {author} {\bibinfo {author} {\bibfnamefont {R.}~\bibnamefont
  {Pausch}}, \bibinfo {author} {\bibfnamefont {A.}~\bibnamefont {Debus}},
  \bibinfo {author} {\bibfnamefont {A.}~\bibnamefont {Huebl}}, \bibinfo
  {author} {\bibfnamefont {U.}~\bibnamefont {Schramm}}, \bibinfo {author}
  {\bibfnamefont {K.}~\bibnamefont {Steiniger}}, \bibinfo {author}
  {\bibfnamefont {R.}~\bibnamefont {Widera}},\ and\ \bibinfo {author}
  {\bibfnamefont {M.}~\bibnamefont {Bussmann}},\ }\bibfield  {title} {\bibinfo
  {title} {Quantitatively consistent computation of coherent and incoherent
  radiation in particle-in-cell codes---a general form factor formalism for
  macro-particles},\ }\href {https://doi.org/10.1016/j.nima.2018.02.020}
  {\bibfield  {journal} {\bibinfo  {journal} {Nucl. Instrum. Methods Phys. Res.
  A}\ }\textbf {\bibinfo {volume} {909}},\ \bibinfo {pages} {419} (\bibinfo
  {year} {2018})}\BibitemShut {NoStop}%
\bibitem [{\citenamefont {Yee}(1966)}]{Yee1966}%
  \BibitemOpen
  \bibfield  {author} {\bibinfo {author} {\bibfnamefont {K.~S.}\ \bibnamefont
  {Yee}},\ }\bibfield  {title} {\bibinfo {title} {Numerical solution of initial
  boundary value problems involving {Maxwell}'s equations in isotropic media},\
  }\href {https://doi.org/10.1109/TAP.1966.1138693} {\bibfield  {journal}
  {\bibinfo  {journal} {IEEE Trans. Antennas Propag.}\ }\textbf {\bibinfo
  {volume} {14}},\ \bibinfo {pages} {302} (\bibinfo {year} {1966})}\BibitemShut
  {NoStop}%
\bibitem [{\citenamefont {Lehe}\ \emph {et~al.}(2013)\citenamefont {Lehe},
  \citenamefont {Lifschitz}, \citenamefont {Thaury}, \citenamefont {Malka},\
  and\ \citenamefont {Davoine}}]{Lehe2013}%
  \BibitemOpen
  \bibfield  {author} {\bibinfo {author} {\bibfnamefont {R.}~\bibnamefont
  {Lehe}}, \bibinfo {author} {\bibfnamefont {A.}~\bibnamefont {Lifschitz}},
  \bibinfo {author} {\bibfnamefont {C.}~\bibnamefont {Thaury}}, \bibinfo
  {author} {\bibfnamefont {V.}~\bibnamefont {Malka}},\ and\ \bibinfo {author}
  {\bibfnamefont {X.}~\bibnamefont {Davoine}},\ }\bibfield  {title} {\bibinfo
  {title} {Numerical growth of emittance in simulations of laser-wakefield
  acceleration},\ }\href {https://doi.org/10.1103/PhysRevSTAB.16.021301}
  {\bibfield  {journal} {\bibinfo  {journal} {Phys. Rev. ST Accel. Beams}\
  }\textbf {\bibinfo {volume} {16}},\ \bibinfo {pages} {021301} (\bibinfo
  {year} {2013})}\BibitemShut {NoStop}%
\bibitem [{\citenamefont {Boris}(1970)}]{Boris1970}%
  \BibitemOpen
  \bibfield  {author} {\bibinfo {author} {\bibfnamefont {J.~P.}\ \bibnamefont
  {Boris}},\ }\bibfield  {title} {\bibinfo {title} {Relativistic plasma
  simulation---optimization of a hybrid code},\ }in\ \href@noop {} {\emph
  {\bibinfo {booktitle} {Proc. Fourth Conf. on Numerical Simulation of
  Plasmas}}},\ \bibinfo {editor} {edited by\ \bibinfo {editor} {\bibfnamefont
  {J.~P.}\ \bibnamefont {Boris}}\ and\ \bibinfo {editor} {\bibfnamefont
  {R.~A.}\ \bibnamefont {Shanny}}}\ (\bibinfo {address} {Naval Research
  Laboratory, Washington, D.C.},\ \bibinfo {year} {1970})\ pp.\ \bibinfo
  {pages} {3--67},\ \bibinfo {note} {nRL Report, available at
  \url{https://apps.dtic.mil/sti/citations/ADA023511}}\BibitemShut {NoStop}%
\bibitem [{\citenamefont {Vay}(2008)}]{Vay2008}%
  \BibitemOpen
  \bibfield  {author} {\bibinfo {author} {\bibfnamefont {J.-L.}\ \bibnamefont
  {Vay}},\ }\bibfield  {title} {\bibinfo {title} {Simulation of beams or
  plasmas crossing at relativistic velocity},\ }\href
  {https://doi.org/10.1063/1.2837054} {\bibfield  {journal} {\bibinfo
  {journal} {Phys. Plasmas}\ }\textbf {\bibinfo {volume} {15}},\ \bibinfo
  {pages} {056701} (\bibinfo {year} {2008})}\BibitemShut {NoStop}%
\bibitem [{\citenamefont {Marcus}\ \emph {et~al.}(2011)\citenamefont {Marcus},
  \citenamefont {Hemsing},\ and\ \citenamefont {Rosenzweig}}]{marcus2011gain}%
  \BibitemOpen
  \bibfield  {author} {\bibinfo {author} {\bibfnamefont {G.}~\bibnamefont
  {Marcus}}, \bibinfo {author} {\bibfnamefont {E.}~\bibnamefont {Hemsing}},\
  and\ \bibinfo {author} {\bibfnamefont {J.}~\bibnamefont {Rosenzweig}},\
  }\bibfield  {title} {\bibinfo {title} {Gain length fitting formula for
  free-electron lasers with strong space-charge effects},\ }\href
  {https://doi.org/10.1103/PhysRevSTAB.14.080702} {\bibfield  {journal}
  {\bibinfo  {journal} {Phys. Rev. ST Accel. Beams}\ }\textbf {\bibinfo
  {volume} {14}},\ \bibinfo {pages} {080702} (\bibinfo {year}
  {2011})}\BibitemShut {NoStop}%
\bibitem [{\citenamefont {Saldin}\ \emph {et~al.}(2004)\citenamefont {Saldin},
  \citenamefont {Schneidmiller},\ and\ \citenamefont
  {Yurkov}}]{saldin2004longitudinal}%
  \BibitemOpen
  \bibfield  {author} {\bibinfo {author} {\bibfnamefont {E.~L.}\ \bibnamefont
  {Saldin}}, \bibinfo {author} {\bibfnamefont {E.~A.}\ \bibnamefont
  {Schneidmiller}},\ and\ \bibinfo {author} {\bibfnamefont {M.~V.}\
  \bibnamefont {Yurkov}},\ }\bibfield  {title} {\bibinfo {title} {Longitudinal
  space charge-driven microbunching instability in the tesla test facility
  linac},\ }\href {https://doi.org/10.1016/j.nima.2004.04.067} {\bibfield
  {journal} {\bibinfo  {journal} {Nucl. Instrum. Methods Phys. Res. A}\
  }\textbf {\bibinfo {volume} {528}},\ \bibinfo {pages} {355} (\bibinfo {year}
  {2004})}\BibitemShut {NoStop}%
\bibitem [{\citenamefont {Venturini}(2008)}]{venturini2008models}%
  \BibitemOpen
  \bibfield  {author} {\bibinfo {author} {\bibfnamefont {M.}~\bibnamefont
  {Venturini}},\ }\bibfield  {title} {\bibinfo {title} {Models of longitudinal
  space-charge impedance for microbunching instability},\ }\href
  {https://doi.org/10.1103/PhysRevSTAB.11.034401} {\bibfield  {journal}
  {\bibinfo  {journal} {Phys. Rev. ST Accel. Beams}\ }\textbf {\bibinfo
  {volume} {11}},\ \bibinfo {pages} {034401} (\bibinfo {year}
  {2008})}\BibitemShut {NoStop}%
\bibitem [{\citenamefont {Wiedemann}(2015)}]{Wiedemann2015PAP}%
  \BibitemOpen
  \bibfield  {author} {\bibinfo {author} {\bibfnamefont {H.}~\bibnamefont
  {Wiedemann}},\ }\href {https://doi.org/10.1007/978-3-319-18317-6} {\emph
  {\bibinfo {title} {Particle Accelerator Physics}}},\ \bibinfo {edition}
  {4th}\ ed.\ (\bibinfo  {publisher} {Springer International Publishing},\
  \bibinfo {address} {Cham},\ \bibinfo {year} {2015})\BibitemShut {NoStop}%
\bibitem [{\citenamefont {Fusco}(2004)}]{fusco2004beam}%
  \BibitemOpen
  \bibfield  {author} {\bibinfo {author} {\bibfnamefont {V.}~\bibnamefont
  {Fusco}},\ }\emph {\bibinfo {title} {Beam dynamics and collective effects in
  {SPARC} project}},\ \href@noop {} {Ph.D. thesis},\ \bibinfo  {school}
  {Universit{\`a} degli Studi di Roma ``La Sapienza''} (\bibinfo {year}
  {2004})\BibitemShut {NoStop}%
\bibitem [{\citenamefont {Berestetskii}\ \emph {et~al.}(1982)\citenamefont
  {Berestetskii}, \citenamefont {Lifshitz},\ and\ \citenamefont
  {Pitaevskii}}]{BerestetskiiLifshitzPitaevskiiQED1982}%
  \BibitemOpen
  \bibfield  {author} {\bibinfo {author} {\bibfnamefont {V.~B.}\ \bibnamefont
  {Berestetskii}}, \bibinfo {author} {\bibfnamefont {E.~M.}\ \bibnamefont
  {Lifshitz}},\ and\ \bibinfo {author} {\bibfnamefont {L.~P.}\ \bibnamefont
  {Pitaevskii}},\ }\href@noop {} {\emph {\bibinfo {title} {Quantum
  Electrodynamics}}},\ \bibinfo {edition} {2nd}\ ed.,\ \bibinfo {series}
  {Course of Theoretical Physics}, Vol.~\bibinfo {volume} {4}\ (\bibinfo
  {publisher} {Butterworth-Heinemann},\ \bibinfo {year} {1982})\BibitemShut
  {NoStop}%
\bibitem [{\citenamefont {Gould}(1989)}]{Gould1989PsPlasma}%
  \BibitemOpen
  \bibfield  {author} {\bibinfo {author} {\bibfnamefont {R.~J.}\ \bibnamefont
  {Gould}},\ }\bibfield  {title} {\bibinfo {title} {Direct positron
  annihilation and positronium formation in thermal plasmas},\ }\href
  {https://doi.org/10.1086/167792} {\bibfield  {journal} {\bibinfo  {journal}
  {Astrophys. J.}\ }\textbf {\bibinfo {volume} {344}},\ \bibinfo {pages} {232}
  (\bibinfo {year} {1989})}\BibitemShut {NoStop}%
\bibitem [{\citenamefont {Agapov}\ and\ \citenamefont
  {Geloni}(2014)}]{agapov2014diffusion}%
  \BibitemOpen
  \bibfield  {author} {\bibinfo {author} {\bibfnamefont {I.}~\bibnamefont
  {Agapov}}\ and\ \bibinfo {author} {\bibfnamefont {G.}~\bibnamefont
  {Geloni}},\ }\bibfield  {title} {\bibinfo {title} {Diffusion effects in
  undulator radiation},\ }\href {https://doi.org/10.1103/PhysRevSTAB.17.110704}
  {\bibfield  {journal} {\bibinfo  {journal} {Phys. Rev. ST Accel. Beams}\
  }\textbf {\bibinfo {volume} {17}},\ \bibinfo {pages} {110704} (\bibinfo
  {year} {2014})}\BibitemShut {NoStop}%
\bibitem [{\citenamefont {Ritus}(1985)}]{ritusJSLR85}%
  \BibitemOpen
  \bibfield  {author} {\bibinfo {author} {\bibfnamefont {V.~I.}\ \bibnamefont
  {Ritus}},\ }\bibfield  {title} {\bibinfo {title} {Quantum effects of the
  interaction of elementary particles with an intense electromagnetic field},\
  }\href {https://doi.org/10.1007/BF01120220} {\bibfield  {journal} {\bibinfo
  {journal} {J. Russ. Laser Res.}\ }\textbf {\bibinfo {volume} {6}},\ \bibinfo
  {pages} {497} (\bibinfo {year} {1985})}\BibitemShut {NoStop}%
\bibitem [{\citenamefont {Baier}\ \emph {et~al.}(1998)\citenamefont {Baier},
  \citenamefont {Katkov},\ and\ \citenamefont {Strakhovenko}}]{Baier-book}%
  \BibitemOpen
  \bibfield  {author} {\bibinfo {author} {\bibfnamefont {V.~N.}\ \bibnamefont
  {Baier}}, \bibinfo {author} {\bibfnamefont {V.~M.}\ \bibnamefont {Katkov}},\
  and\ \bibinfo {author} {\bibfnamefont {V.~M.}\ \bibnamefont {Strakhovenko}},\
  }\href@noop {} {\emph {\bibinfo {title} {Electromagnetic Processes at High
  Energies in Oriented Single Crystals}}}\ (\bibinfo  {publisher} {World
  Scientific},\ \bibinfo {address} {Singapore},\ \bibinfo {year}
  {1998})\BibitemShut {NoStop}%
\bibitem [{\citenamefont {Landau}\ and\ \citenamefont
  {Lifshitz}(1975)}]{landaulifshitz_vol2}%
  \BibitemOpen
  \bibfield  {author} {\bibinfo {author} {\bibfnamefont {L.~D.}\ \bibnamefont
  {Landau}}\ and\ \bibinfo {author} {\bibfnamefont {E.~M.}\ \bibnamefont
  {Lifshitz}},\ }\href@noop {} {\emph {\bibinfo {title} {The Classical Theory
  of Fields}}},\ \bibinfo {edition} {2nd}\ ed.\ (\bibinfo  {publisher}
  {Elsevier},\ \bibinfo {address} {Oxford},\ \bibinfo {year}
  {1975})\BibitemShut {NoStop}%
\end{thebibliography}%

\sectionmajor{Acknowledgements}
\smallskip

\noindent
This article comprises parts of the PhD thesis work of \c{C}a\u{g}r{\i} Erciyes, to be submitted to the Heidelberg University, Germany. We thank Kevin Zink for his support in running large-scale simulations on the institute’s high-performance computing cluster.

\sectionmajor{Author Contributions}
\smallskip

\noindent
M.T. conceived the research. \c{C}.E. and M.T. developed the methodology and designed the theoretical framework. \c{C}.E. implemented the simulation software, performed the formal analysis and investigation, curated the data, and prepared all visualizations. \c{C}.E. and M.T. wrote the original draft. All authors contributed to review and editing of the manuscript. C.H.K. and M.T. provided supervision.

\sectionmajor{Competing interests}
\smallskip

\noindent
The authors declare no competing interests.

\end{document}